**New self-organized benzo[b]thiophene-based materials for GHz applications**

Agnieszka Mieczkowska, Jakub Herman, Natan Rychłowicz, Monika Zając and Piotr Harmata

Institute of Chemistry, Faculty of Advanced Technologies and Chemistry, Military University of Technology, Warsaw, Poland

Corresponding author: Piotr Harmata

Kaliskiego 2

00-908 Warsaw

Poland

piotr.harmata@wat.edu.pl

# New self-organized benzo[b]thiophene-based materials for GHz applications

Abstract

This research delves into the synthesis and characterization of novel liquid crystal compounds derived from benzo[b]lthiophene cores, focusing on their potential applications in microwave technology. Two synthetic strategies were developed to construct rigid cores, resulting in the successful synthesis of ten compounds with varied terminal groups and lateral substituents. Correlations between molecular structure and mesomorphic properties were elucidated through extensive comparative analysis. Birefringence measurements and quantum chemical calculations further provided insights into the optical properties and polarizability anisotropy of the synthesized compounds. The results highlight the influence of structural diversity on the compounds' suitability for microwave applications. Specifically, compounds featuring carbon-carbon triple bonds and polar terminal groups demonstrated enhanced birefringence and polarizability values, indicating their potential in microwave device fabrication. This study underscores the importance of molecular design in optimizing liquid crystal materials for advanced technological applications.

**Introduction**

The development of self-organizing materials can be divided into two main streams, which are closely related to their possible application. The first pertains to the range of visible light of electromagnetic radiation, which emerged due to the development of LC display (LCD) manufacturing technology[1-3]. Among the numerous potential applications, the rapid growth of the display industry has generated significant demand for new materials, leading to widespread advancements in research in this field. The spectrum of applications has expanded as more materials have been developed. It's noteworthy that the spectrum of applications expanded as more materials developed. The second stream of the development of liquid crystals concerns all applications beyond visible light (so-called beyond displays)[4], of which the most popular wavelengths are infrared (IR)[5-9], millimeter wave[10-15].

In recent years, there has been significant interest in using LCs in the microwave range[13,16-21]. Alternatives to semiconductor technology-based devices are being explored in this range. Devices operating in the mm-wave range using liquid crystal technology include Reconfigurable Intelligent Surfaces (RIS)[22-24], Reflectarray Antennas (RA)[25-27], Smart Antennas[21,28,29], and Metasurfaces[30-32]. The 10-60 GHz range seems to be particularly interesting. This frequency range is commonly used for various wireless communication applications, including point-to-point links, satellite communication, radar systems, and some wireless local area networks (WLANs). To operate in the 10-60 GHz range, liquid crystal cells need to be designed to accommodate the higher frequencies. The design should consider factors such as the thickness of the liquid crystal layer, the choice of materials, and the fabrication techniques to minimize signal loss and optimize performance in this frequency range. The main challenge therefore is the proper organic material selection. Manabe from Merck KGaA examined various classes of LC materials for use in GHz[19]. It has been defined that the key parameters determining the suitability of a liquid crystal are tunability $\tau$ (1) and dielectric losses $\tan\delta$ (2) measured in the GHz.

$$(1)\ \tau = \frac{\varepsilon_{\parallel} - \varepsilon_{\perp}}{\varepsilon_{\parallel}} \qquad (2)\ tan\delta_{\varepsilon} = \frac{\varepsilon''_{\parallel} - \varepsilon''_{\perp}}{\varepsilon'_{\parallel} - \varepsilon'_{\perp}}$$ where:

$\varepsilon_{\parallel}$ – the electric permittivity measured parallel to the optical axis of the molecule

$\varepsilon_{\perp}$ – the electric permittivity measured perpendicular to the optical axis of the molecule

$\varepsilon'$ – the real part of the electric permittivity

$\varepsilon''$ – the imaginary part of the electric permittivity

From an application standpoint, maximizing tunability is desired, a parameter often easily predictable due to its correlation with birefringence. However, minimizing dielectric losses proves challenging, given the lack of a clear relationship between molecular structure and loss tangent. Considering the above, promising compounds for GHz applications tend to have long conjugated structures, exemplified by derivatives of bistolane[33], as confirmed by Manabe. Notably, materials utilized in GHz applications so far are compounds initially developed for the visible spectrum. This lack of dedicated materials directly impacts application possibilities, hindering the exploitation of the significant potential inherent in self-organizing materials.

The design of molecules exhibiting high birefringence is well-established , with our research contributing significantly to these studies[34-40]. Worldwide, thousands of compounds have been designed and synthesized using various cores, aimed at enhancing not only birefringence but also other selected parameters. However, these studies have predominantly focused on improving material parameters for devices operating at the visible light spectrum. In this spectral range, it has become obvious that materials cannot be coloured, thus limiting the scope for using alternative cores capable of further enhancing birefringence. One such core, benzo[b]thiophene[41-43], appears promising, despite its relative lack of popularity in liquid crystal area so far. Notably, this core finds extensive use in medicinal chemistry[44], contributing to the synthesis of antibiotics[45], anti-inflammatory drugs, analgesics, anticancer agents[46], antidiabetic medications, and even treatments for neurological conditions such as Parkinson's[47], Alzheimer's[48].

Preliminary studies focused on molecular modelling of the designed structures, enabling the comparison of polarizability anisotropy. This factor, deemed important and theoretically predictable, allowed for evaluating both previously used and known structures (black) and those proposed by us (blue and green).

Table 1. The influence of the core structure on the polarizability anisotropy

| Core structure | Polarizability anisotropy Δα = [au] |
|---|---|
| 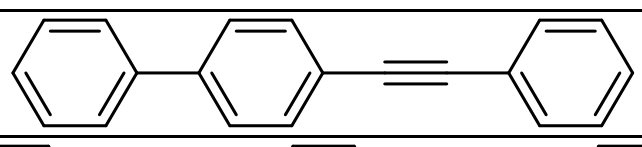 | 354 |
| 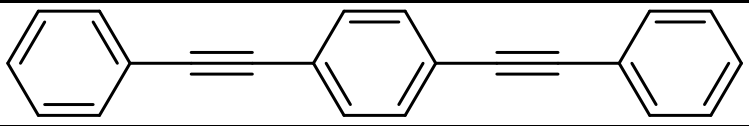 | 528 |
| 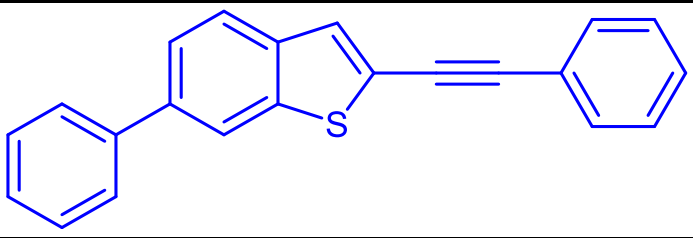 | 463 |
| 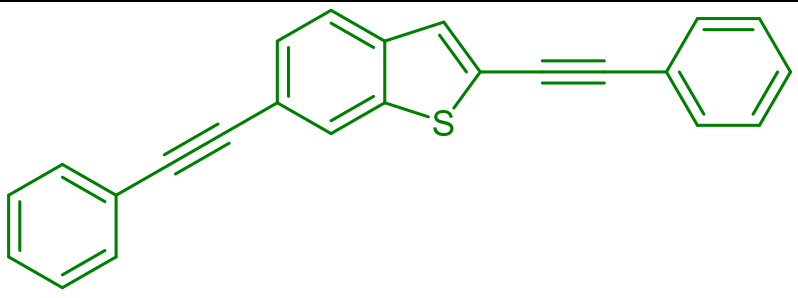 | 642 |

The calculations indicate that the designed core will allow to obtain compounds with even higher birefringence values, which should improve the previously described key parameters for GHz applications. The selection of this core was not random. Above all, we wanted to maintain the most linear molecule possible, as is the case with classical cores, where rings or their linkers consistently occupy 1,4 positions (bistolanes or phenyltolanes)[43]. Therefore, we decided to functionalize the 2,6 rather than 2,5 position of benzo[b]thiophene, as 2,5 derivatives give a more bent structure, affecting mesogenity and birefringence[41]. Additionally, combining benzene and thiophene into a condensed ring allows for greater electron conjugation than when these rings are connected by a bond (phenylthiophene). The selection of the heteroatom in the ring was also analyzed. Oxygen or sulfur atoms were considered. However, thiophene derivatives are much more stable than furan, and sulfur atoms have greater polarizability than oxygen. Given the extensive pi-electron system and the inclusion of a sulfur atom, it was expected that the resulting compounds would exhibit absorption within the visible spectrum. However, this characteristic does not limit their use in the GHz.

## 1. Materials

6-bromobenzo[b]thiophene was purchased from Angene International Limited (Nanjing, China) and used as received. 1-bromo-3-fluoro-4-(trifluoromethoxy)benzene, 1-bromo-2-fluoro-4-(trifluorometoksy)benzene, 2-fluoro-4-iodobenzonitrile, 4-iodoaniline, 2,6-difluoro-4-iodoaniline were purchased from Trimen Chemicals (Lodz, Poland), Magnesium for Grignard reactions (turnings), 1,8-Diazabicyclo(5.4.0)undec-7-ene (DBU), n-Butyllithium solution in hexane (C=2.5M), were purchased from Acros-Organics, (Geel, Belgium) and used as received. Toluene, heptane, hexane, acetonitrile, chloroform, butanone, acetone, anhydrous ethanol, methanol, ethanol, 1,3-propanediol, dichloromethane, Iodine, anhydrous caesium carbonate, calcium carbonate, copper(I) iodide, diisopropylamine, sodium sulfite, hydrochloric acid concentrate were purchased from Avantor Performance Materials Poland S.A (Gliwice, Poland) and used as received. Palladium(II) acetate, XPhos, bis(triphenylphosphine)palladium(II) chloride, trimethylsilylacetylene, tripropyl borate, sodium hydride (60 % dispersion in mineral oil), thiophosgene were purchased from Sigma-Aldrich Sp. z.o.o (Poznan, Poland) and used as received. Triethylamine, magnesium sulfate, 2-methyl-but-3-yn-2-ol were purchased from Chempur (Piekary Slaskie, Poland) and used as received. THF was distilled from sodium/benzophenone under nitrogen atmosphere prior to use. 2-(4-pentylphenyl)-1,3,2-dioxaborate, 4-propylphenylboronic acid, 2-[3-fluoro-4-(trifluoromethoxy)phenyl]-1,3,2-dioxaborinan, 2-(3,5-difluoro-4-(trifluoromethoxy)phenyl)-1,3,2-dioxaborinane were synthesised in our laboratory and all the details of the synthesis and purification are gathered in work[6]. 4-ethynyl-2-fluoro-1-(trifluoromethoxy)benzene was synthesised in our laboratory, and all the details of the synthesis and purification are gathered in the work[7]. 2-(3-fluoro-4-cyanophenyl)-1,3,2-dioxaborinane, 4-(1,3,2-dioxaborinan-2-yl)-2,6-difluorobenzonitrile were synthesised in our laboratory but the synthesis of these compounds has never been published.

## 2. Synthesis

In this work, ten new materials were synthesized and characterized. Compounds can be divided into two main groups. The first are compounds where 2-phenylbenzo[b]thiophene was used as the core (Group I). The second group are compounds where 2-(phenylethynyl)benzo[b]thiophene (Group II) was used as the core. The compounds in the groups differ in: the type of terminal groups, and the amount and position of the lateral substituent (fluorine atom). Due to their structure, it was expected that the synthesized compounds would exhibit liquid crystalline properties. Detailed synthetic procedures and spectra are included in the Supplementary Information.

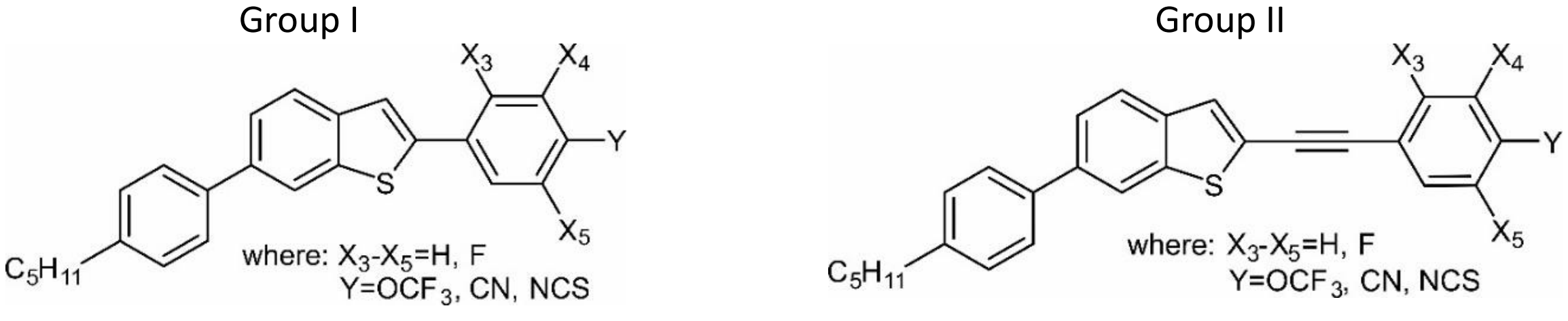

Figure 1 General structure of synthesized compounds

### 3.1 Synthetic strategy

Due to the structure of benzo[b]thiophene and its possible functionalization, two synthetic strategies for building a rigid core were developed (Figure 2). The proposed routes do not cover different types of lateral substitutions but focus on possible approaches to building a core of LC. Variant I was based on the use of the so-called "slow core growth", used primarily for the synthesis of unsymmetrically terminated compounds. In the first step, the Suzuki-Miyaura cross-coupling reaction is carried out. In the second stage, the core was functionalized with proper halogen. The selection of intermediates was carried out in such a way that the aryl halides are formed from substrates with a higher number of rings, with the simultaneous use of mono-ring phenylboronic or terminal acetylene derivatives. In this way, the possibility of emerging homo-coupling side reaction impurities was minimized. This was followed by another Suzuki-Miyaura coupling reaction (variant IA) or a Sonogashira coupling reaction (variant IB).

Variant II was designed to emphasise unsymmetric compounds but allowed the synthesis of both symmetrically and unsymmetrically ended compounds. In the first stage, the core functionalization reaction was carried out to obtain a bifunctional derivative, which allowed the synthesis of symmetrical compounds. However, at this stage, difficulties were encountered in selectively introducing the halogen atom. In the next step, a selective Suzuki-Miyaura coupling reaction was carried out. In the final stage, as in Variant I, a Suzuki-Miyaura (Variant IIA) or Sonogashira (IIB) protocol was planned to be carried out. However, this synthetic approach was ultimately not used due to the difficulties mentioned earlier. It is necessary to optimize the conditions of the reaction in which the benzotiophene unit is functionalized.

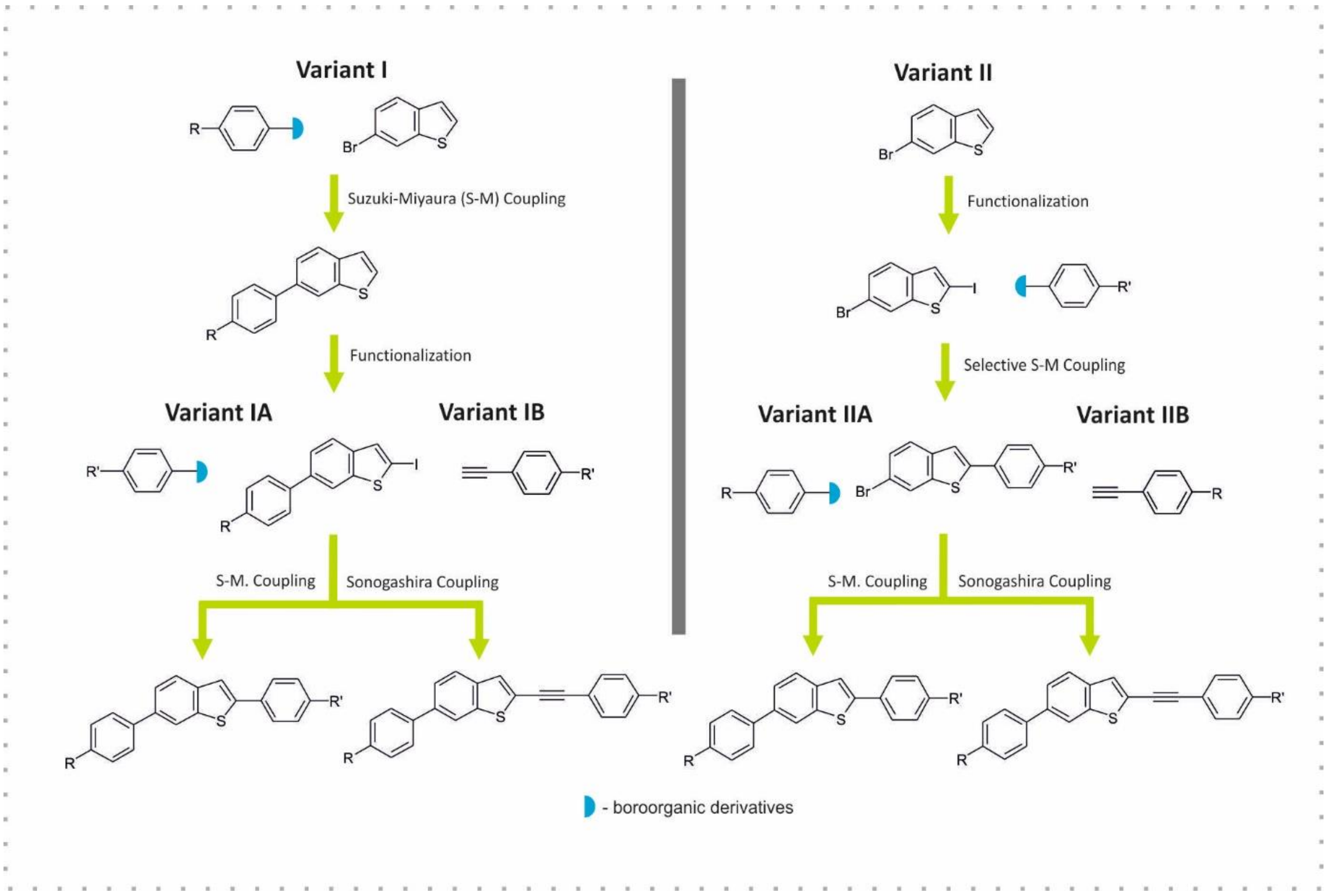

Figure 2. Variants for the synthesis of liquid crystal benzo[b]thiophene derivatives.

### 3.2 The synthesis of Group I materials

The first stage of synthesis (Figure 3) is the Suzuki-Miyaura coupling reaction involving commercially available 6-bromobenzotiophene **1** and 2-(4-pentylphenyl)-1,3,2-dioxaborinane **2** at the 6-position of the benzo[b]thiophene core. The resulting intermediate **3** was then subjected to halide-functionalization. For this purpose, an ortho-directed metalation reaction was carried out using Lithium diisopropylamide (LDA) followed by iodine treatment in dry THF at -78°C. The compound **4** obtained in this way participated in all subsequent coupling reactions. The Suzuki-Miyaura coupling reactions were performed using either the appropriate esters or boronic acids **5a** – **5e**, resulting in the compounds **6a** - **6e**.

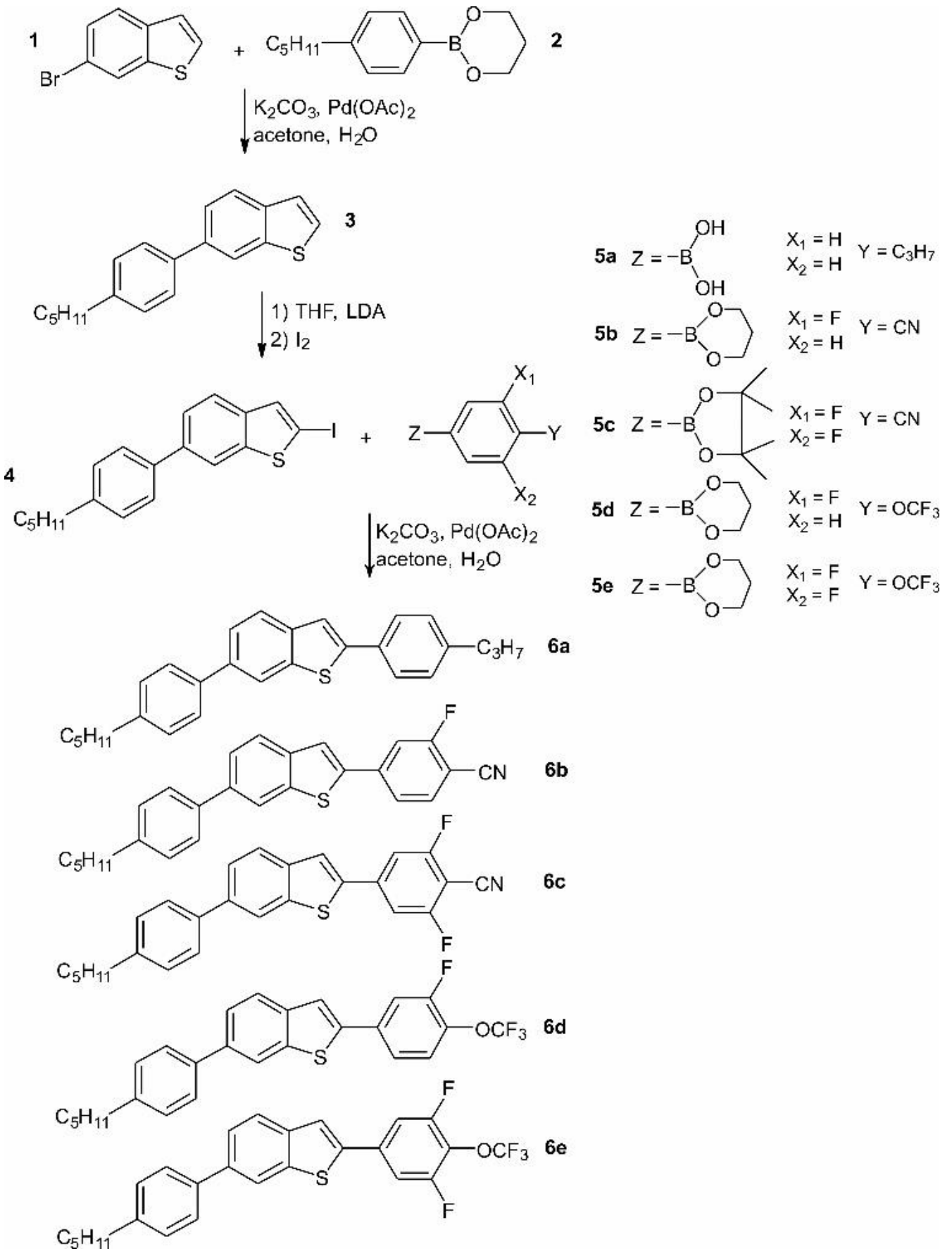

Figure 3. Synthesis of final compounds from group I

### 3.3 The synthesis of compounds from the Group II

The synthesis of the final compounds from group II is shown in Figures 4 and 5. A different approach in the synthesis resulted from the different methods of obtaining terminal acetylene derivatives. The choice of the method depended on the stability and sensitivity of the terminal group under the conditions of carbon carbon triple bond deprotection. In the first step of this synthesis (Figure 4), a mono-protected acetylene derivative is introduced into the aromatic ring. In the case of synthesized compounds, the source of the triple bond is 2-methyl-but-3-yn-2-ol, which is found to be a suitable choice among commercially available mono-protected acetylene derivatives. The second step is deprotection of the obtained derivative by the hydrolysis reaction using a catalytic amount of a base, in this case sodium hydride. After obtaining terminal acetylenes **10a** and **10b**, the previously synthesized 6-(4-pentylphenyl)-2-iodobenzo[b]thiophene was used to obtain the final compounds **11a** and **11b**. In the second method, ethynyltrimethylsilane was used as the source of the triple bond unit. The main advantage of this method is the mild conditions for the deprotection of terminal acetylene, which is carried out at room temperature. In the first stage of the synthesis (Figure 5), the Sonogashira coupling reaction was carried out between the appropriate iodo derivative **13a** - **13c** and ethynyltrimethylsilane **12**. The obtained trimethyl(phenylethynyl)silane derivatives **14a** - **14c** were subjected to deprotection. Three terminal acetylenes **15a** - **15c** were obtained. Amine derivatives (**15b** and **15c**) are sensitive to visible light, therefore standard solutions were prepared and stored in the dark under refrigeration (podaj temp). In the next step, the obtained terminal acetylenes and the previously synthesized intermediate **4** were used and coupled. This allowed to obtain the final compound **16** and two derivatives of aromatic amines **17a** and **17b**. Due to the instability at high temperatures and light sensitivity, reactions involving aniline derivatives were carried out at room temperature or not exceeding 40°C. The amino terminated compounds **17a**, **17b** were further converted into corresponding isothiocyanates **18a**, **18b** (NCS group), using $CSCl_2$ and $CaCO_3$.

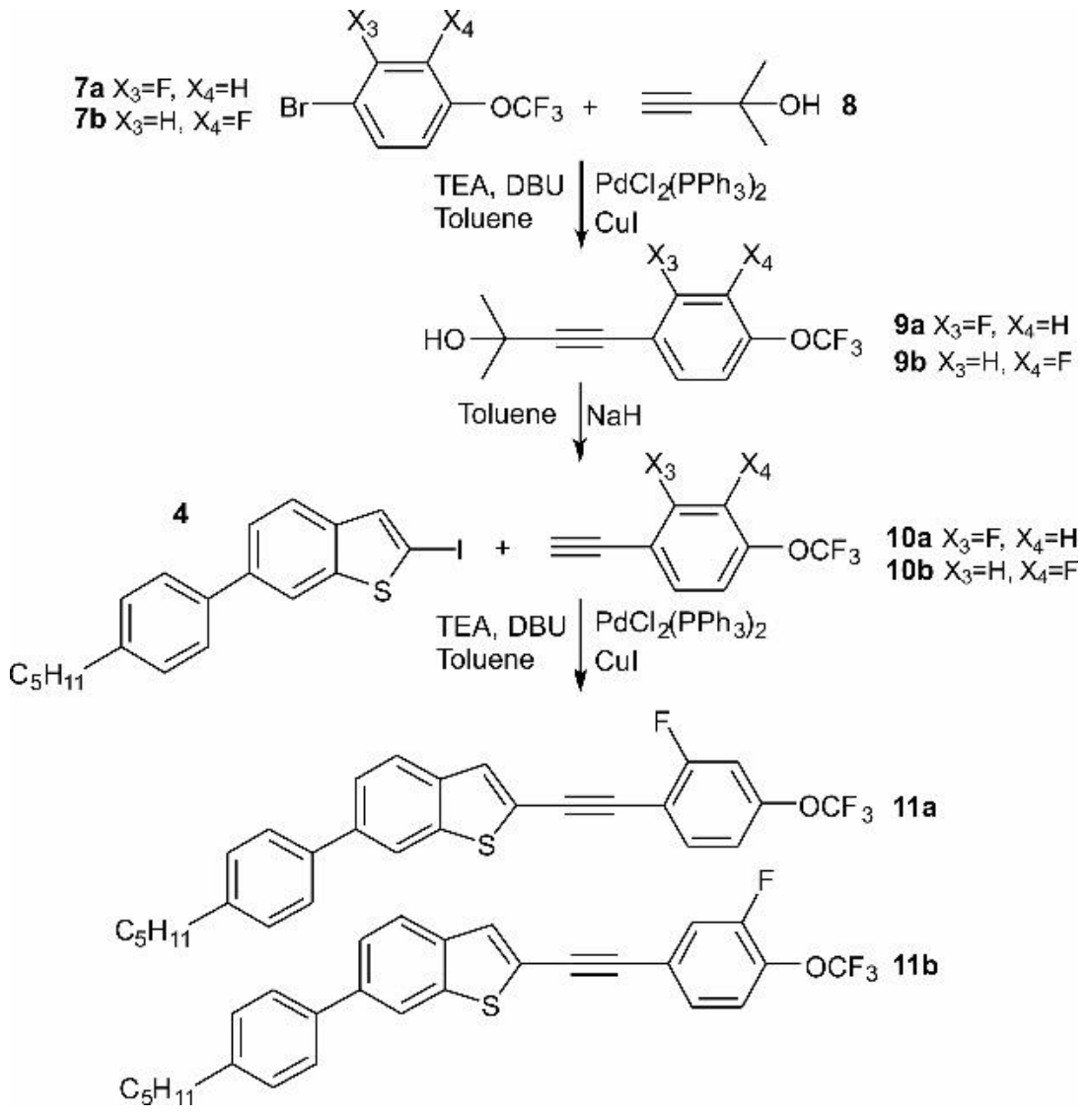

Figure 4. Synthesis of final compounds from group II terminally substituted with trifluoromethoxy group

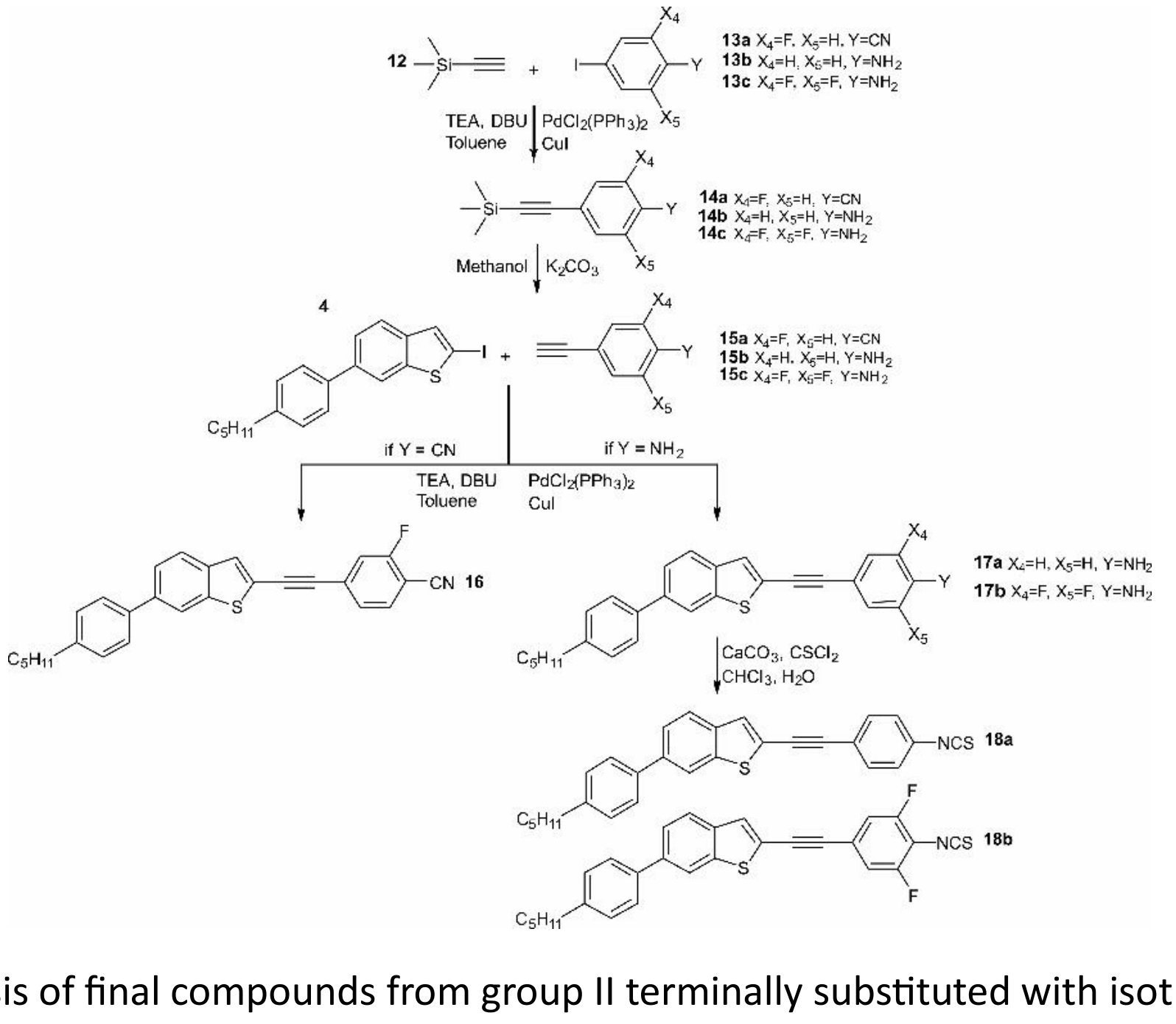

Figure 5. Synthesis of final compounds from group II terminally substituted with isothiocynate group

## 3. Characterization and measurements – analytical instrumentation

Synthesis progress and purity of synthesised compounds were determined using SHIMADZU GCMS-QP2010S (Shimadzu, Kyoto, Japan) series gas chromatograph equipped with quadrupole mass analyser MS(EI), high-performance liquid chromatography HPLC-PDA-MS (APCI-ESI dual source) Shimadzu LCMS 2010 EV (Shimadzu, Kyoto, Japan) equipped with a polychromatic UV–VIS detector (Shimadzu, Kyoto, Japan). and by thin layer chromatography (silica gel on aluminum). Proton (1H) and carbon (13C) nuclear magnetic resonance (NMR) spectra in CDCl3 were collected using a Bruker, model Avance III spectrometer (Bruker, Billerica, MA, USA). The phase transition temperatures and enthalpy data were determined by polarising optical microscopy (POM) with an OLYMPUS BX51(Olympus, Shinjuku, Tokyo, Japan) equipped with a Linkam hot stage THMS-600 (Linkam Scientific Instruments Ltd., Tadworth, United Kingdom) and differential scanning calorimeter SETARAM DSC 141 (KEP Technologies Group's DNA, Montauban, France) during heating/cooling cycles (with rate 2°/min). Refractive indices of the multicomponent nematic mixtures were measured by The Metricon Model 2010/M Prism Coupler (Metricon Corporation, Pennington, NJ, USA) equipped with 443 nm, 636 nm, and 1550 nm lasers. Samples of liquid crystals were placed on Kapton® MT polyimide film (DuPont, Wilmington, DE, USA). Ordinary refractive index no and extraordinary refractive index ne were measured separately using different polarization of incident beams. Samples were measured at room temperatures (25 °C).

## 4. Mesomorphic properties and discussion

Temperatures and enthalpies of phase transitions of synthesized compounds are listed in Table 1

Table 2 Temperatures and enthalpies of phase transitions for synthesized compounds, taken from the first heating.

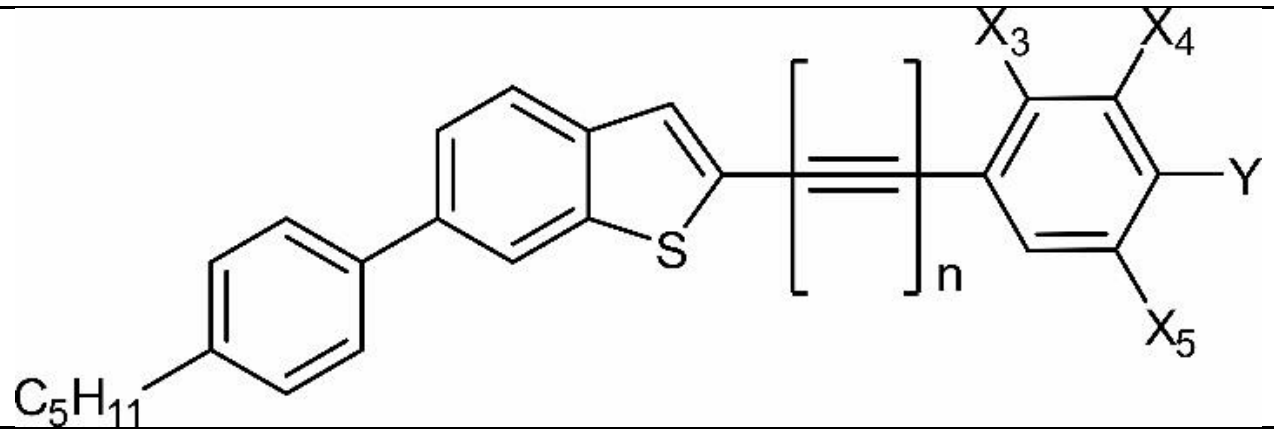

| Compund number | n | $X_3$ | $X_4$ | $X_5$ | Y | Temperatures and enthalpies of phase transitions | | | | | | | | | | | | | | | |
|---|---|---|---|---|---|---|---|---|---|---|---|---|---|---|---|---|---|---|---|---|---|
| | | | | | | Cr I | [°C] | [kJ/mol] | Cr II | [°C] | [kJ/mol] | SmC | [°C] | [kJ/mol] | SmA | [°C] | [kJ/mol] | N | [°C] | [kJ/mol] | Iso |
| **6a** | 0 | H | H | H | $C_3H_7$ | * | 162.3 | 9.85 | - | - | - | - | - | - | * | 210.4 | 1.77 | * | 223.4 | 0.53 | * |
| **6b** | 0 | F | H | H | CN | * | 110.7 | 11.98 | * | 113.6 | 12.36 | - | - | - | | - | - | * | 228.2 | 0.26 | * |
| **6c** | 0 | F | F | H | CN | * | 138.4 | 24.85 | - | - | - | - | - | - | | - | - | * | 175.6 | 0.10 | * |
| **6d** | 0 | F | H | H | $OCF_3$ | * | 140.6 | 12.50 | - | - | - | - | - | - | * | 223.6 | 5.86 | | - | - | * |
| **6e** | 0 | F | F | H | $OCF_3$ | * | 113.6 | 16.86 | - | - | - | - | - | - | * | 179.0 | 3.88 | | - | - | * |
| **11a** | 1 | F | H | H | $OCF_3$ | * | 64.7 | 18.27 | - | - | - | * | 70.4 | 6.58 | * | 204.4 | 4.59 | | - | - | * |
| **11b** | 1 | H | H | F | $OCF_3$ | * | 57.7 | 17.51 | - | - | - | * | 74.4 | 7.82 | * | 214.6 | 5.28 | | - | - | * |
| **16** | 1 | F | H | H | CN | * | 120.4 | 26.70 | - | - | - | - | - | - | | - | - | * | 224.3 | 0.37 | * |
| **18a** | 1 | H | H | H | NCS | * | 138.8 | 30.61 | - | - | - | - | - | - | * | 194.2 | 0.057 | * | 250.2 | 0.14 | * |
| **18b** | 1 | F | F | H | NCS | * | 96.2 | 22.65 | - | - | - | - | - | - | | - | - | * | 219.0 | 0.35 | * |

As a result of conducted research, 10 compounds with a 2,6-benzo[b]thiophene core were obtained. All the final compounds had a 4-pentylphenyl group at position 6 of the core but differed in the functional group substituted at position 2 of the core. Introducing different functional groups aimed to examine the physicochemical differences of the obtained products and conduct a comparative analysis. The intention was to find the following correlations that affect mesomorphic properties:

• Influence of the terminal substituent

• Influence of the number of lateral substituents

• Influence of the location of lateral substituents

• Core structure - shape anisotropy.

To better illustrate the occurring correlations, the results are presented graphically in the Figures 6, 7, 8 and 9. Figure 6 shows the influence of the number and position of fluorine on the mesomorphic properties for selected compounds.

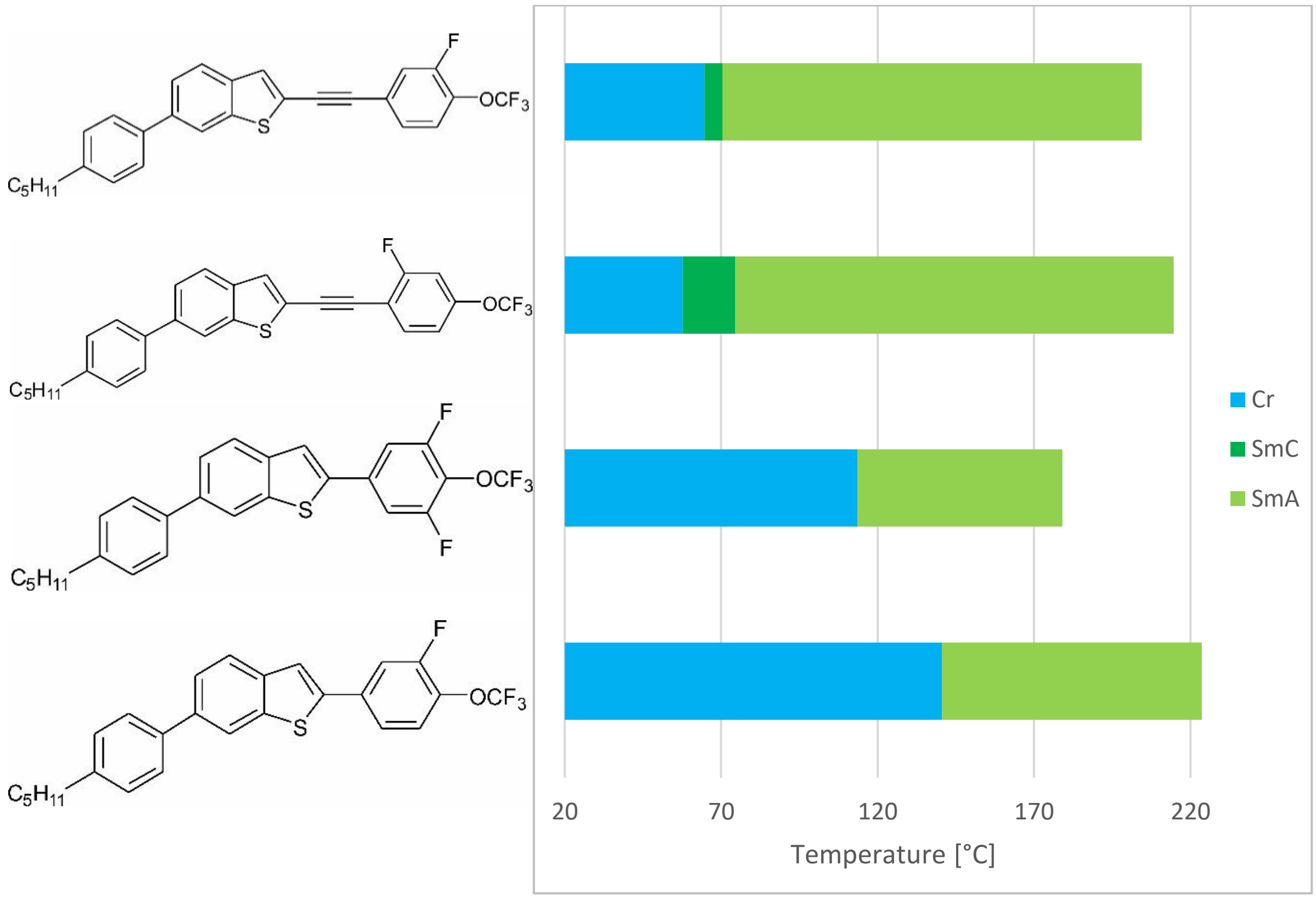

Figure 6. Mesomorphic properties of compounds containing the $OCF_3$ terminal group

Figure 6 presents a comparison of all obtained compounds containing a terminal trifluoromethoxy ($OCF_3$) group. All compounds exhibit smectic phases (SmA and SmC). Compounds **6d** and **6e** were compared in terms of the quantity of lateral substituents, and it can be observed that the introduction of an additional fluorine atom lowers the melting and clearing temperature but does not affect the type of phase sequence. Next, the properties of compounds **6d** and **11b** were examined regarding the presence of an ethynyl linkage in the molecular core. Its presence significantly lowered the melting temperature, broadened the temperature range of the highly ordered smectic phase SmA, and resulted in the appearance of a narrow-temperature-range smectic phase SmC. Subsequently, compounds **11a** and **11b** were compared to determine the influence of the position of the lateral substituent on the mesomorphic properties. Upon comparing both products, it can be concluded that the position of the lateral substituent does not significantly affect the melting and clearing temperatures, as well as the type of mesomorphic phases present. Due to the fact that compounds with a terminal $OCF_3$ group do not exhibit the desired nematic phase, this precludes their utilization as components of mixtures.

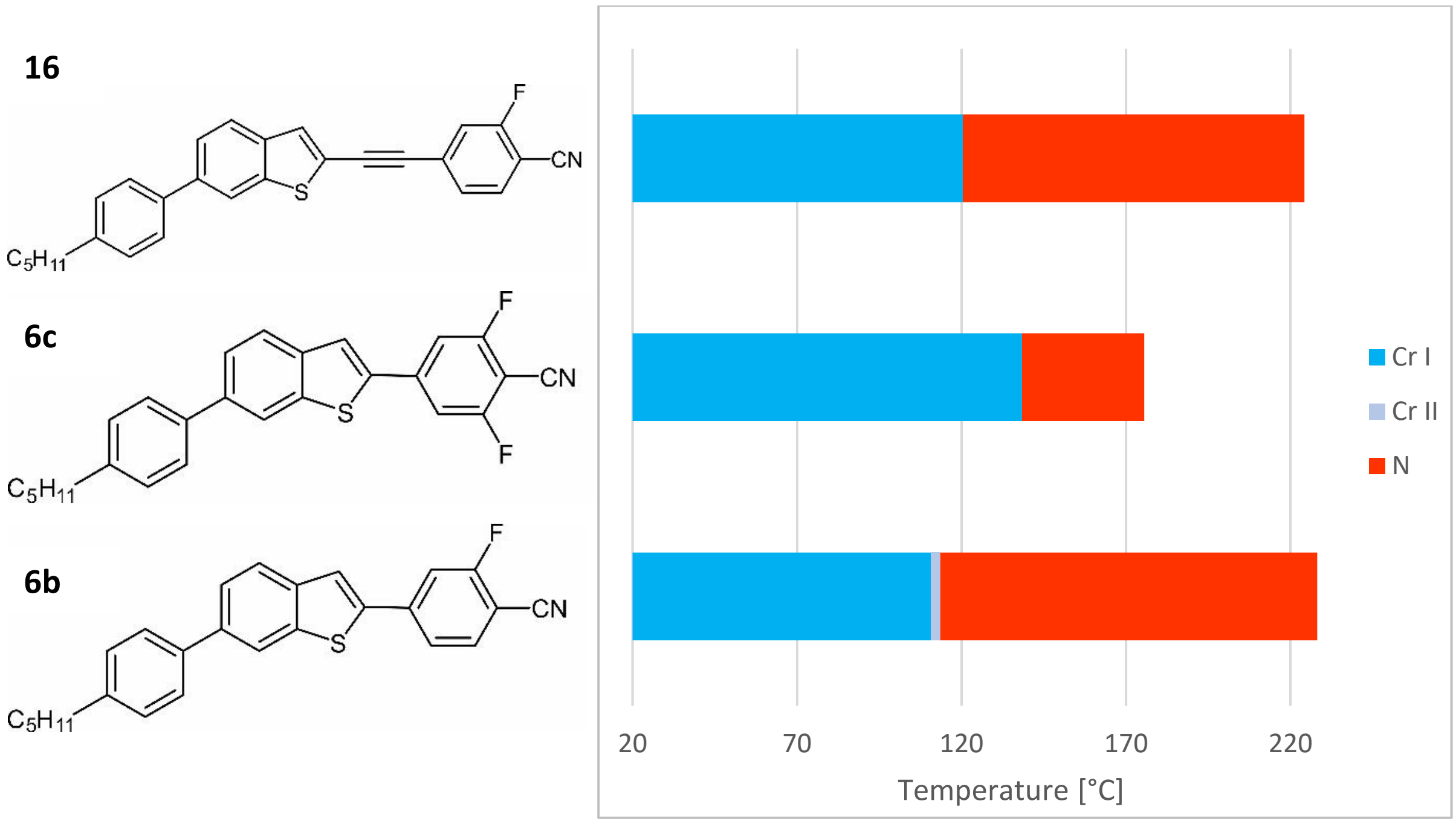

Figure 7. Mesomorphic properties of compounds containing the CN terminal group

For further comparative analysis of the final compounds, Figure 7 presents a compilation of all obtained products with a terminal cyano (CN) group. Analyzing structures **6b** and **6c** in terms of the quantity of lateral substituents, it can be observed that the introduction of an additional fluorine atom leads to an increase in the melting temperature, a decrease in the clearing temperature, and a narrowing of the nematic phase temperature range. Compound **6b**, unlike the others, exhibits a crystal-to-crystal transition. Comparing the final products **6b** and **8c** based on the core structure, specifically the presence of an ethynyl linkage, it does not introduce significant changes in temperature range of the mesophase nor the transition temperatures itself. The elongation of the core was expected to increase the temperature range of the nematic phase.

**18b**

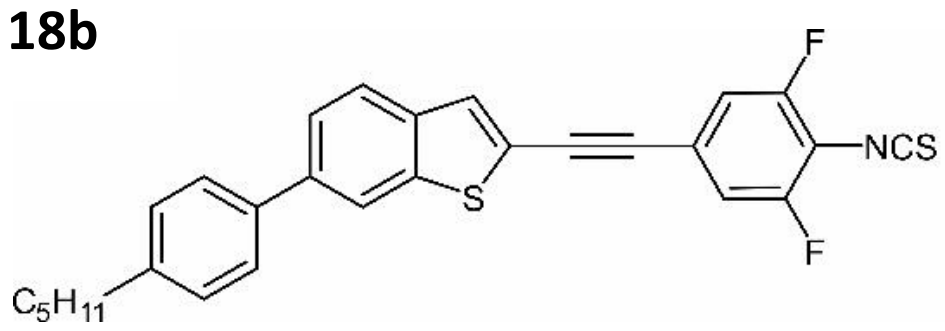

**18a**

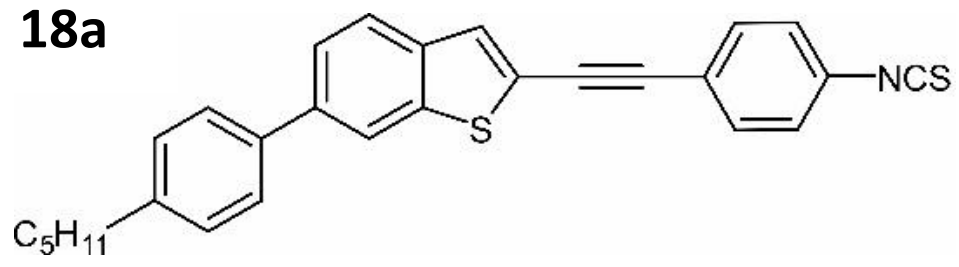

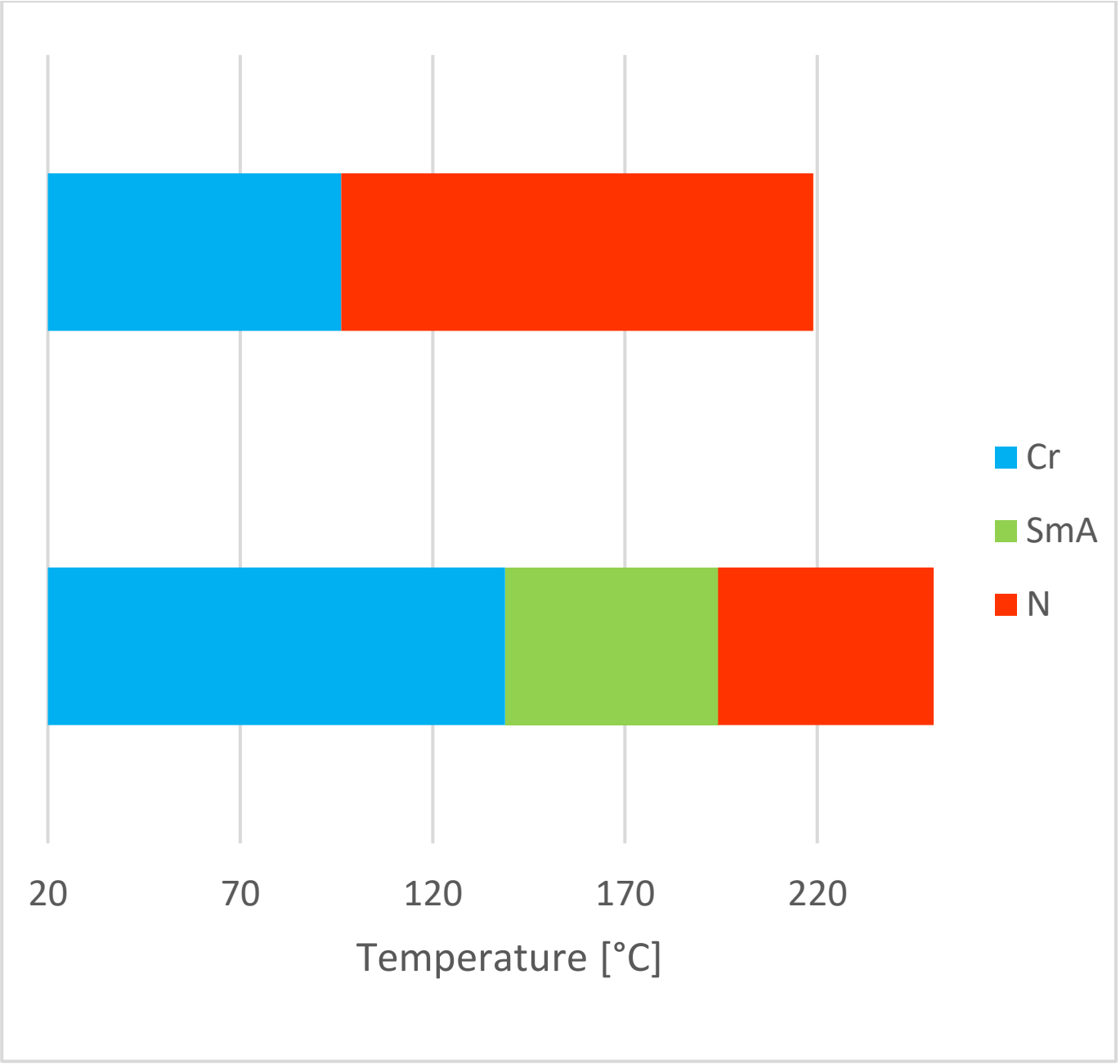

Figure 8. Mesomorphic properties of compounds containing the NCS terminal group

The next group of compounds contain an NCS group; their mesomorphic behavior is presented in Figure 8. Comparing the influence of the quantity of lateral substituents in compounds **18a** and **18b**, it can be concluded that the introduction of two fluorine atoms decreases both the melting and clearing temperatures, destabilising the smectic A phase, with simultaneous significant broadening of the nematic phase. Compound **18b**, in terms of mesomorphic properties, is the best among the obtained compounds and could be used as a base component for formulating a nematic mixture for applications in microwave technology.

**16**

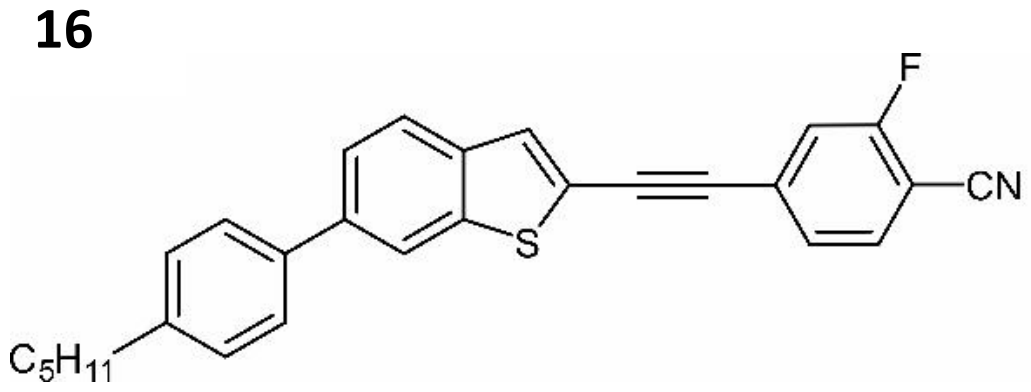

**11b**

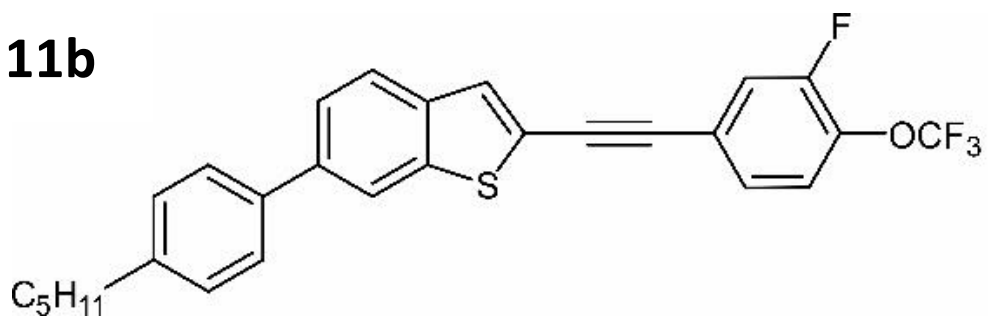

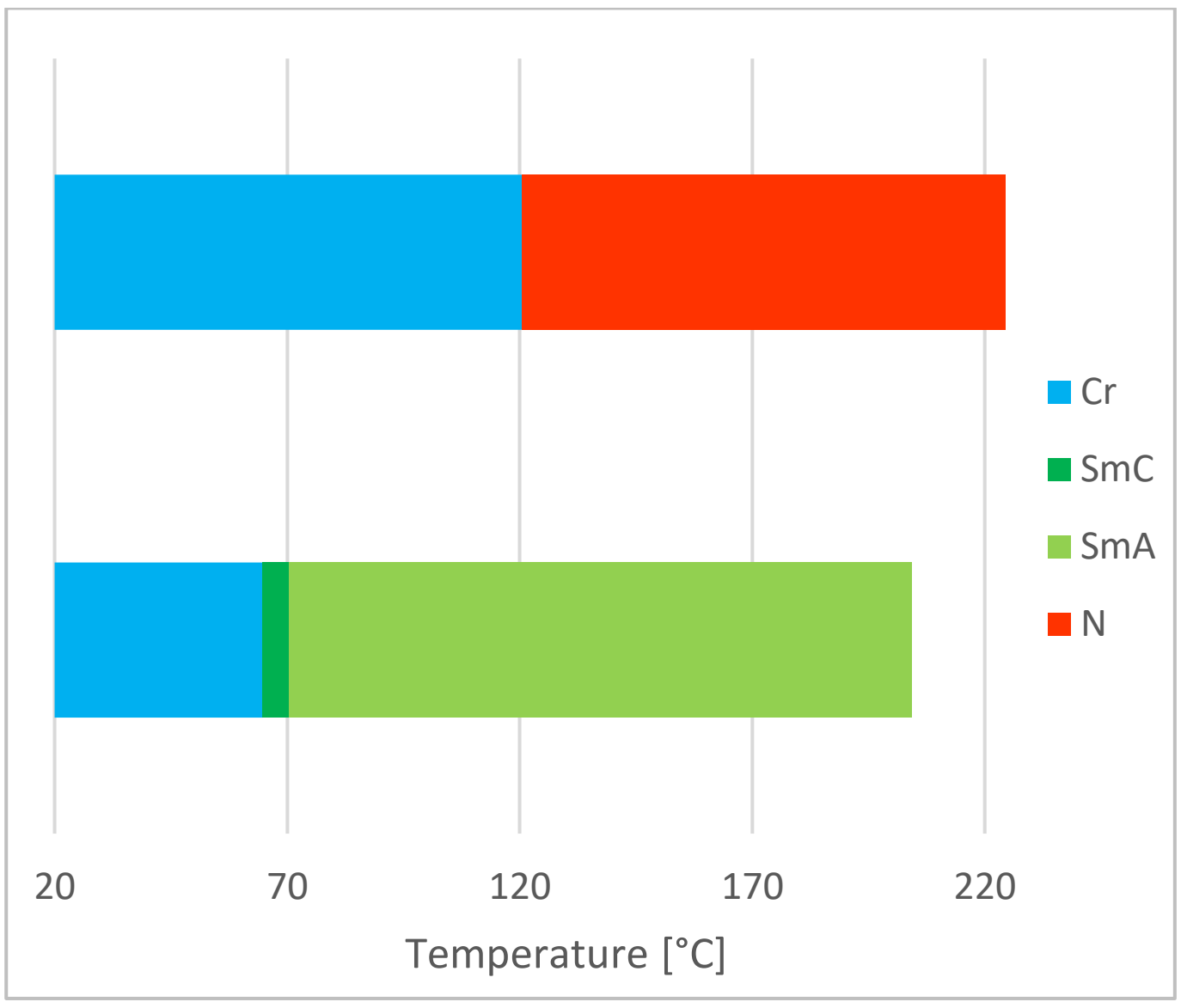

Figure 9. Influence of the terminal group on mesomorphic properties

Among all the compounds, it is also possible to compare those with an identical core structure but differing in the terminal group. Compounds **11b** and **16** were analyzed, and from Figure 9, it can be inferred that the introduction of a cyano (CN) group into the core results in a significant increase in the melting temperature, a slight increase in the clearing temperature, and total change of mesomorphism observed, i.e. destabilization of the highly ordered smectic phases. Attempts were made to synthesize a compound with such a core but terminated with an NCS group. Despite performing two syntheses under the same conditions as for compounds **18a** and **18b**, the desired product could not be obtained. Terminal acetylene appears to be decomposed during the reaction.

## 5. Birefringence measurements

Birefringence measurements were conducted on three selected compounds, each incorporated into low-birefringent nematic material (presented in Table 2) at compositions ranging from 2% to 10% weight percentage. The values of birefringence for the pure liquid crystalline compounds were determined by extrapolation from the equation (3):

$$(\Delta n)gh = x(\Delta n)g + (1-x)(\Delta n)h \qquad (3)$$

where Δn is the birefringence value:
gh- of the test mixture,
g- the base mixture,
h- the test compound,
and x is the concentration of the test compound added to the base mixture given in molar fractions.

The nematic base is a ternary mixture of compounds from the 4n-alkyloxyphenyl trans-4n-cyclohexylcarboxylate family—see Table 3. It has the following properties: isotropisation temperature $T_{N\text{-}Izo}$ = 63.6°C, melting point Tm < -20°C, viscosity η = 21.2 mPa*s and birefringence Δn = 0.07 measured at T = 22°C for 636nm. Table 4 shows the values of birefringence for the tested structures.

Table 3. Composition of nematic base mixture.

| Compound | n | wt % | Properties |
|---|---|---|---|
| 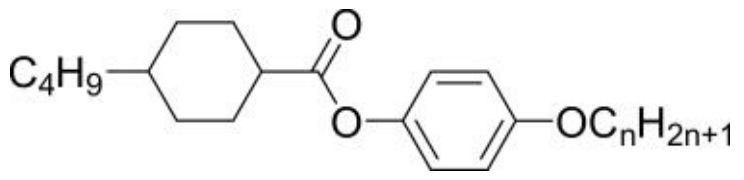 | 1 | 32 | $T_{N\text{-}Iso}$ = 63.6 °C<br>$T_{Cr\text{-}N}$ < −20 °C<br>η = 21.2 mPa*s<br>Δn = 0.07; 20 °C (636 nm)<br>Δε = −1.3; 20 °C (1 kHz) |
| | 2 | 31 | |
| | 5 | 37 | |

Table 4. Birefringence values extrapolated from low birefringent nematic host for three compounds **6a**, **16** and **18b** at wavelengths - 443nm, 636nm and 1550nm at 25°C.

| Compound | Δn | | |
|---|---|---|---|
| | 443nm | 636nm | 1550nm |
| **6b** | 0.6057 | 0.3431 | 0.2516 |
| **16** | 0.6788 | 0.4091 | 0.2834 |
| **18b** | 0.7946 | 0.4612 | 0.3825 |

With the values of the birefringence for any three wavelengths, using the methods of mathematical model fitting to the experimental results, we determined the values of the three coefficients of the Cauchy equation $A_{e,o}$, $B_{e,o}$ and $C_{e,o}$ (4):

$$n_{e,o} = A_{e,o} + \frac{B_{e,o}}{\lambda^2} + \frac{C_{e,o}}{\lambda^4} \quad (4)$$

Having Cauchy coefficients, we further calculated the birefringence dispersion in a large range of the electromagnetic spectrum – See Figure 10.

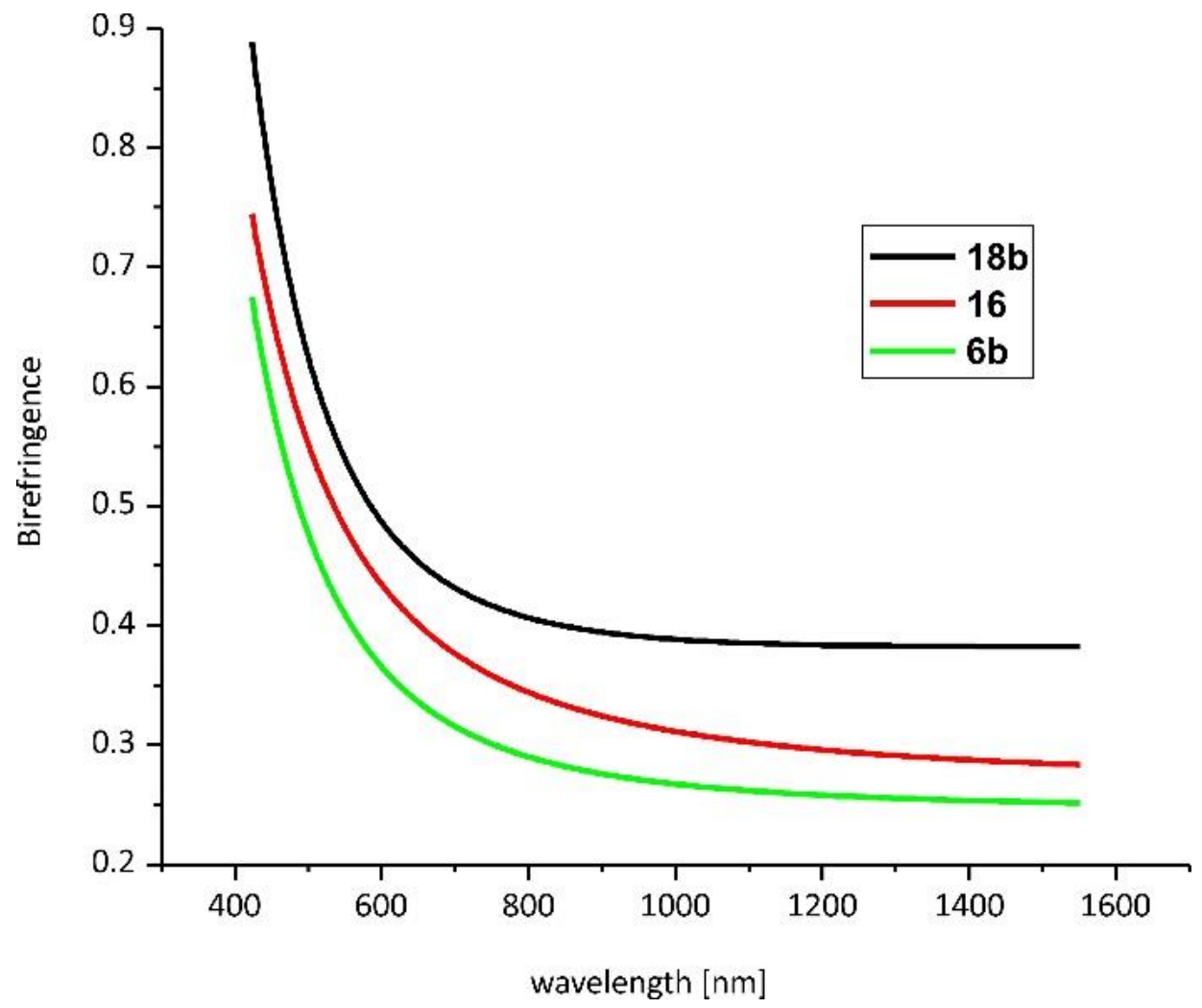

Figure 10. Birefringence dispersion determined for tested compounds.

The goal was to extrapolate birefringence values for individual compounds within the mixture. These compounds were chosen for their structural diversity, possessing either a cyano (CN) polar group or an isothiocyanate (NCS) group, with variations in the connectivity of aromatic units—either through a direct aromatic-aromatic connection or linked by a carbon-carbon triple bond. The resulting birefringence values at three different wavelengths (443nm, 636nm, and 1550nm) shed light on the influence of these structural units on the fundamental optical parameter. Notably, compound **6b**, characterized by the absence of carbon-carbon triple bonds, exhibited the lowest birefringence values across the entire investigated wavelength spectrum. In contrast, compound **16**, featuring an additional carbon-carbon triple bond compared to **6b**, demonstrated higher delta n values, emphasizing the impact of triple bonds on optical response. Furthermore, compound **18b**, with both a carbon-carbon triple bond and an additional highly polarizable NCS polar group, displayed the highest birefringence values across all wavelengths. This underscores the cumulative effect of the selected structural elements on birefringence, providing valuable insights into the nuanced interplay between molecular composition and optical properties. Using mixtures with a low-birefringent nematic material allowed for a comprehensive understanding of how these compounds contribute to birefringence in practical applications.

## 6. Calculations based on density functional theory (DFT)

To investigate the properties of the obtained compounds in more detail, quantum chemical calculations were performed based on DFT model to obtain polarizability anisotropy values. The obtained results enable the identification of correlations between the molecular structure (diversity of terminal groups, quantity and location of lateral groups) and their properties.

Quantum-chemical calculations were performed using the Gaussian16 software package (Revision C.01)[49]. Ground state optimization was performed with B3LYP[50,51] functional and 6-311+G(d,p)[52,53] basis set. After the optimization, the molecular vibrational frequencies were calculated (without changing the functional and basis set) to verify that the stationery points correspond to the true minima. For the optimized geometry, single-point energy calculations were performed along with polarizability calculations (Table 5).

Table 5. The calculated values of polarizability anisotropy for synthesized compounds

| Compound | The value of polarizability anisotropy $\Delta\alpha$ [au] |
|---|---|
| **6a** | 422 |
| **6b** | 461 |
| **6c** | 464 |
| **6d** | 415 |
| **6e** | 418 |
| **11a** | 588 |
| **11b** | 583 |
| **16** | 638 |
| **18a** | 765 |
| **18b** | 753 |

Polarizability anisotropy parameter is expressed in $Bohr^3$(with 1 Bohr = 0.52917 Å)

Obtained values of polarizability anisotropy were analyzed. The results of compounds containing an ethynyl linkage in their molecule (**11a**, **11b**, **16**, **18a**, **18b**) are noticeably higher than the results of the other products. Carbon – carbon triple bond strongly participates in the π-electron conjugation of the core and simultaneously elongates it. The compounds containing the polar terminal group NCS (**18a**, **18b**) exhibit the highest polarizability values. In the case of both discussed compounds (**11a** and **11b**), the weak polarizability of the carbon-fluorine bond does not have a significant contribution to the polarizability of the entire molecule, thus it does not significantly affect their polarizability anisotropy.

## 7. Conclusions:

The synthesis and characterization of liquid crystal compounds derived from benzo[b]thiophene cores have yielded valuable insights into their potential applications, particularly in microwave technology. Designing molecules based on previously known materials for the GHz region of the electromagnetic spectrum enhances their polarizabilities, thus increasing birefringence, and potentially enabling their use in GHz. Incorporating the benzo[b]thiophene, an unconventional rigid core building block, resulted in higher birefringence of synthesized compounds. A systematic analysis of the influence of lateral and terminal substituents allowed for tailoring the compound to achieve the desired phase sequence. The most promising candidates for the mentioned application are compounds with an acetylene group in the rigid core, bearing polar groups as terminal substituents. These findings underscore the significance of molecular design in tailoring liquid crystal materials for advanced technological applications, paving the way for further exploration and innovation in this field.

**Founding:** The project was funded by grant number 2022/06/X/ST5/00700 from the National Science Centre.

**Acknowledgments:** We gratefully acknowledge Polish high-performance computing infrastructure PLGrid (HPC Centers: ACK Cyfronet AGH, WCSS) for providing computer facilities and support within computational grant no. PLG/2023/016670.

**ORCID:**

Piotr Harmata 0000-0003-2318-6293

Agnieszka Mieczkowska 0009-0003-9745-6644

Natan Rychłowicz 0000-0002-1929-0410

Monika Zając 0000-0002-7707-8143

Jakub Herman 0000-0002-5549-7283

# Supplementary Information

# New self-organized benzo[b]thiophene-based materials for GHz applications

**Agnieszka Mieczkowska, Jakub Herman, Natan Rychłowicz, Monika Zając and Piotr Harmata***

Institute of Chemistry, Faculty of Advanced Technologies and Chemistry,
Military University of Technology, ul. gen. S. Kaliskiego 2, 00-908 Warsaw, Poland

email: piotr.harmata@wat.edu.pl

6-(4-pentylphenyl)benzothiophene **3**

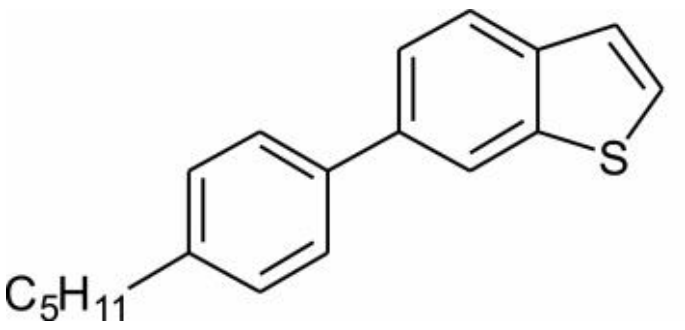

The solution of 6-bromobenzothiophene (10.00g - 0.047mol) **1**, 2-(4-pentylphenyl)-1,3,2-dioxaborate (10.19g - 0.049mol) **2,** anhydrous caesium carbonate $Cs_2CO_3$ (53.60 g - 0.16 mol), acetone (250cm$^3$) and water (100cm$^3$) were added to a flask and stirred under an $N_2$ atmosphere and heated up to the boiling temperature and kept under this conditions for 10 minutes. Then it was cooled slowly to 35°C and palladium(II) acetate (0.1%mol) with XPhos (0.05%mol) was added. Next, mixture was stirred under reflux for 2h. When the reaction was finished acetone was evaporated and obtained mixture was filtered under reduced pressure and washed by 100cm$^3$ of toluene. Crude product was extracted with toluene. Organic layer was washed two times with water, separated, dried over magnesium sulfate $MgSO_4$ and solvent evaporated. Solid product was purified by column chromatography on silica gel using hexane as eluent and crystallized from anhydrous ethanol.

Yield: 10,55g (80,16%).

2-iodo-6-(4-pentylphenyl)benzothiophene **4**

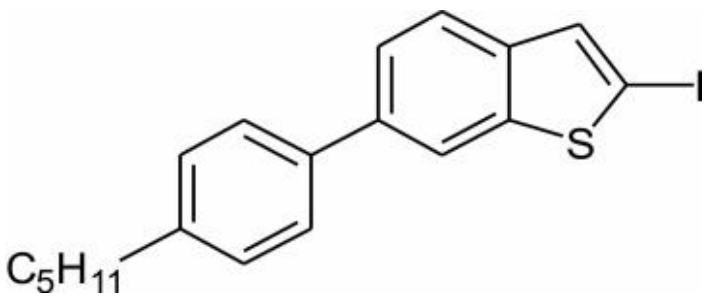

Anhydrous tetrahydrofuran THF (500cm$^3$) and diisopropylamine (19.88 cm$^3$ - 0.14mol) were introduced into a flask and stirred under an $N_2$ atmosphere. The reaction mixture was cooled to -78°C in a dry ice - acetone bath. Then n-Butyllithium (56cm$^3$ - 2.50mol/dm$^3$) was added dropwise (temperature of reaction mixture increased by 8°C), followed by 6-(4-pentylphenyl)benzothiophene (26.40g - 0.094mol) dissolved in anhydrous THF (30cm$^3$). After 10 minutes, iodine (35.56g - 0.14mol) dissolved in 100cm$^3$ of anhydrous THF was added dropwise to the reaction mixture (temperature of reaction mixture increased by 7°C). Thirty minutes after the addition was complete, the cooling bath was removed to allow the solution to reach room temperature. Sodium sulfite was added to the mixture for discoloration. Everything was transferred to a separatory funnel. Organic layer separated and solvent evaporated. Crude product was extracted with dichloromethane Organic layer was washed two times with water, separated, dried over magnesium sulfate $MgSO_4$ and solvent evaporated Solid product was crystallized from ethanol.

Yield: 32g (83,6%)

4-propylphenylboronic acid **5a**

**<u>All the details are in the article[1]</u>**

6-(4-pentylphenyl)-2-(4-propylphenyl)benzothiophene **6a**

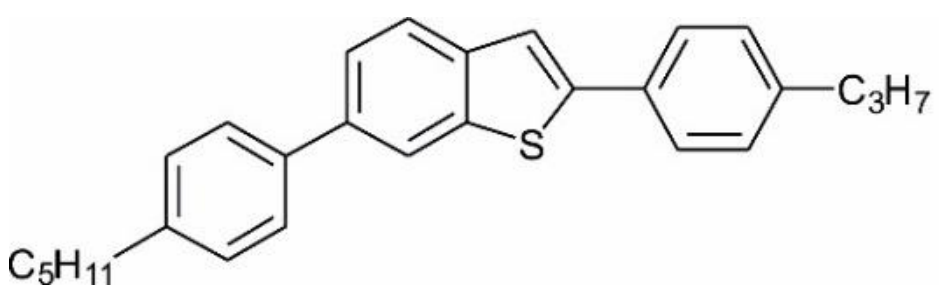

The synthesis was carried out in the same way as in the case of the synthesis of compound **3** from 6-(4-pentylphenyl)-2-iodobenzothiophene (1g - 0.0025mol) **4**, 4-propylphenylboronic acid **5a** (0.40g - 0.0025mol), acetone (45cm$^3$), water (15cm$^3$), anhydrous potassium carbonate $K_2CO_3$ (0.86g - 0.0063mol). Water was added to the solution. Inorganic layer was separated, organic layer was washed two times with water, dried over magnesium sulfate $MgSO_4$, and solvent evaporated. Solid product was crystallized from ethanol. MS(EI) m/z: 398(M+); 369; 341; 325; 312; 297; 221; 156; 41; 29. 1H NMR (500 MHz, CHLOROFORM-D) d ppm 0.9 (m, 3 H) 1.0 (t, J=7.3 Hz, 3 H) 1.4 (ddd, J=6.9, 3.8, 3.6 Hz, 4 H) 1.7 (m, J=7.4, 7.4, 7.4, 7.4 Hz, 4 H) 2.7 (m, 4 H) 7.2 (d, J=8.2 Hz, 2 H) 7.3 (d, J=7.9 Hz, 2 H) 7.5 (s, 1 H) 7.6 (d, J=7.9 Hz, 3 H) 7.7 (d, J=8.2 Hz, 2 H) 7.8 (d, J=8.2 Hz, 1 H) 8.0 (s, 1 H). 13C NMR (126 MHz, CHLOROFORM-D) d ppm 14.1, 14.3, 22.8, 24.7, 31.4, 31.8, 35.8, 38.0, 118.8, 120.5, 123.8, 124.3, 126.6, 127.3, 129.1, 129.3, 132.0, 137.7, 138.6, 140.0, 140.3, 142.3, 143.3, 144.8.

Yield: 0,3g (33,4%)

2-(3-fluoro-4-cyanophenyl)-6-(4-pentylphenyl)benzothiophene **6b**

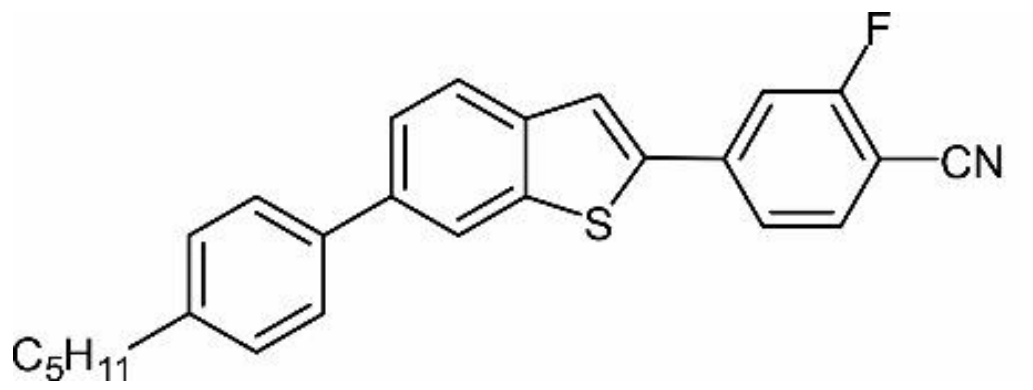

The synthesis was carried out in the same way as in the case of the synthesis of compound **3** from 6-(4-pentylphenyl)-2-iodobenzothiophene (1.60g - 0.0040mol) **4**, 2-(3-fluoro-4-cyanophenyl) -1,3,2-dioxaborinan (0.90g - 0.0044mol) **5b**, anhydrous potassium carbonate $K_2CO_3$ (1.93g - 0.014mol), acetone (80cm$^3$) and water (20cm$^3$). The solid product was crystallised from the ethanol. Then the compound was purified by column chromatography on silica gel using dichloromethane and hexane in ratio 1:2 as the eluent and recrystallized from ethanol. MS(EI) m/z: 399(M+); 355; 342; 327; 308; 221; 171; 41; 29. 1H NMR (500 MHz, CHLOROFORM-D) d ppm 0.9 (m, 3 H) 1.4 (m, 4 H) 1.7 (qd, J=7.5, 7.3 Hz, 2 H) 2.7 (m, 2 H) 7.3 (d, J=7.9 Hz, 2 H) 7.5 (dd, J=9.9, 1.4 Hz, 1 H) 7.5 (m, 3 H) 7.6 (m, 3 H) 7.8 (d, J=8.2 Hz, 1 H) 8.0 (s, 1 H). 13C NMR (126 MHz, CHLOROFORM-D) d ppm 14.3, 22.8, 31.4, 31.8, 35.8, 100.3, 113.9, 114.1, 120.5, 122.5, 122.7, 124.7, 125.0, 127.3, 129.2, 134.1, 138.0, 139.2, 140.5, 141.1, 141.5, 141.6, 142.9, 164.6.

Yield: 1g (63,7%)

2-[3,5-difluoro-4-cyanophenyl]-6-(4-pentylphenyl)benzothiophene **6c**

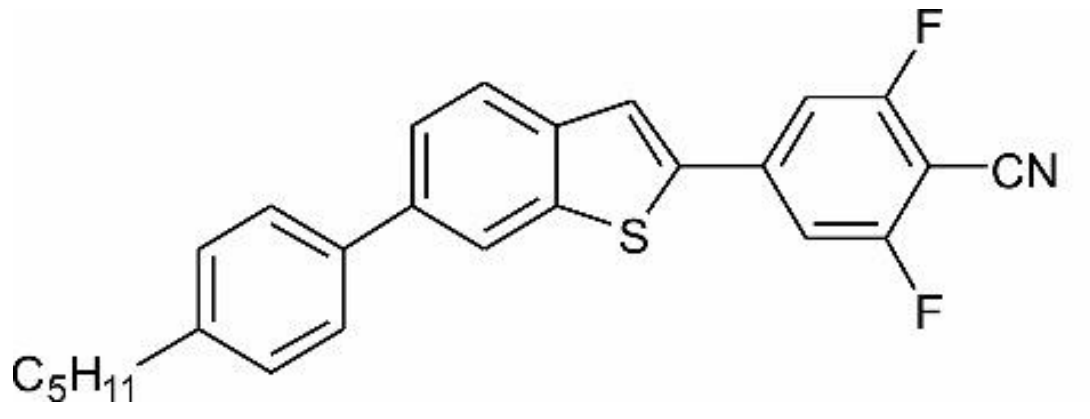

The synthesis was carried out in the same way as in the case of the synthesis of compound **3** from 6-(4-pentylphenyl)-2-iodobenzothiophene (3.00g - 0.0074mol) **4**, 4-(1,3,2-dioxaborinan-2-yl)-2,6-difluorobenzonitrile (2.15g - 0.0081mol) **5c**, acetone (80cm$^3$), water (20cm$^3$), anhydrous potassium carbonate $K_2CO_3$ (3.48g - 0.025mol). The obtained mixture was poured into the beaker along with 40cm$^3$ of 2-3% of hydrochloric acid, and then dissolved in $CH_2Cl_2$. Inorganic layer was separated, organic layer was washed by water two times, dried over magnesium sulfate $MgSO_4$, and then solvent evaporated. The solid product was crystallised from the ethanol. Then the compound was purified by column chromatography on silica gel using $CH_2Cl_2$ as the eluent and

recrystallized from ethanol. MS(EI) m/z: 417(M+); 373; 360; 345; 326; 313; 221; 180; 41; 29. 1H NMR (500 MHz, CHLOROFORM-D) d ppm 0.9 (m, 3 H) 1.4 (m, 4 H) 1.7 (dd, J=15.1, 7.5 Hz, 2 H) 2.7 (m, 2 H) 7.3 (dd, J=13.7, 8.2 Hz, 4 H) 7.6 (d, J=8. 2 Hz, 2 H) 7.6 (m, 2 H) 7.8 (d, J=8.5 Hz, 1 H) 8.0 (s, 1 H). 13C NMR (126 MHz, CHLOROFORM-D) d ppm 14.3, 22.8, 31.3, 31.8, 35.8, 109.5, 109.7, 120.5, 123.5, 125.0, 125.1, 127.3, 129.3, 137.8, 139.0, 139.5, 139.8, 141.2, 142.3, 143.0, 162.5.

Yield: 1,8g (58,4%)

2-[3-fluoro-4-(trifluoromethoxy)phenyl]-6-(4-pentylphenyl)benzothiophene **6d**

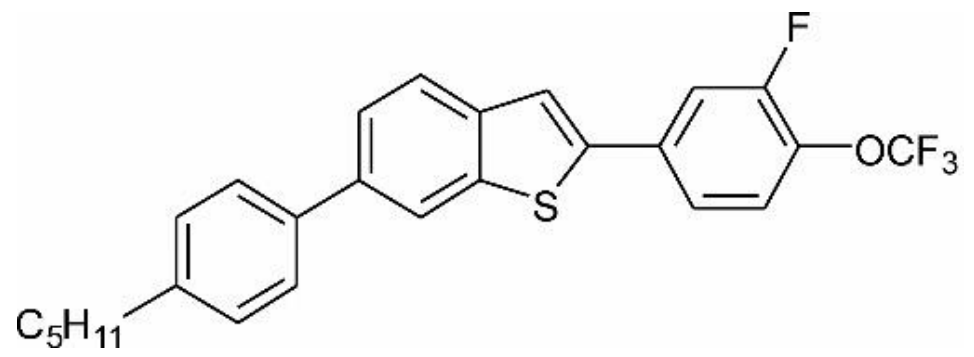

The synthesis was carried out in the same way as in the case of the synthesis of compound **6c** from 6-(4-pentylphenyl)-2-iodobenzothiophene (3.00g - 0.0074 mol) **4**, 2-[3-fluoro-4-(trifluoromethoxy)phenyl]-1,3,2-dioxaborinane (2.44g - 0.0074 mol) **5d**, acetone (50cm$^3$), water (30cm$^3$), anhydrous potassium carbonate $K_2CO_3$ (3.58g - 0.026mol). The obtained solid product was crystallised from the acetonitrile. Then the compound was purified by column chromatography on silica gel using hot hexane as the eluent and recrystallized from methanol. MS(EI) m/z: 458(M+); 414; 401; 373; 332; 315; 304; 270; 221; 69; 41; 29. 1H NMR (500 MHz, CHLOROFORM-D) d ppm 0.9 (m, 3 H) 1.4 (m, 4 H) 1.7 (m, J=7.4, 7.4, 7.4, 7.4 Hz, 2 H) 2.7 (m, 2 H) 7.3 (d, J=8.2 Hz, 2 H) 7.3 (t, J=8.1 Hz, 1 H) 7.5 (d, J=8.5 Hz, 1 H) 7.5 (m, 2 H) 7.6 (d, J=8.2 Hz, 2 H) 7.6 (dd, J=8.4, 1.4 Hz, 1 H) 7.8 (d, J=8.2 Hz, 1 H) 8.0 (s, 1 H). 13C NMR (126 MHz, CHLOROFORM-D) d ppm 14.0, 22.6, 31.1, 31.6, 35.6, 114.9, 115.1, 120.3, 120.7, 122.4, 124.1, 124.5, 127.1, 129.0, 135.0, 138.0, 138.4, 139.2, 140.5, 141.2, 142.4, 153.6, 155.7.

Yield: 2,55g (75,4%)

2-[3,5-difluoro-4-(trifluoromethoxy)phenyl]-6-(4-pentylphenyl)benzothiophene **6e**

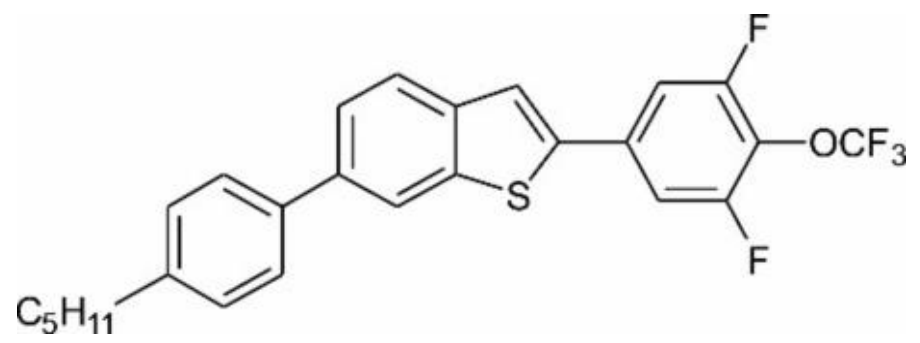

The synthesis was carried out in the same way as in the case of the synthesis of compound **3** from 6-(4-pentylphenyl)-2-iodobenzothiophene (3.00g - 0.0074 mol) **4**, 2-[3,5-difluoro-4-(trifluoromethoxy)phenyl]-1,3,2-dioxaborinane (2.08g - 0.0074 mol) **5e**, acetone (90cm$^3$), water (30cm$^3$), anhydrous potassium carbonate $K_2CO_3$ (3.58g - 0.026mol). The obtained mixture was poured into the beaker containing 200cm$^3$ of water and 50cm$^3$ of 2-3% of hydrochloric acid, stirred for approximately 10 minutes. The precipitate was separated from the liquid using reduced pressure filtration, then dissolved in dichloromethane, dried over magnesium sulfate $MgSO_4$, concentrated and crystallised from the acetone. Due to the low yield of the isolated product, it was decided to change the solvent to acetonitrile. Then the compound was purified by column chromatography on silica gel using hot hexane as the eluent and recrystallized from acetonitrile. MS(EI) m/z: 476(M+); 432; 419; 391; 350; 333; 322; 288; 221; 69; 41; 29. 1H NMR (500 MHz, CHLOROFORM-D) d ppm 0.9 (m, 3 H) 1.4 (m, 4 H) 1.7 (m, 2 H) 2.7 (m, 2 H) 7.3 (d, J=7.9 Hz, 2 H) 7.3 (m, 2 H) 7.5 (s, 1 H) 7.6 (d, J=7.9 Hz, 2 H) 7.6 (dd, J=8.2, 1.5 Hz, 1 H) 7.8 (d, J=8.2 Hz, 1 H) 8.0 (s, 1 H). 13C NMR (126 MHz, CHLOROFORM-D) d ppm 14.3, 22.8, 31.4, 31.8, 35.8, 110.4, 120.6, 121.7, 124.5, 124.9, 127.4, 129.2, 135.4, 135.4, 138.1, 139.1, 139.3, 140.4, 140.8, 142.8, 155.4, 157.5.

Yield: 1,4g (40%)

4-(2-fluoro-4-(trifluoromethoxy)phenyl)-2-methylbut-3-yn-2-ol **9a**

1-bromo-2-fluoro-4-(trifluorometoksy)benzen **7a** (70.00g; 0.27mol), triethylamine (TEA) (37.40cm$^3$ - 0.27mol), 1,8-diazabicyclo[5.4.0]undec-7-ene (DBU) (40cm$^3$ - 0.27mol), $PdCl_2(PPh_3)_2$ (0.3%mol), CuI (0.15%mol), and 500cm$^3$ of toluene were added to a flask and stirred under an $N_2$ atmosphere, heated up to the boiling temperature and kept under this conditions for 10 min. Then the mixture was cooled to 60$^o$C, 2-methyl-but-3-yn-2-ol **8** (29cm$^3$ - 0.30mol) in 60cm$^3$ of toluene was added dropwise. After the addition of 15% of the solution, the mixture became turbid and a precipitate formed. The remaining solution was added dropwise, and the reaction mixture was refluxed for 4 hours. Then the mixture was cooled to room temperature. The resulting precipitate was filtered under reduced pressure. The filtrate was poured into a large amount of water and stirred for 15 minutes. Than the layers were separated, the organic layer was washed three times with water, separated and dried over magnesium sulfate $MgSO_4$. The product was purified by distillation under reduced pressure.

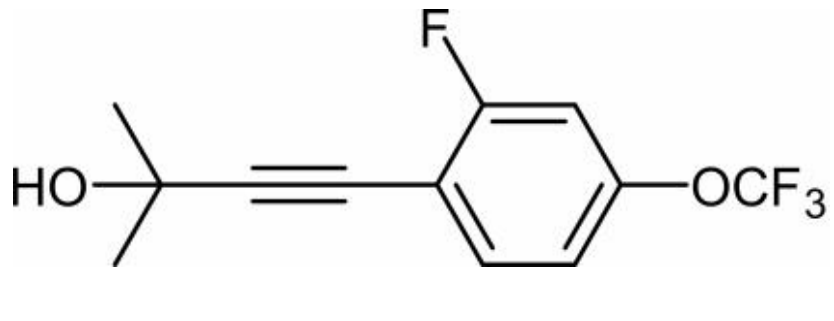

b.p. 80°C at 0.4 mmHg, yield: 54.2g (77.0%)

1-ethynyl-2-fluoro-4-(trifluoromethoxy)benzene **10a**

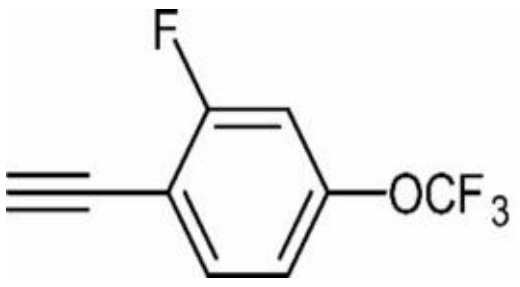

10a

4-(2-fluoro-4-(trifluoromethoxy)phenyl)-2-methylbut-3-yn-2-ol **9a** (54.20g; 0.21mol), sodium hydride NaH (10%mol) and 300cm$^3$ of toluene were added to a flask. The reaction mixture was slowly heated to reflux. Acetone was collected as the reaction progressed. After one hour GC-MS analysis showed complete reaction. The reaction mixture was cooled to room temperature. All was filtered through a filter plate. The organic layer was washed three times with water, separated and dried over magnesium sulfate $MgSO_4$. It was then purified by distillation under reduced pressure.

b.p. 104°C at 200mmHg, yield: 36.5g (86.4%)

2-fluoro-4-((trimethylsilyl)ethynyl)benzonitrile **14a**

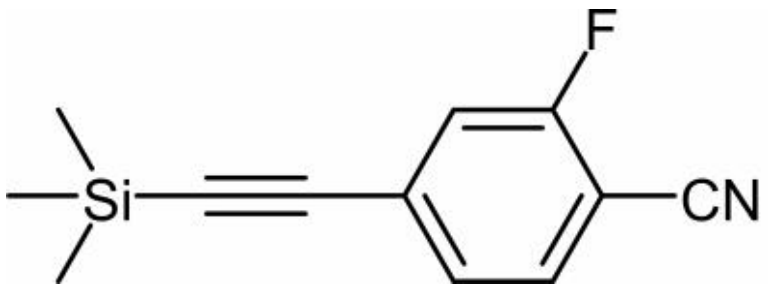

2-fluoro-4-iodobenzonitrile **13a** (30.00g - 0.12mol), triethylamine TEA (18.80cm$^3$ - 0.13mol), 1,8-diazabicyclo[5.4.0]undec-7-ene (DBU) (18.0cm$^3$ - 0.13mol), $PdCl_2(PPh_3)_2$ (0.3%mol), CuI (0.15%mol), and 350cm$^3$ of toluene were added to a flask and stirred under an $N_2$ atmosphere, heated up to the boiling temperature and kept under this conditions for 10 min. Then the mixture was cooled to 50°C, trimethylsilylacetylene (19cm$^3$ - 0.13mol) in 50cm$^3$ of toluene was added dropwise. After the addition of 15% of the solution, the mixture became turbid and a precipitate formed. The remaining solution was added dropwise, and the reaction mixture was kept at 50 °C for 3 hours. Then the mixture was cooled to room temperature. 200cm$^3$ of water was added and reaction mixture was filtered under reduced pressure. All was transferred to a separatory funnel. The organic layer was washed 3 times with water, the inorganic layer was washed once with toluene. The combined organic layers were dried over magnesium sulfate $MgSO_4$ and the solvent was evaporated. The crude product was used for further reaction assuming 100% conversion.

4-ethynyl-2-fluorobenzonitrile **15a**

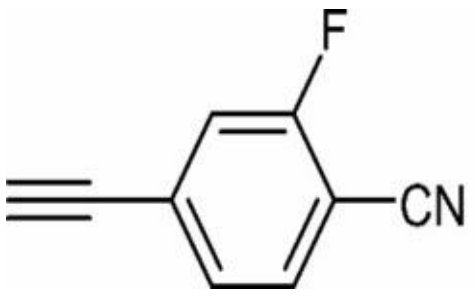

15a

2-fluoro-4-((trimethylsilyl)ethynyl)benzonitrile (26.50g - 0.12mol), anhydrous potassium carbonate $K_2CO_3$ (1.70g - 0.012mol), 300cm$^3$ methanol and 100cm$^3$ of anhydrous THF were added to a flask and stirred. The whole was stirred for 3 hours at room temperature. The whole was poured into 200 cm$^3$ of water in a beaker. Stirred for about 30 minutes. Then 25cm$^3$ of concentrated hydrochloric acid was added and stirred for 5 minutes. The whole was transferred to a separatory funnel and 150 cm$^3$ of dichloromethane was added. The organic layer was washed three times with water. The inorganic layer was washed once with dichloromethane. Dry the combined organic layers over magnesium sulfate, the solvent was evaporated. Crystallized from acetone.

Yield: 13.8g (81%)

4-((trimethylsilyl)ethynyl)aniline **14b**

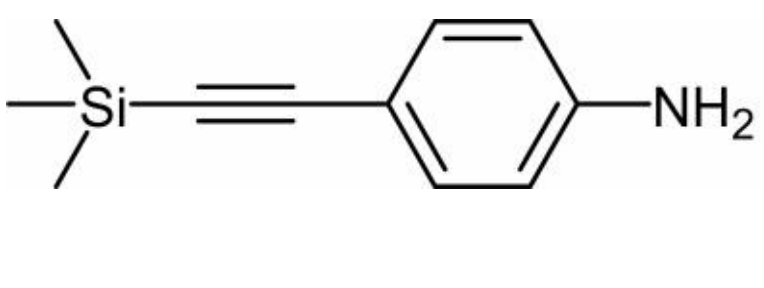

The synthesis was carried out in the same way as in the case of the synthesis of compound **14a**

(2-fluoro-4-((trimethylsilyl)ethynyl)benzonitrile) from 4-iodoaniline (30.00g - 0.14mol) **13b,** triethylamine (TEA) (20.60cm$^3$ - 0.15mol), 1,8-diazabicyclo[5.4.0]undec-7-ene (DBU) (19.30cm$^3$ - 0.15mol), $PdCl_2(PPh_3)_2$ (0.3%mol), CuI (0.15%mol), and 250cm$^3$ of toluene were added to a flask and stirred under an $N_2$ atmosphere, heated up to 40°C. Then trimethylsilylacetylene (22.70cm$^3$ - 0.15mol) in 50cm$^3$ of toluene was added dropwise. After the addition of 15% of the solution, the mixture became turbid and a precipitate formed. The remaining solution was added dropwise, and the reaction mixture was kept at 40°C for 2 hours. The post-reaction treatment was performed analogously to **14a** (2-fluoro-4-((trimethylsilyl)ethynyl)benzonitrile). Crystallized from hexane.

Yield: 21.5g (82%)

4-ethynylaniline **15b**

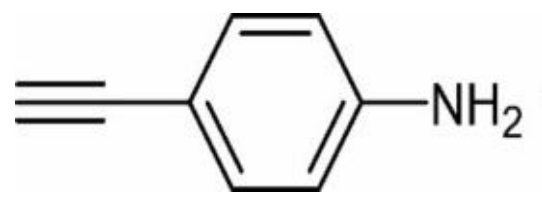

15b

The synthesis was carried out in the same way as in the case of the synthesis of compound **15a** (4-ethynyl-2-fluorobenzonitrile) from 4-((trimethylsilyl)ethynyl)aniline (21.50g - 0.11mol) **14b,** anhydrous potassium carbonate $K_2CO_3$ (1.89g - 0.013mol), 300cm$^3$ methanol and 100cm$^3$ of anhydrous THF were added to a flask and stirred. The whole was stirred for 3 hours at room temperature. 50cm$^3$ of water was added to the reaction mixture and stirred for 20 minutes. The whole was transferred to a separatory funnel and organic layer was dried with $MgSO_4$. Then crude product was dissolved in 100cm$^3$ of toluene. Filtered through a pad of activated carbon under reduced pressure. The product was poured into a 250 cm$^3$ volumetric flask and made up to the mark. Stored in the freezer. 100% overreaction assumed. C=0.45mol/dm$^3$

2,6-difluoro-4-((trimethylsilyl)ethynyl)aniline **14c**

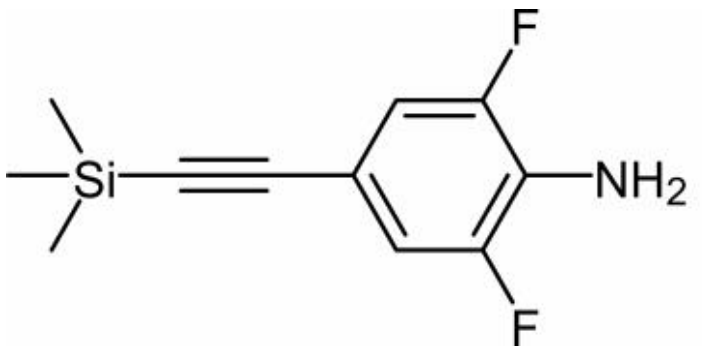

The synthesis was carried out in the same way as in the case of the synthesis of compound **14b** (4-((trimethylsilyl)ethynyl)aniline) from 2,6-difluoro-4-iodoaniline (30.00g - 0.12mol) **13c**, triethylamine (TEA) (17.60cm$^3$ - 0.13mol), 1,8-diazabicyclo[5.4.0]undec-7-ene (DBU) (16.50cm$^3$ - 0.13mol), $PdCl_2(PPh_3)_2$ (0.3%mol), CuI (0.15%mol), and 250cm$^3$ of toluene were added to a flask and stirred under an $N_2$ atmosphere, heated up to 40°C. Then trimethylsilylacetylene (18.30cm$^3$ - 0.13mol) in 50cm$^3$ of toluene was added dropwise. After the addition of 15% of the solution, the mixture became turbid and a precipitate formed. The remaining solution was added dropwise, and the reaction mixture was kept at 40°C for 2 hours. Then the mixture was cooled to room temperature. 200cm$^3$ of water was added and reaction mixture was filtered under reduced pressure. All was transferred to a separatory funnel. The organic layer was washed 3 times with water, the inorganic layer was washed once with toluene. The combined organic layers were dried over magnesium sulfate $MgSO_4$ and the solvent was evaporated. The crude product was purified by column chromatography on silica gel using hexane as the eluent. It was not possible to evaporate all the hexane using a rotary vacuum evaporator. It was decided to assume 100% reaction for the next step.

4-ethynyl-2,6-difluoroaniline **15c**

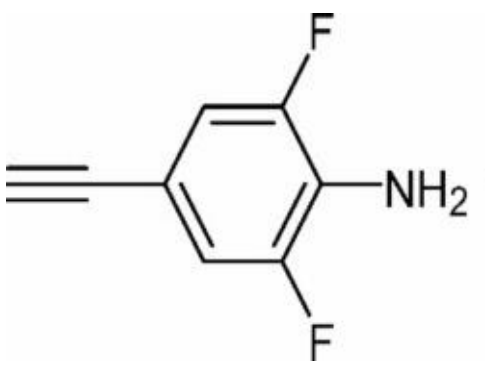

15c

The synthesis was carried out in the same way as in the case of the synthesis of compound **15b** (4-ethynylaniline) from 2,6-difluoro-4-((trimethylsilyl)ethynyl)aniline (26.50g - 0.12mol) **14c**, anhydrous potassium carbonate $K_2CO_3$ (1.60g - 0.012mol), 300cm$^3$ methanol and 100cm$^3$ of anhydrous THF were added to a flask and stirred. The post-reaction treatment was performed analogously to **15b** (4-ethynylaniline). C = 0.47mol/dm$^3$

2-{[2-fluoro-4-(trifluoromethoxy)phenyl]ethynyl}-6-(4-pentylphenyl)benzothiophene **11a**

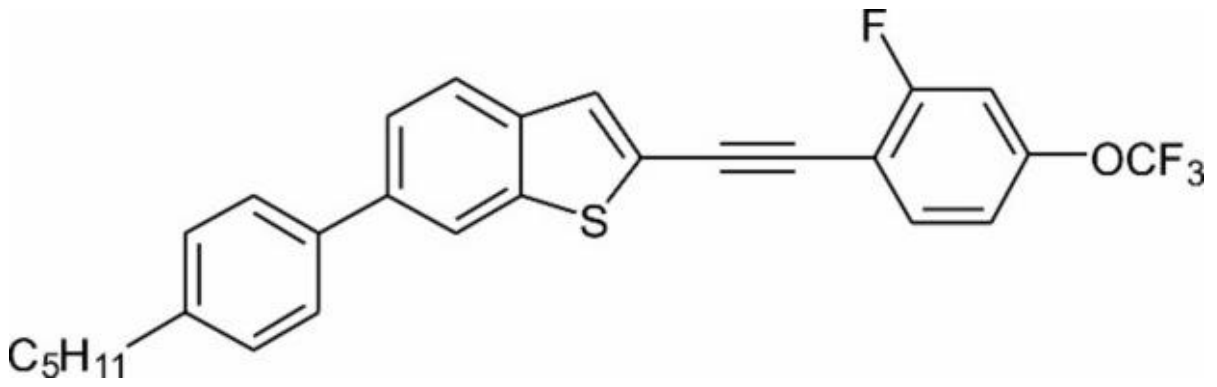

6-(4-pentylphenyl)-2-iodobenzothiophene (2.00g - 0.0049mol) **4**, triethylamine (TEA) (0.76cm$^3$ - 0.726g/cm$^3$), 1,8-diazabicyclo[5.4.0]undec-7-ene (DBU) (0.81cm$^3$ - 1.019g/cm$^3$), $PdCl_2(PPh_3)_2$, CuI, and toluene (50cm$^3$) were added to a flask and stirred under an $N_2$ atmosphere, heated up to the boiling temperature and kept under this conditions for 10 min. Then the mixture was cooled to room temperature, and 1-ethynyl-2-fluoro-4-(trifluoromethoxy)benzene (1.10g - 0.0054mol) **10a** dissolved in 10cm$^3$ of toluene was added dropwise. After the addition of 1/3 of the solution, the mixture became turbid and a precipitate formed. The remaining solution was added dropwise, and the reaction mixture was refluxed for 5 hours. Water was added to the flask and stirred for 20 minutes, then the contents were transferred to a separatory funnel and washed with water. The organic layer was washed 3 times with water, the inorganic layer was washed once with toluene. The combined organic layers were dried over magnesium sulfate $MgSO_4$ and the solvent was evaporated. Then the compound was purified by column chromatography on silica gel using hexane as the eluent and recrystallized from acetonitrile. MS(EI) m/z: 482(M+); 438; 425; 397; 356; 339; 328; 294; 212;163; 69; 41; 29. 1H NMR (500 MHz, CHLOROFORM-D) d ppm 0.9 (t, J=6.7 Hz, 3 H) 1.4 (m, 4 H) 1.7 (m, J=7.4, 7.4, 7.4, 7.4 Hz, 2 H) 2.7 (m, 2 H) 7.0 (d, J=8.2 Hz, 2 H) 7.3 (d, J=7.9 Hz, 2 H) 7.6 (t, J=7.9 Hz, 4 H) 7.6 (dd, J=8.2, 1.5 Hz, 1 H) 7.8 (d, J=8.2 Hz, 1 H) 8.0 (s, 1 H). 13C NMR (126 MHz, CHLOROFORM-D) d ppm 14.3, 22.8, 31.4, 31.8, 35.8, 87.0, 89.1, 109.2, 110.5, 116.7, 119.5, 120.2, 121.5, 122.5, 124.3, 124.7, 127.4, 129.2, 129.6, 134.3, 138.2, 139.3, 141.6, 142.7, 161.8, 163.8.

Yield: 1,15g (65,7%)

2-{[3-fluoro-4-(trifluoromethoxy)phenyl]ethynyl}-6-(4-pentylphenyl)benzothiophene **11b**

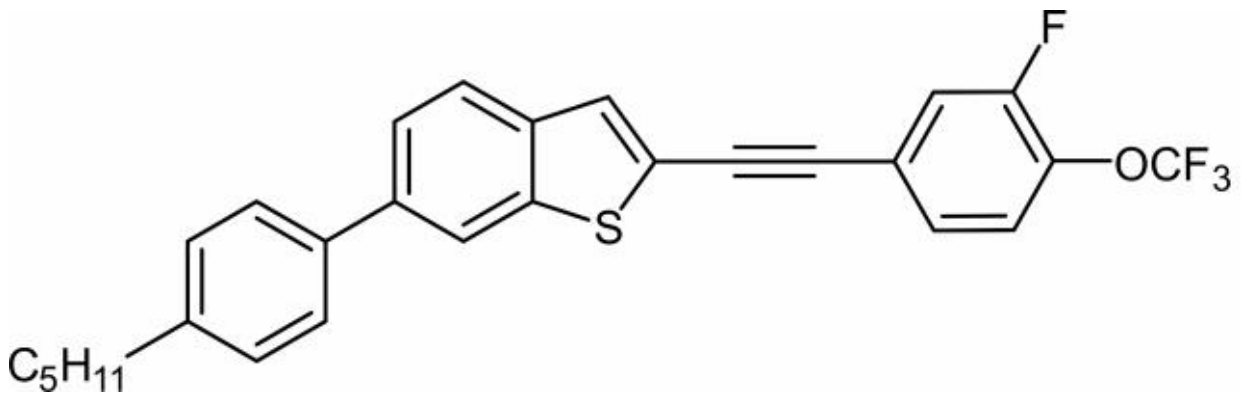

The synthesis was carried out in the same way as in the case of the synthesis of compound **11a** from 6-(4-pentylphenyl)-2-iodobenzothiophene (3.00g - 0.0074mol) **4**, triethylamine (TEA) (1.55cm$^3$ - 0.726g/cm$^3$) **4**, 1,8-diazabicyclo[5.4.0]undec-7-ene (DBU) (1.66cm$^3$ - 1.019g/cm$^3$), $PdCl_2(PPh_3)_2$, CuI, toluene (50cm$^3$) and 1-ethynyl-3-fluoro-4-(trifluoromethoxy)benzene (1.1g - 0.0054mol) **10b** dissolved in 15cm$^3$ of toluene. The mixture was refluxed for 4 hours. The post-reaction treatment was performed analogously to **11a** (2-{[2-fluoro-4-(trifluoromethoxy)phenyl]ethynyl}-6-(4-pentylphenyl)benzothiophene). Then the compound was purified by column chromatography on silica gel using hexane as the eluent and recrystallized from acetonitrile. MS(EI) m/z: 482(M+); 438; 425; 397; 356; 339; 328; 294; 212; 163; 69; 41; 29. 1H NMR (500 MHz, CHLOROFORM-D) d ppm 0.9 (m, 3 H), 1.4 (m, 4 H), 1.7 (m, 2 H), 2.7 (m, 2 H), 7.3 (m, 4 H), 7.4 (dd, J=10.4, 1.8 Hz, 1 H), 7.5 (s, 1 H), 7.6 (d, J=8.2 Hz, 2 H), 7.6 (dd, J=8.5, 1.5 Hz, 1 H), 7.8 (d, J=8.2 Hz, 1 H), 8.0 (s, 1 H). 13C NMR (126 MHz, CHLOROFORM-D) d ppm 14.3, 22.8, 31.4, 31.8, 35.8, 85.2, 92.5, 120.2, 120.4, 120.5, 122.3, 123.4, 123.9, 124.3, 124.8, 127.4, 128.2, 128.2, 129.2, 129.6, 138.1, 139.4, 141.5, 142.8, 153.3, 155.3.

Yield: 1,45g (82,9%)

2-{[3-fluoro-4-(trifluoromethoxy)phenyl]ethynyl}-6-(4-pentylphenyl)benzothiophene **16**

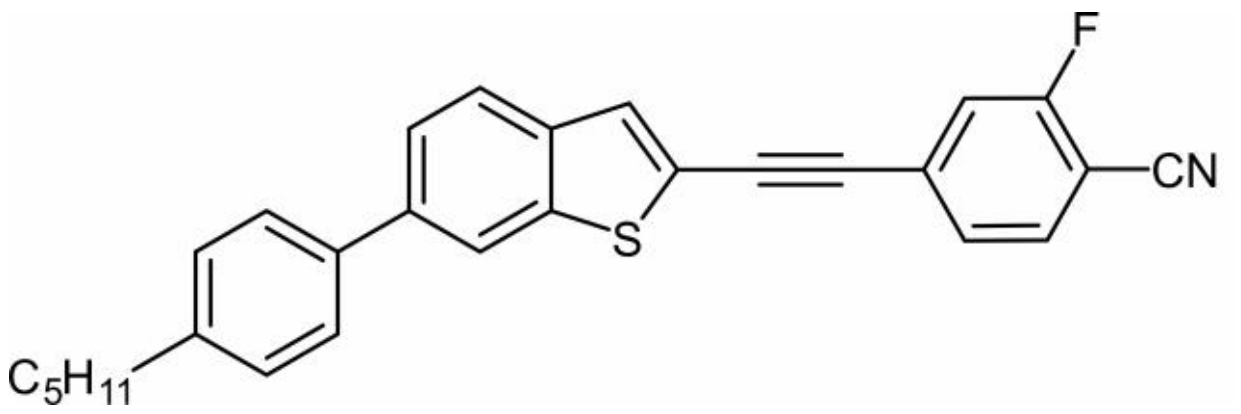

The synthesis was carried out in the same way as in the case of the synthesis of compound **11a** from 6-(4-pentylphenyl)-2-iodobenzothiophene (3.00g - 0.0074mol) **4**, triethylamine (TEA) (1.55cm$^3$ - 0.726g/cm$^3$), 1,8-diazabicyclo[5.4.0]undec-7-ene (DBU) (1.66cm$^3$ - 1.019g/cm$^3$), $PdCl_2(PPh_3)_2$, CuI, toluene (50cm$^3$) and 4-ethynyl-2-fluorobenzonitrile (1.61g - 0.0011mol) **15a** dissolved in 15cm$^3$ of toluene. The mixture was refluxed for 3 hours. The post-reaction treatment was performed analogously to **11a** (2-{[2-fluoro-4-(trifluoromethoxy)phenyl]ethynyl}-6-(4-pentylphenyl)benzothiophene) Then the compound was purified by column chromatography on silica gel using hexane as the eluent and recrystallized from acetonitrile. MS(EI) m/z: 423(M+); 379; 366; 351; 332; 319; 245; 183; 57; 41; 29. 1H NMR (500 MHz, CHLOROFORM-D) d ppm 0.9 (m, 3 H), 1.4 (m, 4 H), 1.7 (m, 2 H), 2.7 (m, 2 H), 7.3 (d, J=8.2 Hz, 2 H), 7.4 (m, 2 H), 7.6 (m, 4 H), 7.6 (dd, J=8.5, 1.5 Hz, 1 H), 7.8 (d, J=8.2 Hz, 1 H), 8.0 (s, 1 H). 13C NMR (126 MHz, CHLOROFORM-D) d ppm 14.2, 22.8, 31.3, 31.8, 35.8, 88.9, 92.2, 101.4, 113.8, 119.0, 120.2, 121.5, 124.5, 124.9, 127.3, 127.9, 129.2, 130.0, 130.5, 133.5, 137.9, 139.7, 141.8, 142.9, 161.9, 163.9.

Yield: 1,85g (59,3%)

4-[6-(4-pentylphenyl)benzothiophen-2-yl]ethynylaniline **17a**

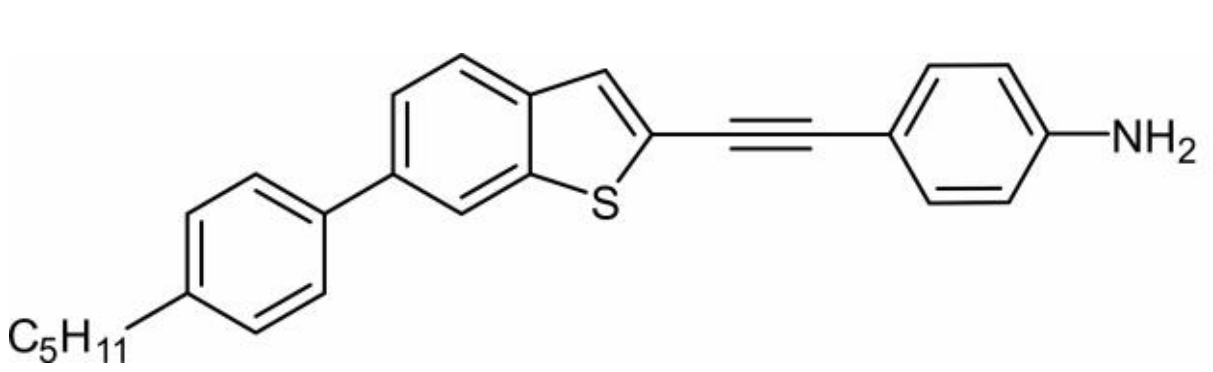

The synthesis was carried out in the same way as in the case of the synthesis of compound **16** from 6-(4-pentylphenyl)- 2-iodobenzothiophene (3.00g - 0.0074mol) **4**, triethylamine (1.09cm$^3$ - 0.726g/cm$^3$), 1,8-diazabicyclo[5.4.0]undec-7-ene (DBU) (1.17cm$^3$ - 1.019g/cm$^3$), $PdCl_2(PPh_3)_2$, CuI, toluene (50cm$^3$) and 4-ethynylaniline (0.91g - 0.0078mol) **15b**. The mixture was heated for 3 hours at a temperature not exceeding 50°C. Water was added to the flask and stirred for a few minutes, then the contents were transferred to a separatory funnel along with toluene. The organic layer was washed two times with water, dried over magnesium sulfate $MgSO_4$ and the solvent was evaporated. The crude product was used for further reaction assuming 100% conversion. MS(EI) m/z: 395(M+); 351; 338; 321; 281; 221; 207; 169; 44; 28.

2,6-difluoro-4-[6-(4-pentylphenyl)benzothiophen-2-yl]ethynylaniline **17b**

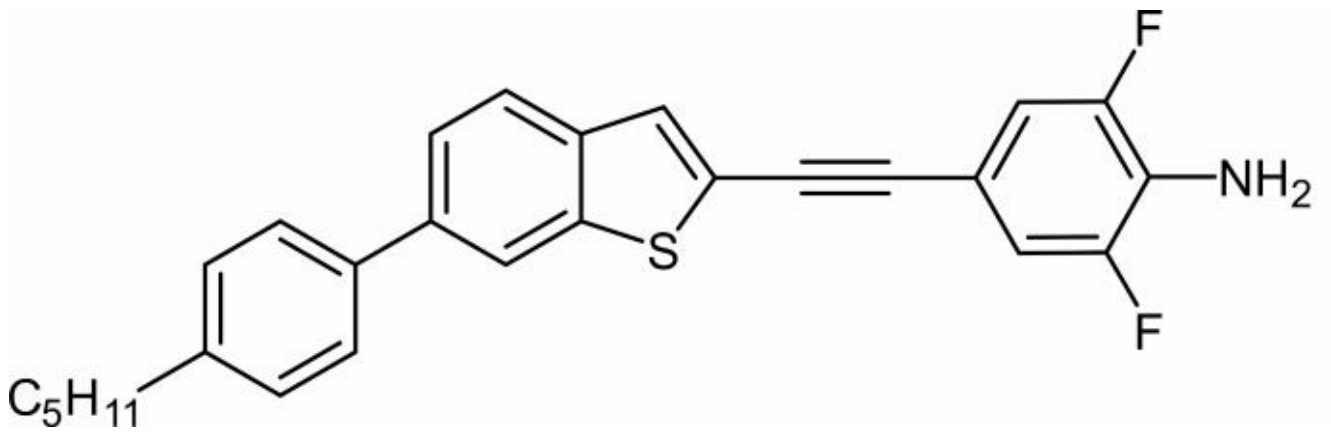

The synthesis was carried out in the same way as in the case of the synthesis of compound **17a** from 6-(4-pentylphenyl)-2-iodobenzothiophene (2g - 0.0049mol) **4**, triethylamine (1.03cm$^3$ - 0.726g/cm$^3$), 1,8-diazabicyclo[5.4.0]undec-7-ene (DBU) (1.11cm$^3$ - 1.019g/cm$^3$), $PdCl_2(PPh_3)_2$, CuI, toluene (50cm$^3$) and 4-ethynyl-2,6-difluoroaniline (1.15g - 0.0074mol) **15c**. The post-reaction treatment was performed analogously to **17a** (4-[6-(4-pentylphenyl)benzothiophen-2-yl]ethynylaniline). The crude product was used for further reaction assuming 100% conversion. MS(EI) m/z: 431(M+); 387; 374; 359; 340; 326; 245; 216; 187; 41; 29.

2-[(4-isothiocyanatophenyl)ethynyl]-6-(4-pentylphenyl)benzothiophene **18a**

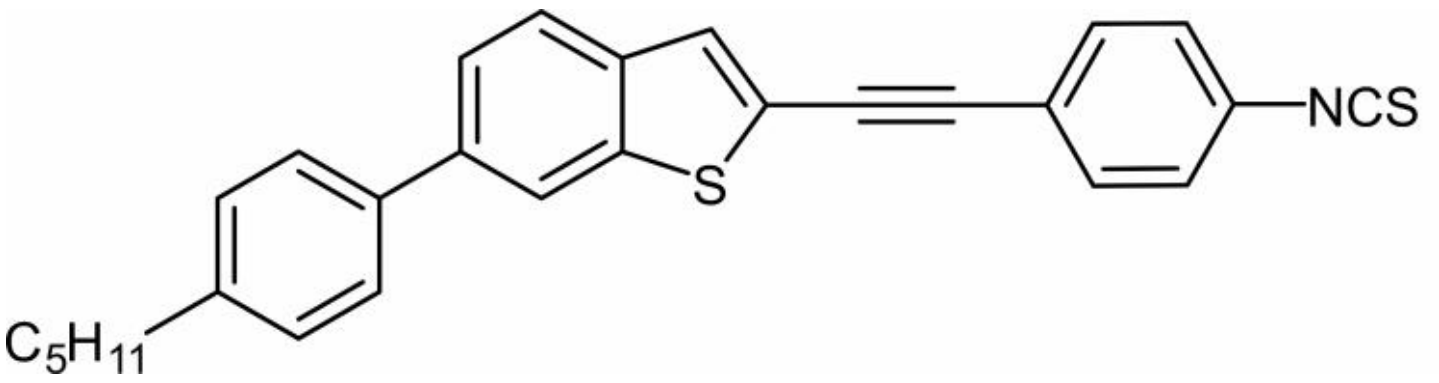

Calcium carbonate $CaCO_3$ (2.22g - 0.022mol), thiophosgene (0.70cm$^3$ - 1.50g/cm$^3$), chloroform (50cm$^3$), water (50cm$^3$), and 4-{[(4-pentylphenyl)benzothiophen-2-yl]ethynyl}aniline (2.92g - 0.0074mol) **17a** dissolved in the minimum amount of chloroform were added to a flask and stirred. The reaction was carried out for 3 hours at room temperature. The resulting mixture was filtered through a filter plate, the organic layer was washed two times with water, dried over magnesium sulfate $MgSO_4$ and the solvent was evaporated. Crystallised from acetone and butanone. Then the compound was purified by column chromatography on silica gel using hot heptane as the eluent and recrystallized from acetone and butanone. MS(EI) m/z: 437(M+); 405; 393; 380; 348; 321; 289; 276; 207; 190; 160; 57; 41; 28. 1H NMR (500 MHz, THF) d ppm 0.9 (m, 3 H), 1.4 (dt, J=7.4, 3.8 Hz, 4 H), 1.7 (m, 2 H), 2.7 (m, 2 H), 7.3 (d, J=7.9 Hz, 2 H), 7.3 (d, J=8.9 Hz, 2 H), 7.6 (m, 5 H), 7.7 (dd, J=8.5, 1.5 Hz, 1 H), 7.8 (d, J=8.5 Hz, 1 H), 8.1 (s, 1 H). 13C NMR (126 MHz, THF) d ppm 14.4, 23.4, 32.1, 32.5, 36.3, 85.6, 94.5, 120.6, 122.6, 123.3, 124.9, 125.1, 126.8, 127.8, 129.7, 130.0, 132.4, 133.6, 134.7, 138.8, 139.1, 139.9, 142.2, 143.1.

Yield: 1,6g (49,5%)

2-[(3,5-difluoro-4-isothiocyanatophenyl)ethynyl]-6-(4-pentylphenyl)benzothiophene **18b**

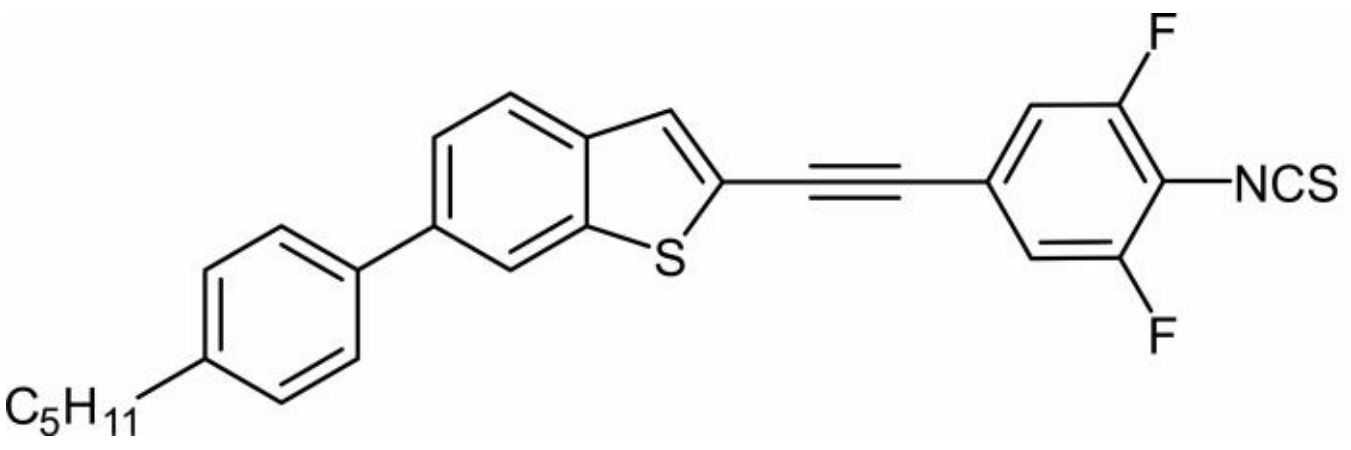

The synthesis was carried out in the same way as in the case of the synthesis of compound **18a** from calcium carbonate $CaCO_3$ (1.50g - 0.015mol), thiophosgene (0.47cm$^3$ - 1.50g/cm$^3$), chloroform (50cm$^3$), water (50cm$^3$), and 2,6-difluoro-4-{6-(4-pentylphenyl)benzothiophen-2-yl}ethynylaniline (2.92g - 0.0074mol) **17b** dissolved in the 10cm$^3$ of chloroform. The post-reaction treatment was performed analogously to **17a** (2-[(4-isothiocyanatophenyl)ethynyl]-6-(4-pentylphenyl)benzothiophene). Crystallised from acetone. Then the compound was purified by column chromatography on silica gel using hot heptane as the eluent and recrystallized from acetone. Yield: 0,4g (17,2%), MS(EI) m/z: 473(M+);429; 416; 401; 382; 357; 325; 245’ 208; 178; 57; 41; 28. 13C NMR (126 MHz, CHLOROFORM-D) d ppm 14.3, 22.8, 31.4, 31.8, 35.8, 87.1, 92.2, 115.1, 115.3, 120.2, 121.9, 122.1, 124.4, 124.9, 127.4, 129.2, 130.0, 138.0, 139.6, 141.7, 142.8, 146.0, 156.9, 158.9.

## Mass spectrum

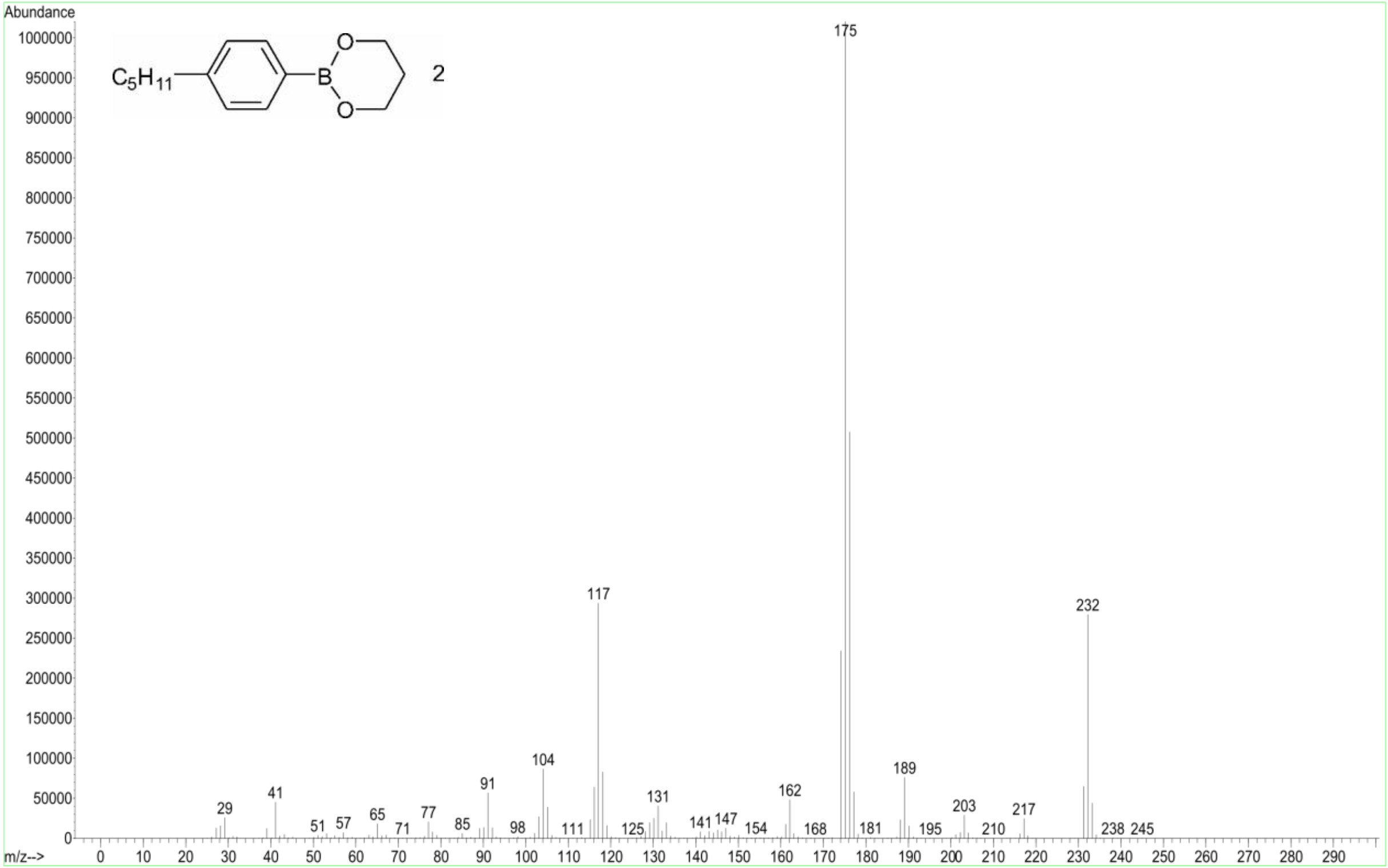

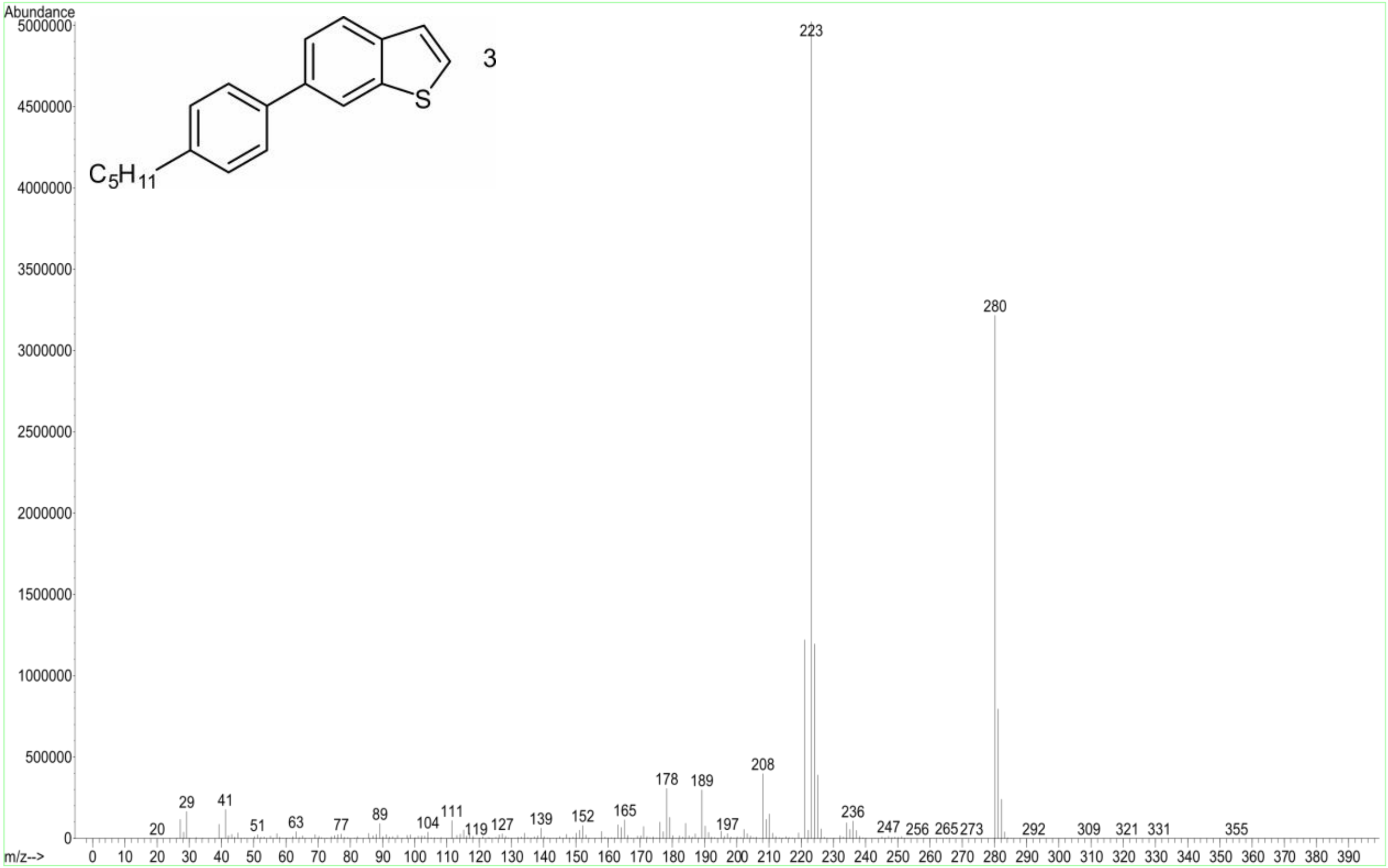

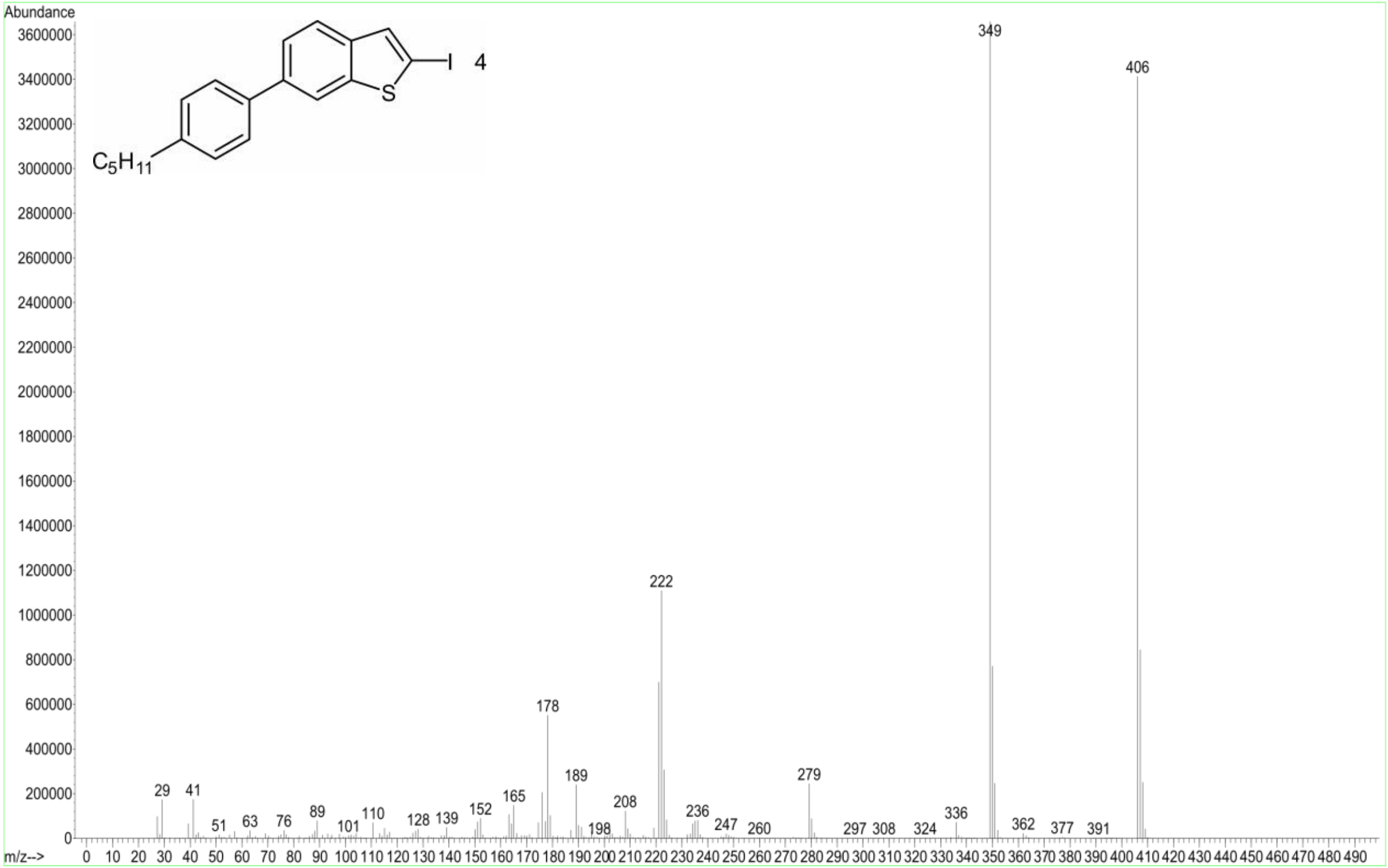
Abundance
C5H11
S
I
4
3600000
3400000
3200000
3000000
2800000
2600000
2400000
2200000
2000000
1800000
1600000
1400000
1200000
1000000
800000
600000
400000
200000
0
349
406
222
178
279
189
208
29
41
89
110
165
152
139
128
101
76
63
51
198
236
247
260
297
308
324
336
362
377
391
m/z-->
0 10 20 30 40 50 60 70 80 90 100 110 120 130 140 150 160 170 180 190 200 210 220 230 240 250 260 270 280 290 300 310 320 330 340 350 360 370 380 390 400 410 420 430 440 450 460 470 480 490

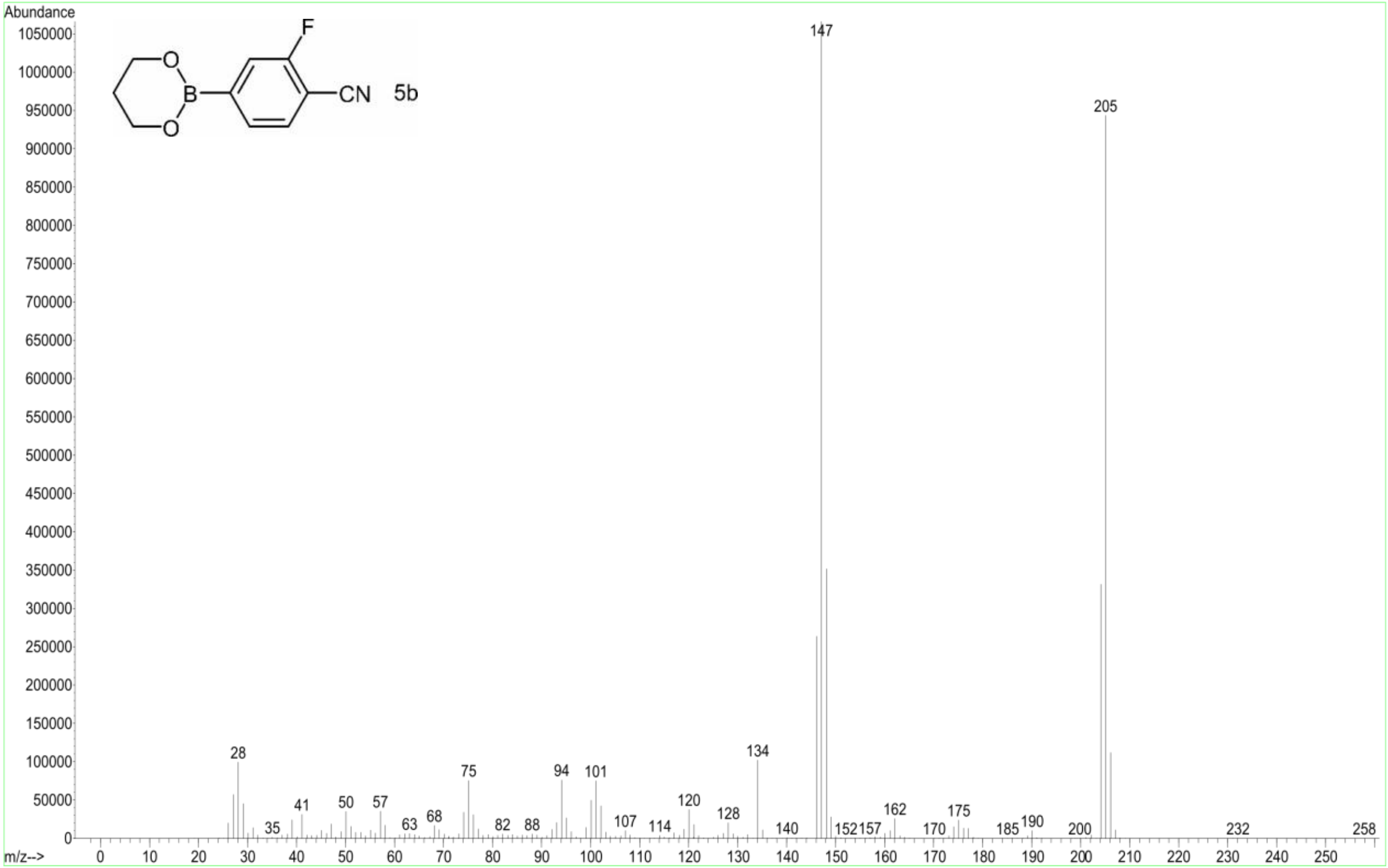
Abundance
O
B
O
F
CN
5b
1050000
1000000
950000
900000
850000
800000
750000
700000
650000
600000
550000
500000
450000
400000
350000
300000
250000
200000
150000
100000
50000
0
147
205
134
28
75
94
101
120
162
175
41
50
57
68
35
63
82
88
107
114
128
140
152157
170
185
190
200
232
258
m/z-->
0 10 20 30 40 50 60 70 80 90 100 110 120 130 140 150 160 170 180 190 200 210 220 230 240 250

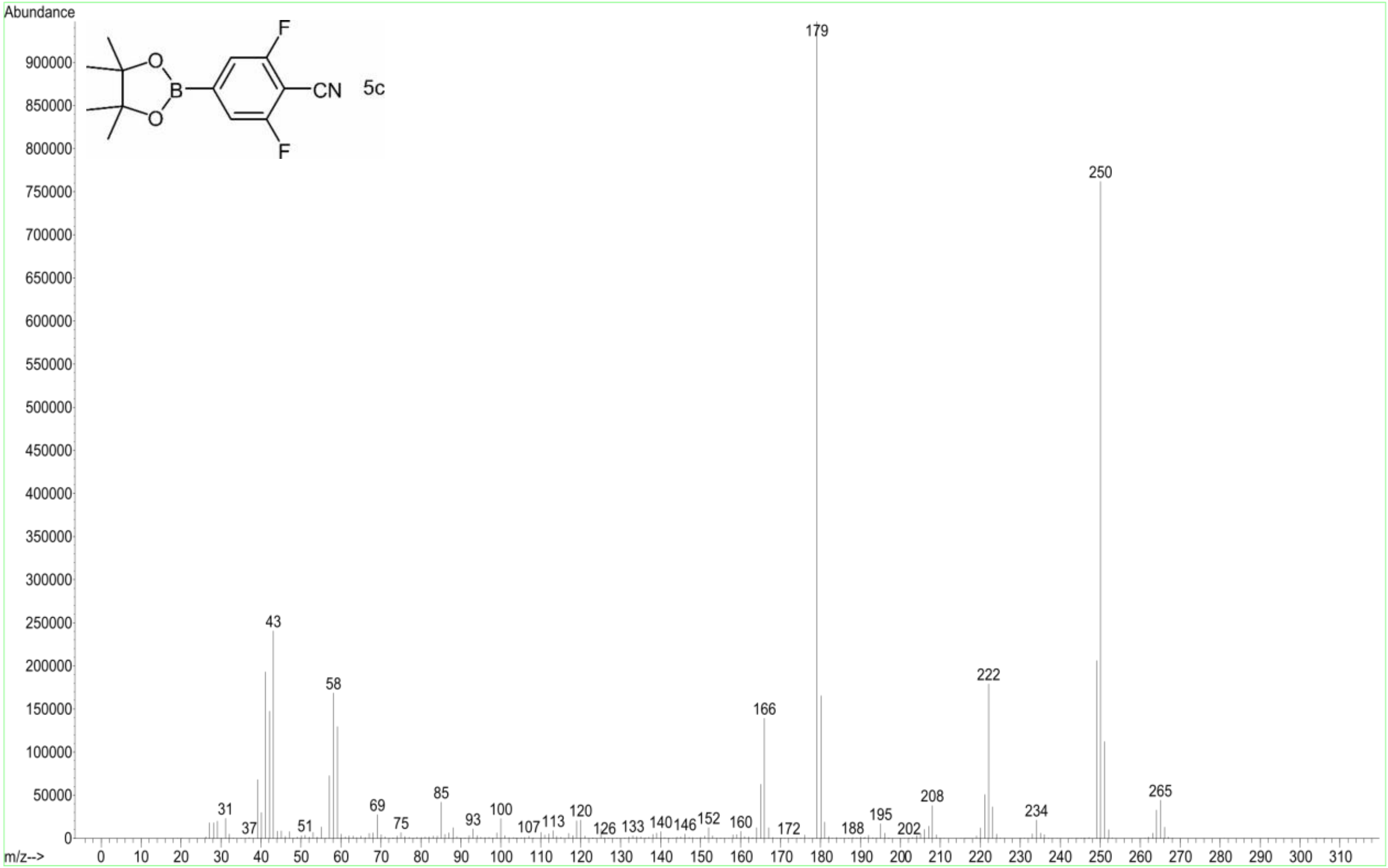

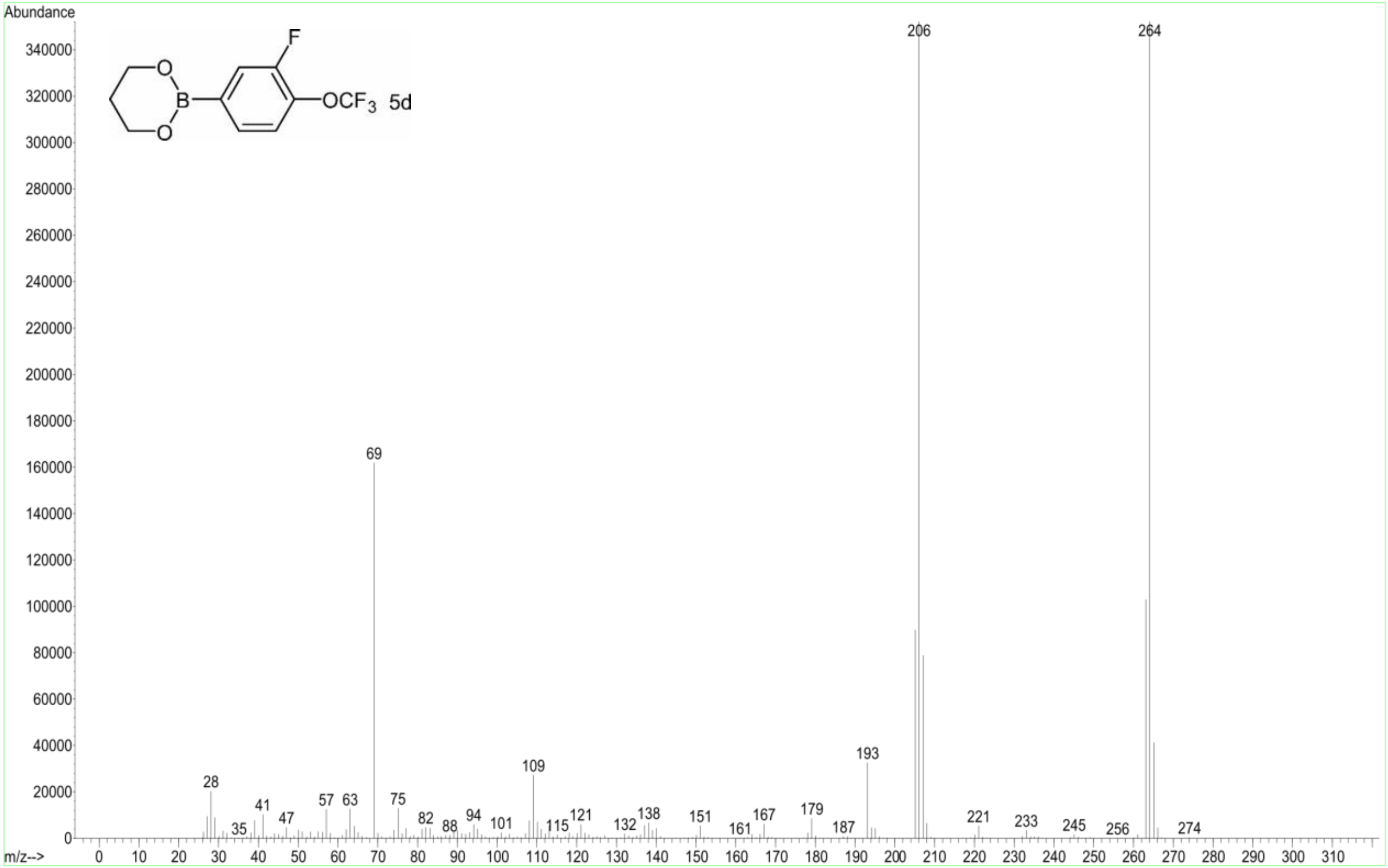

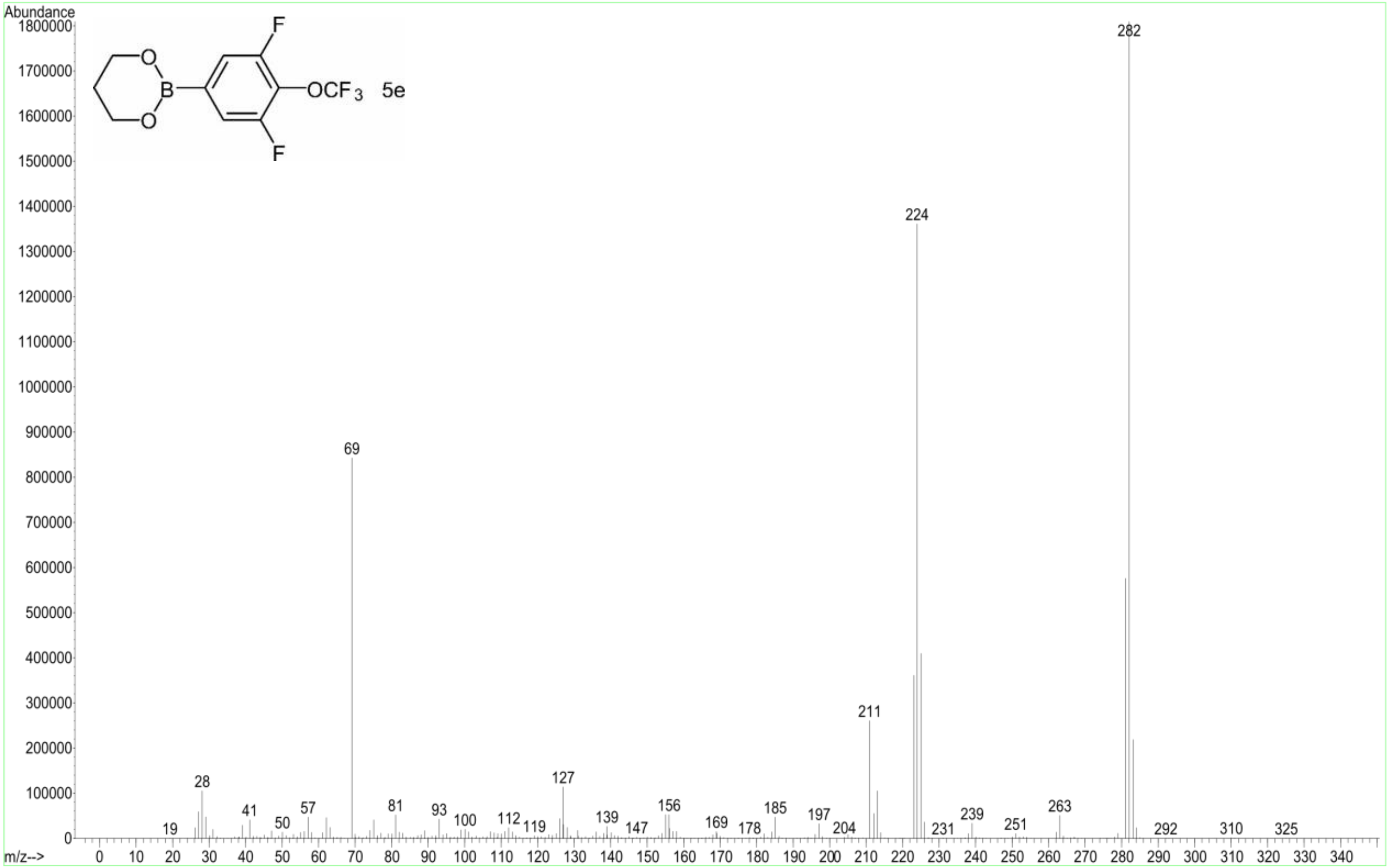

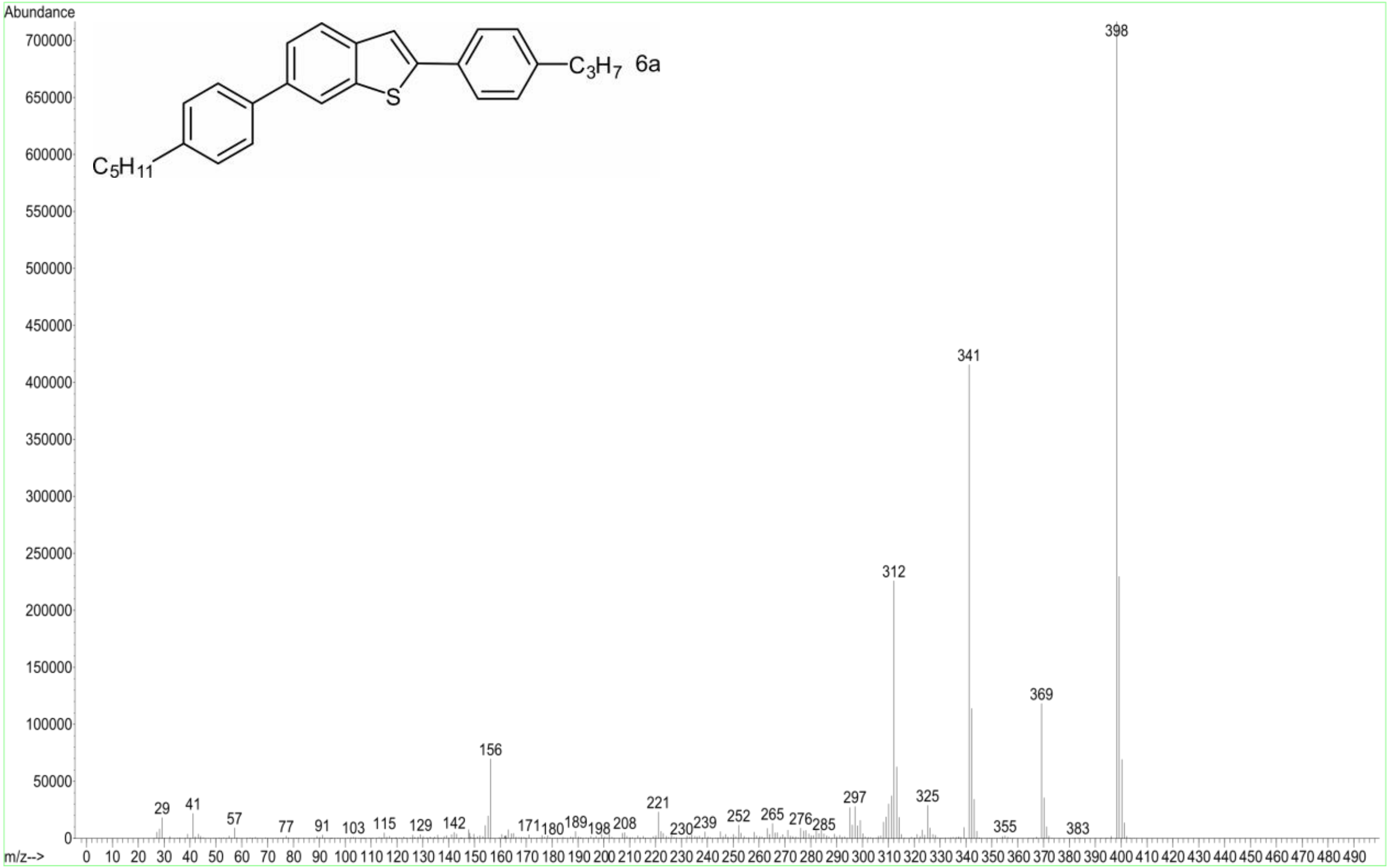

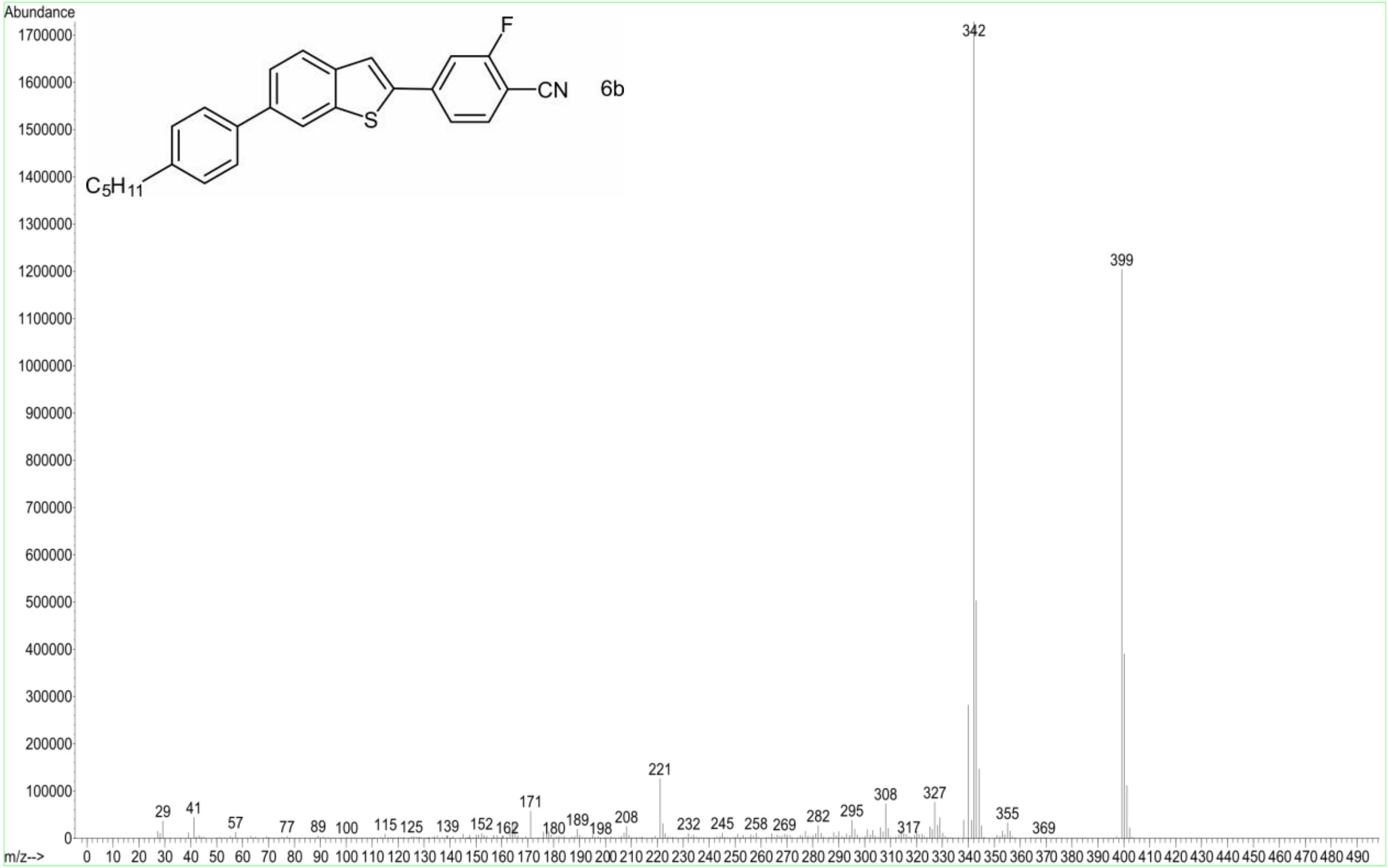

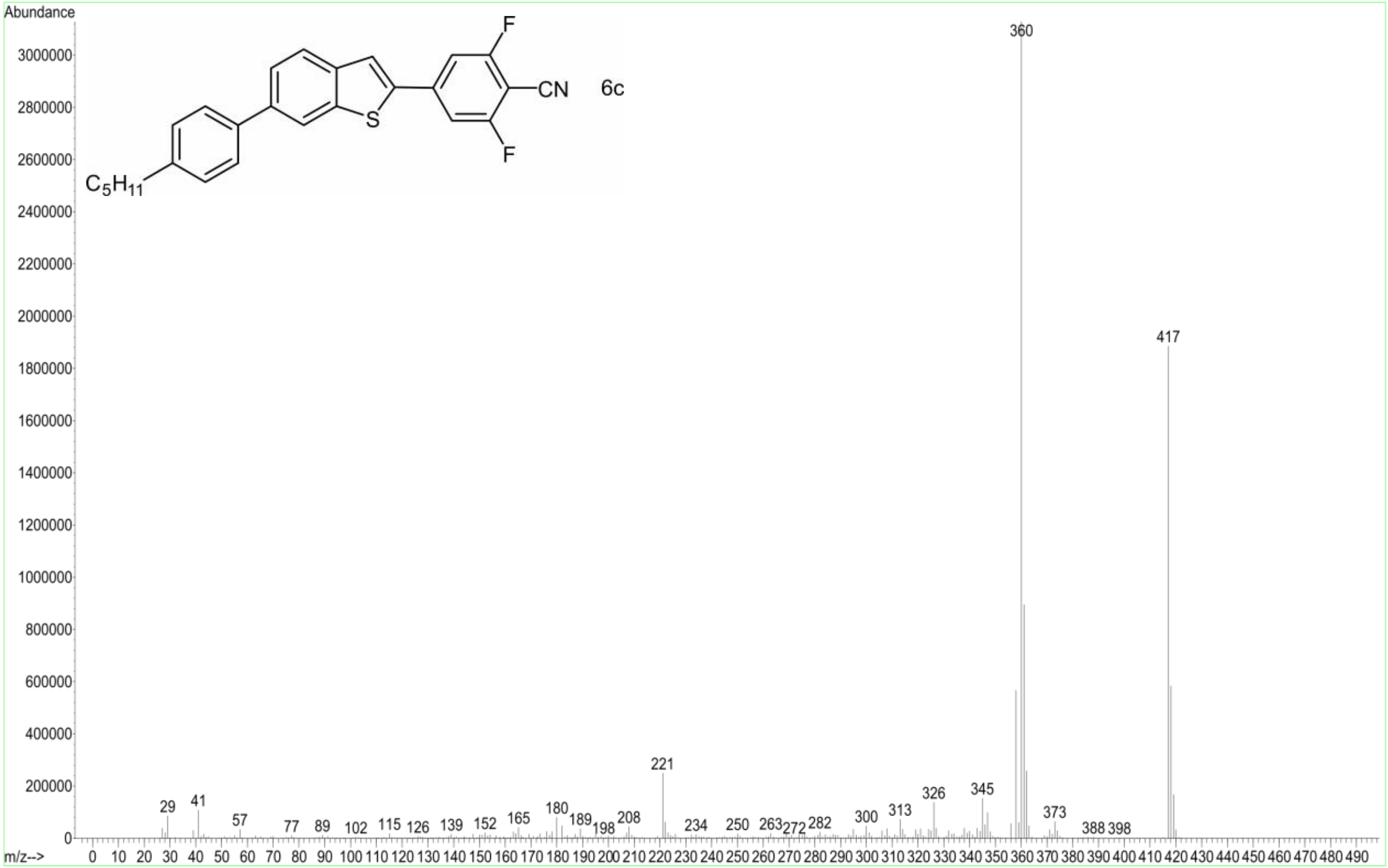

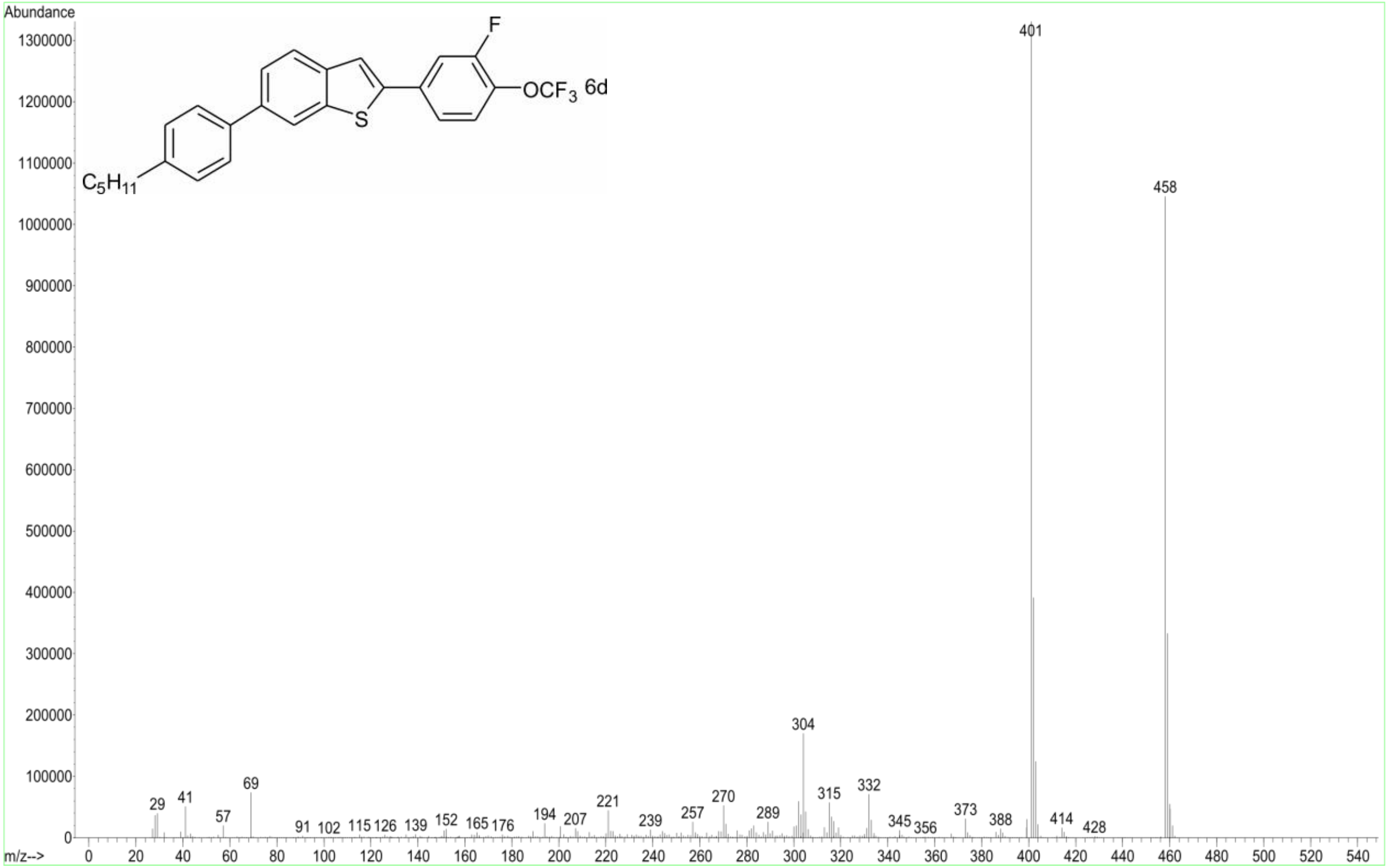

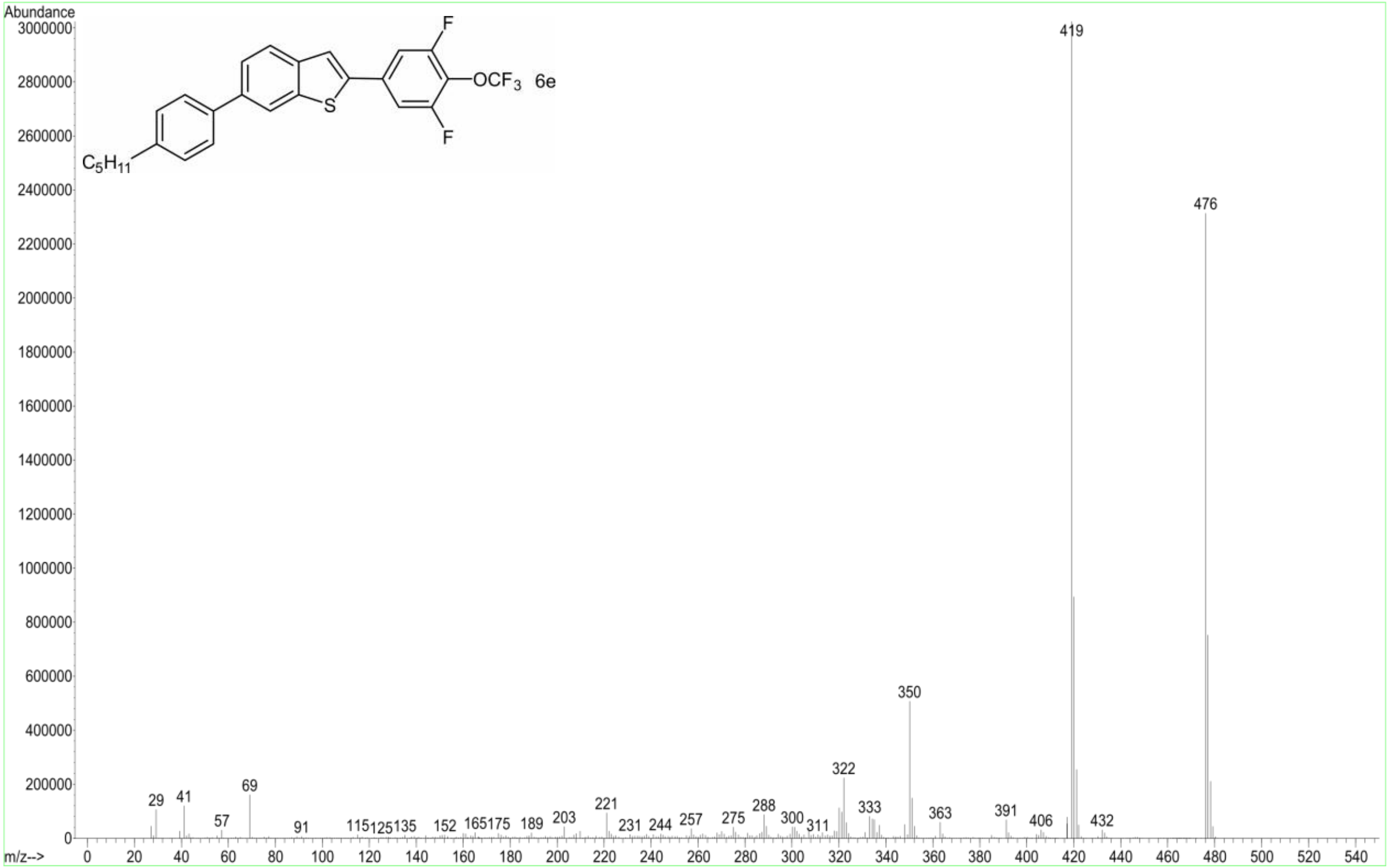

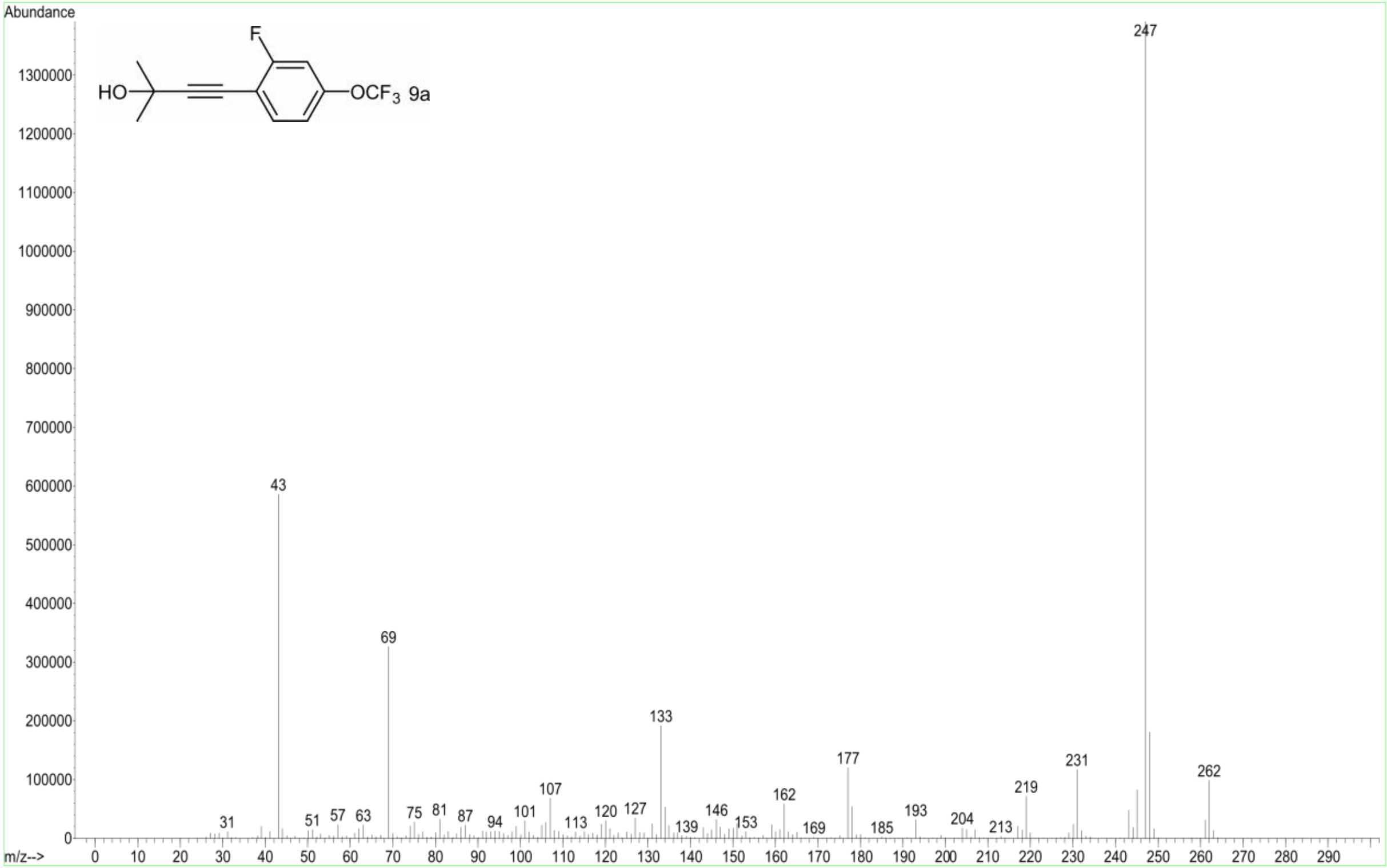

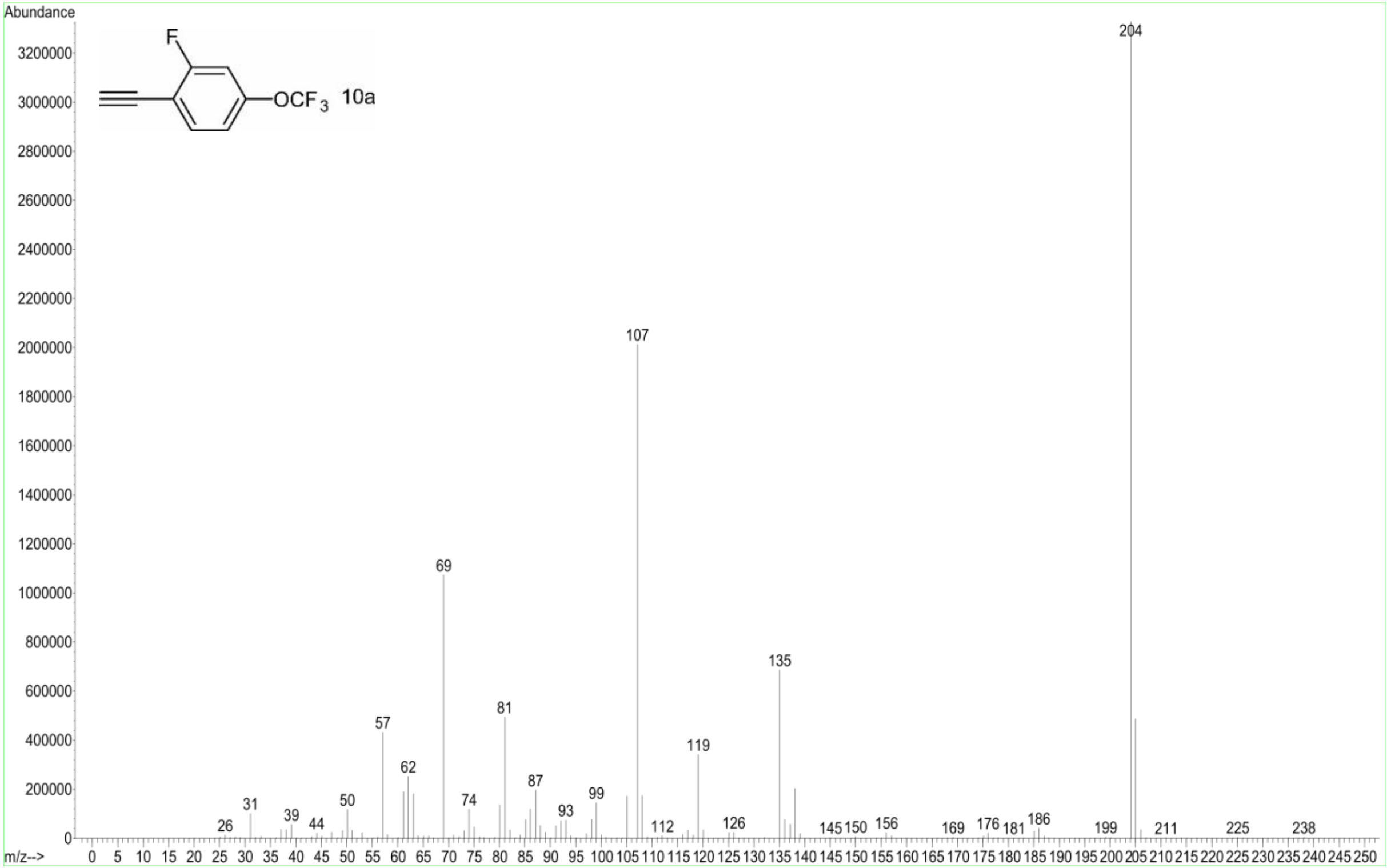

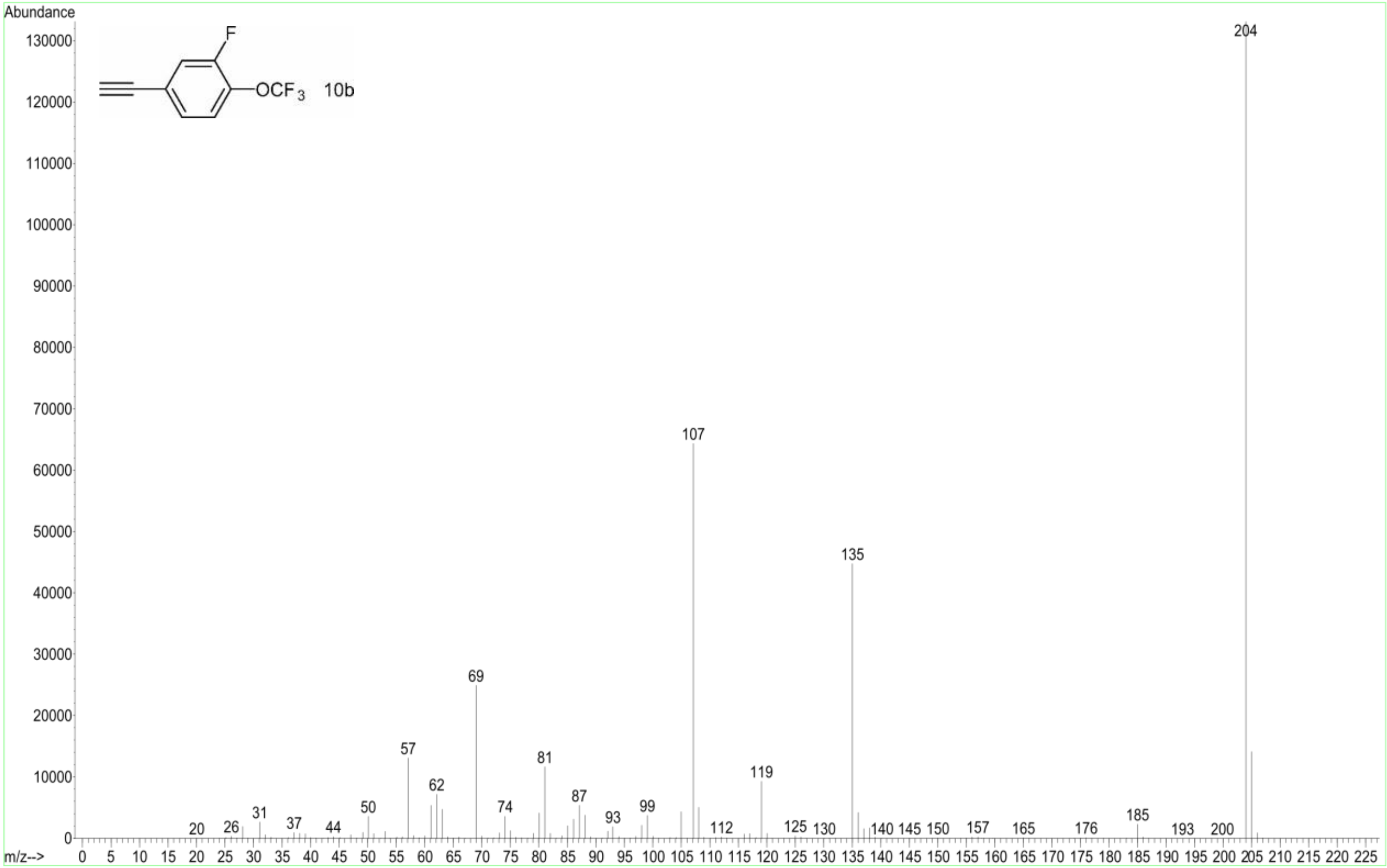

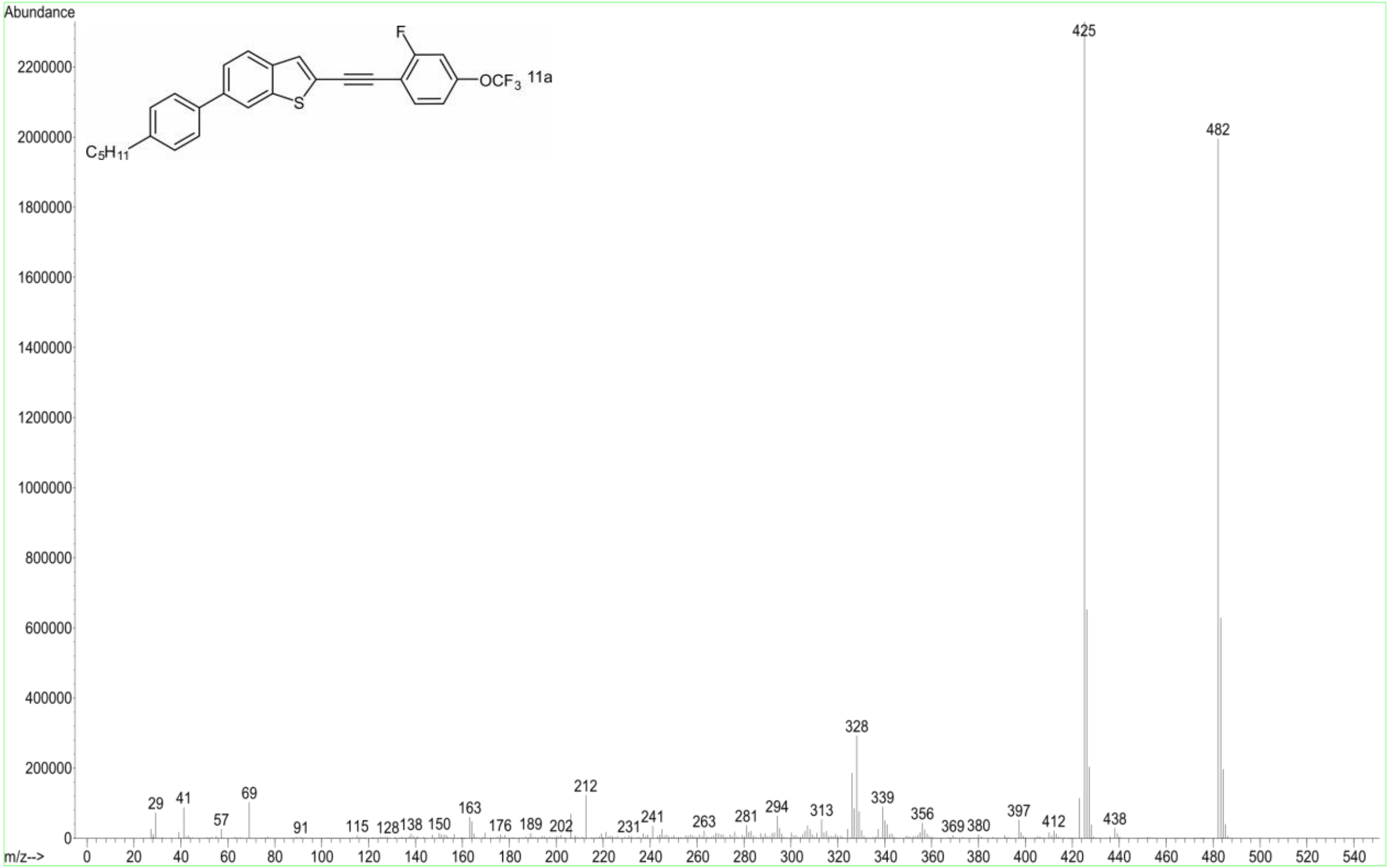

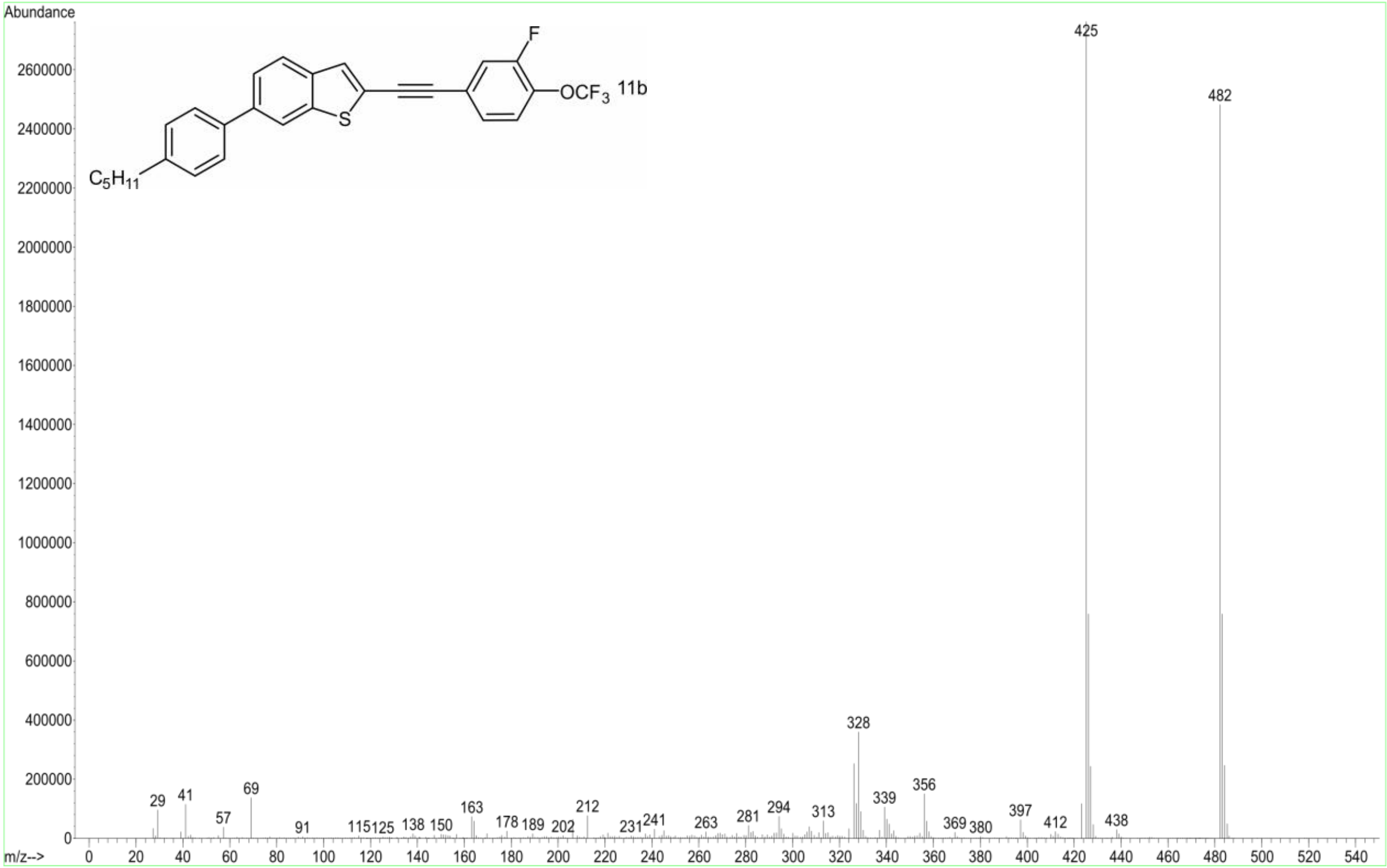

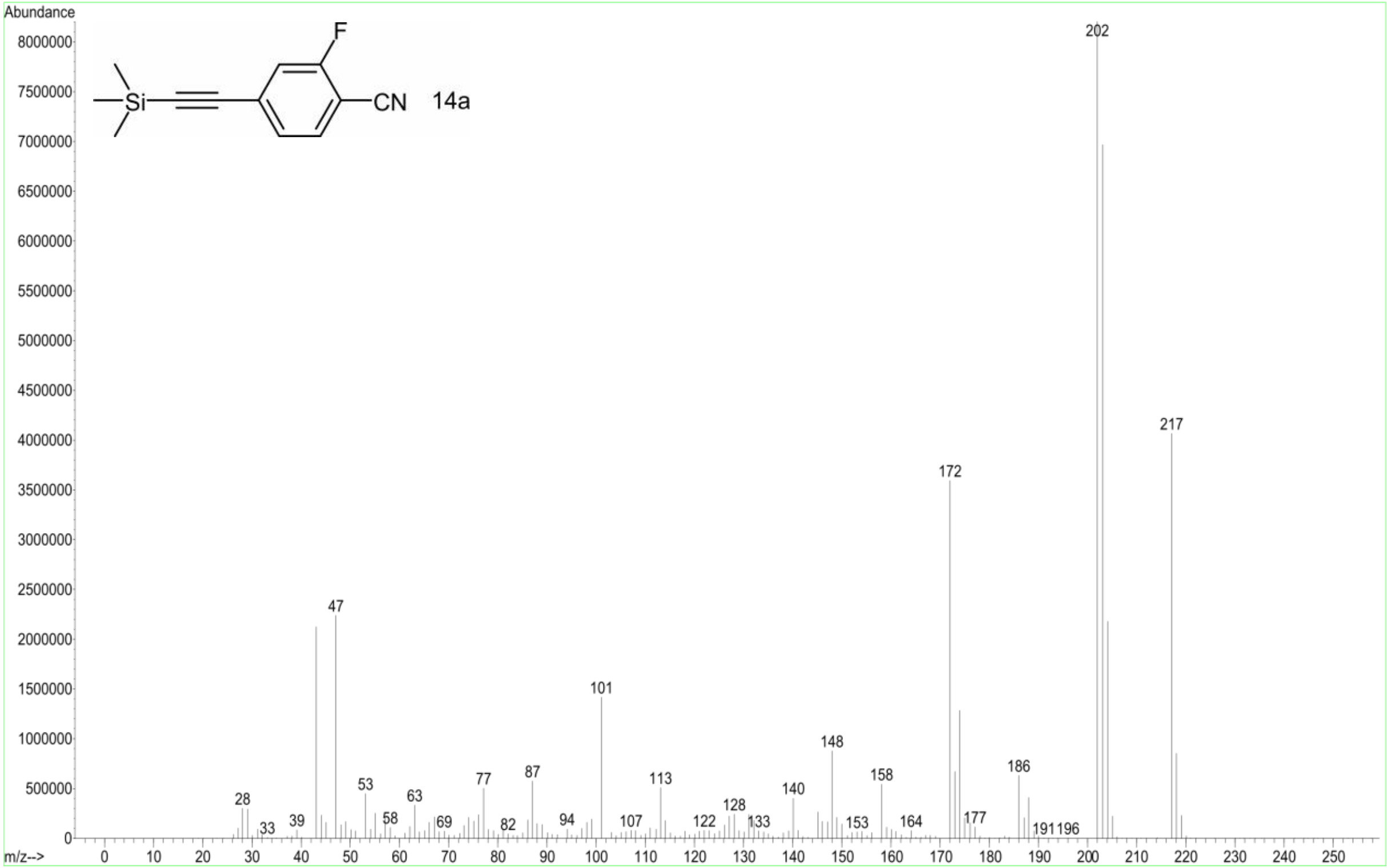

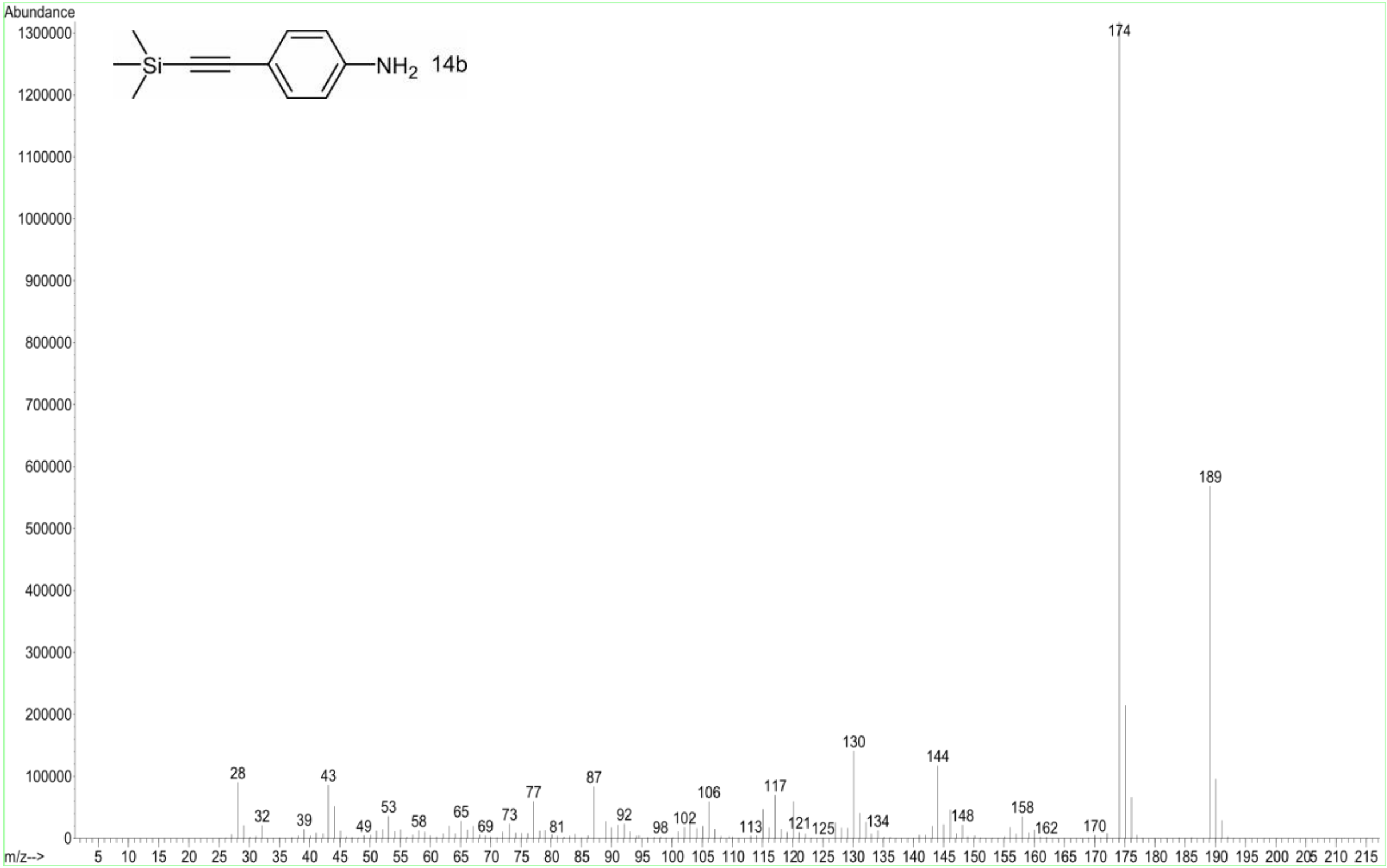

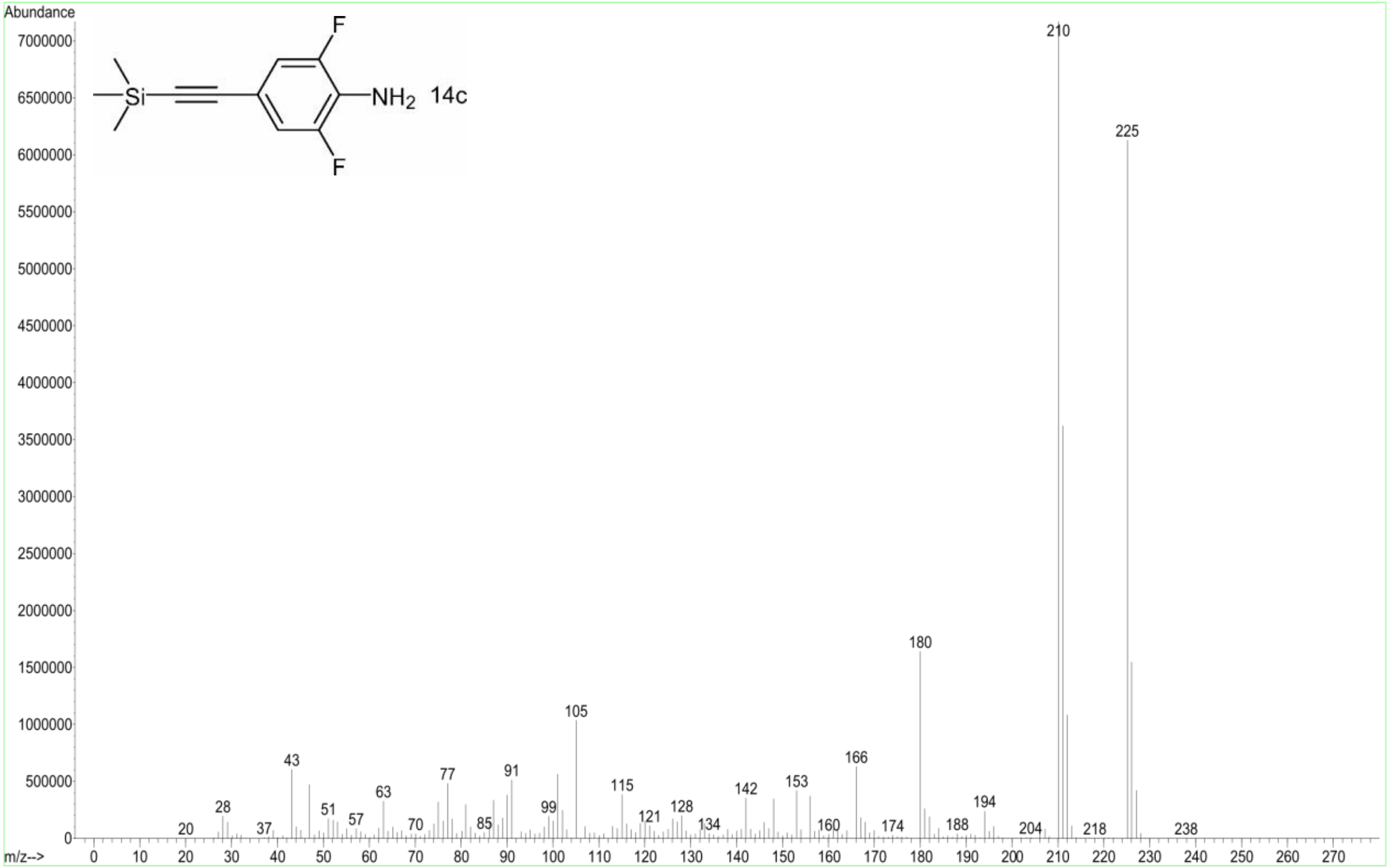

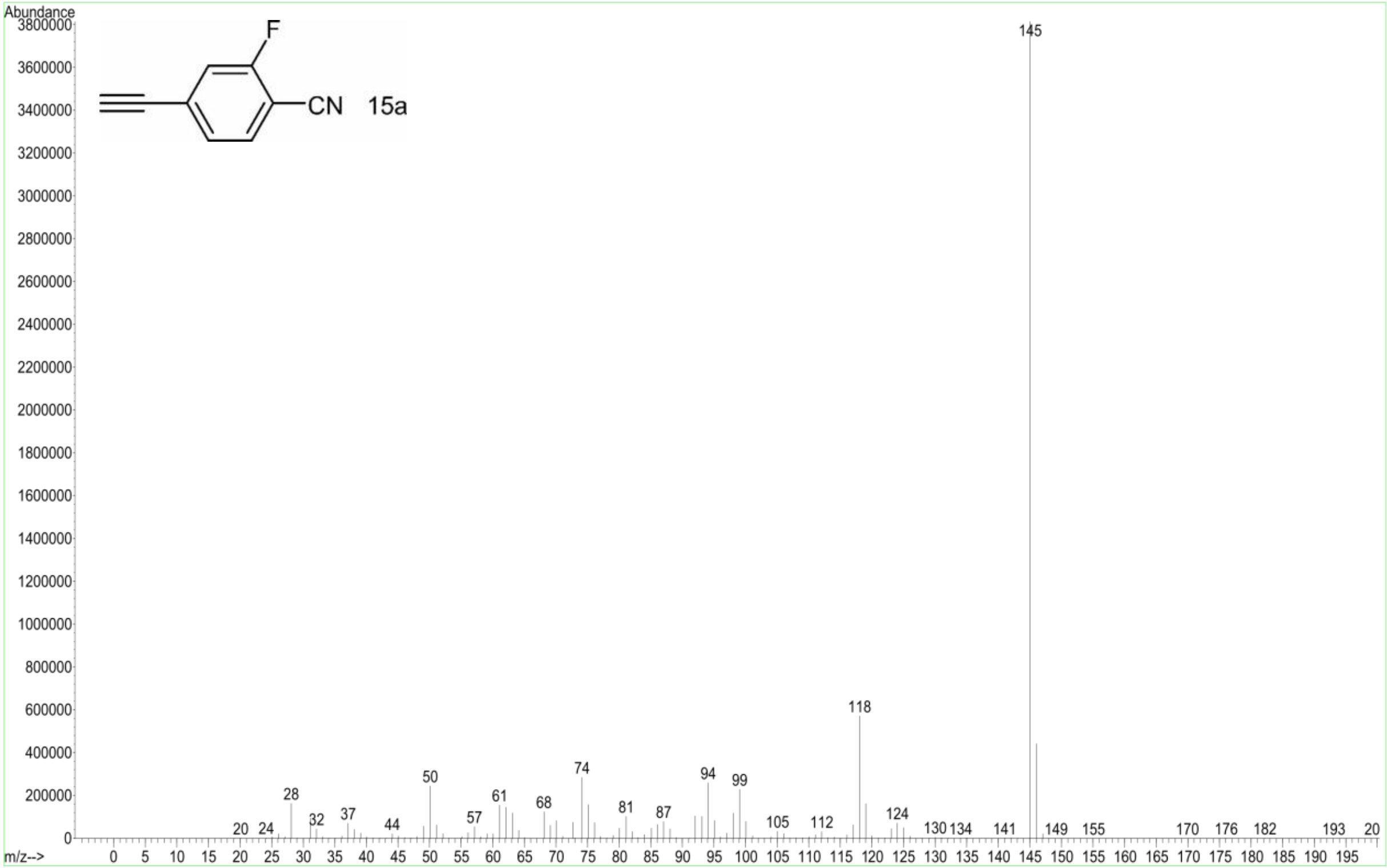

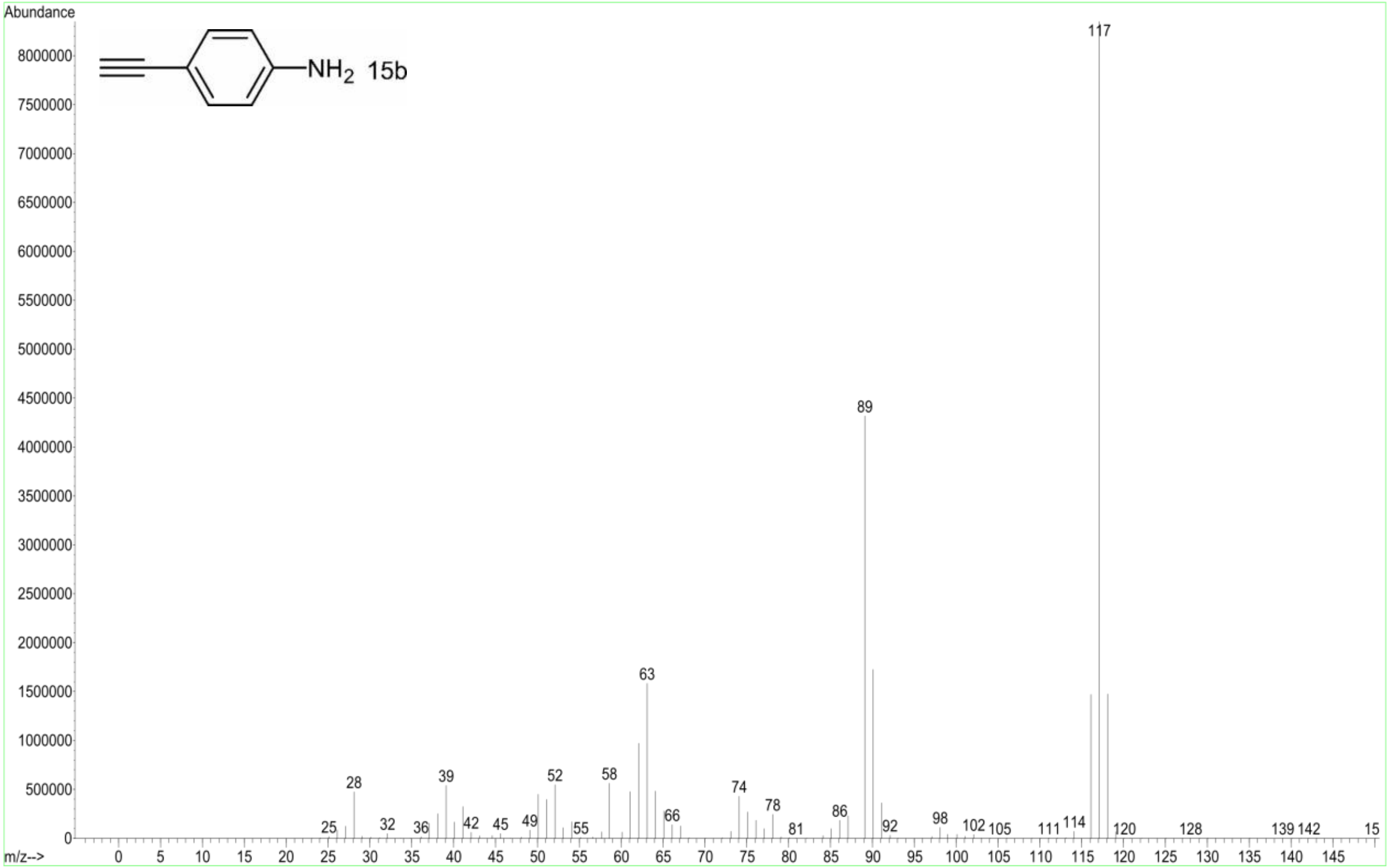

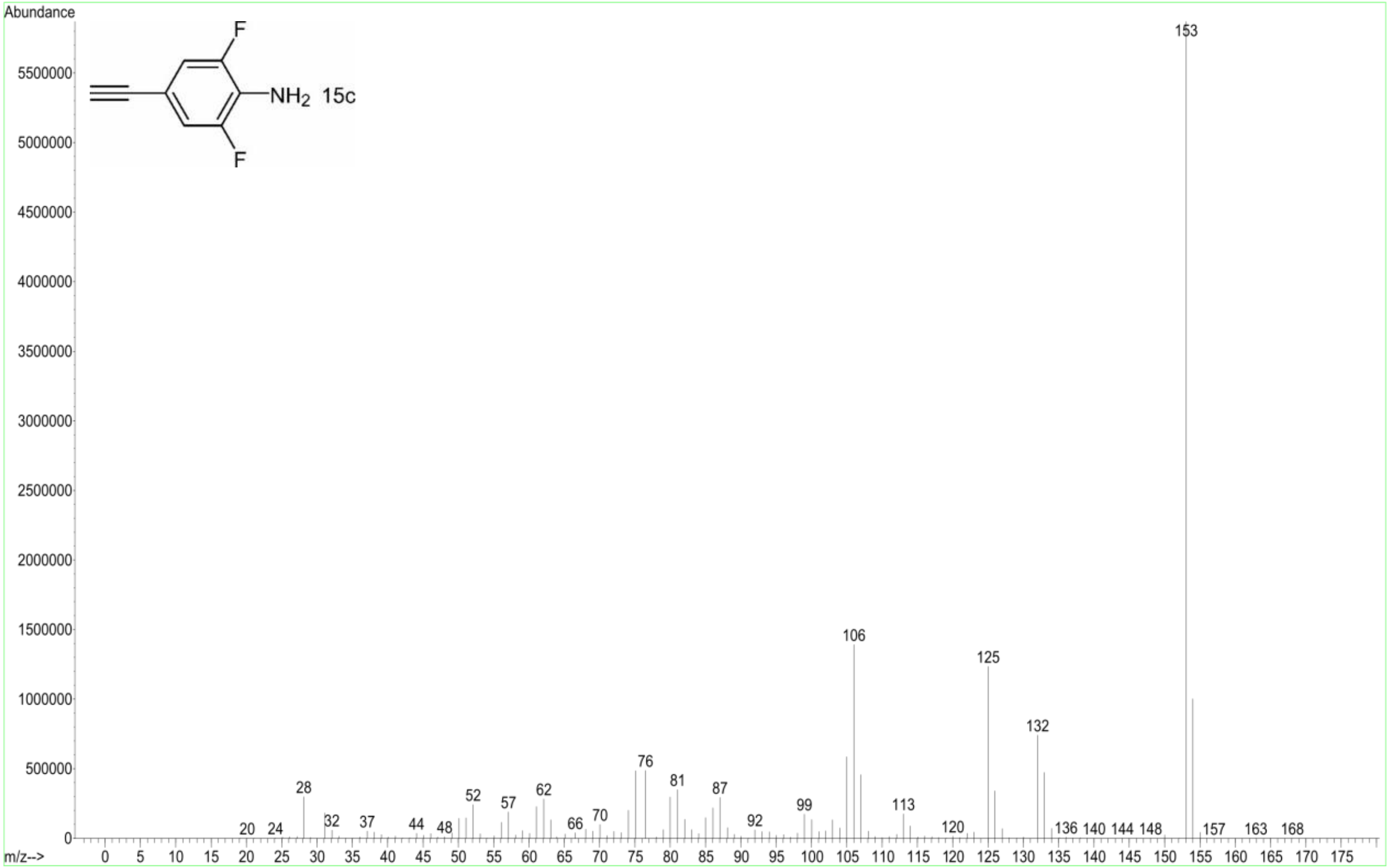

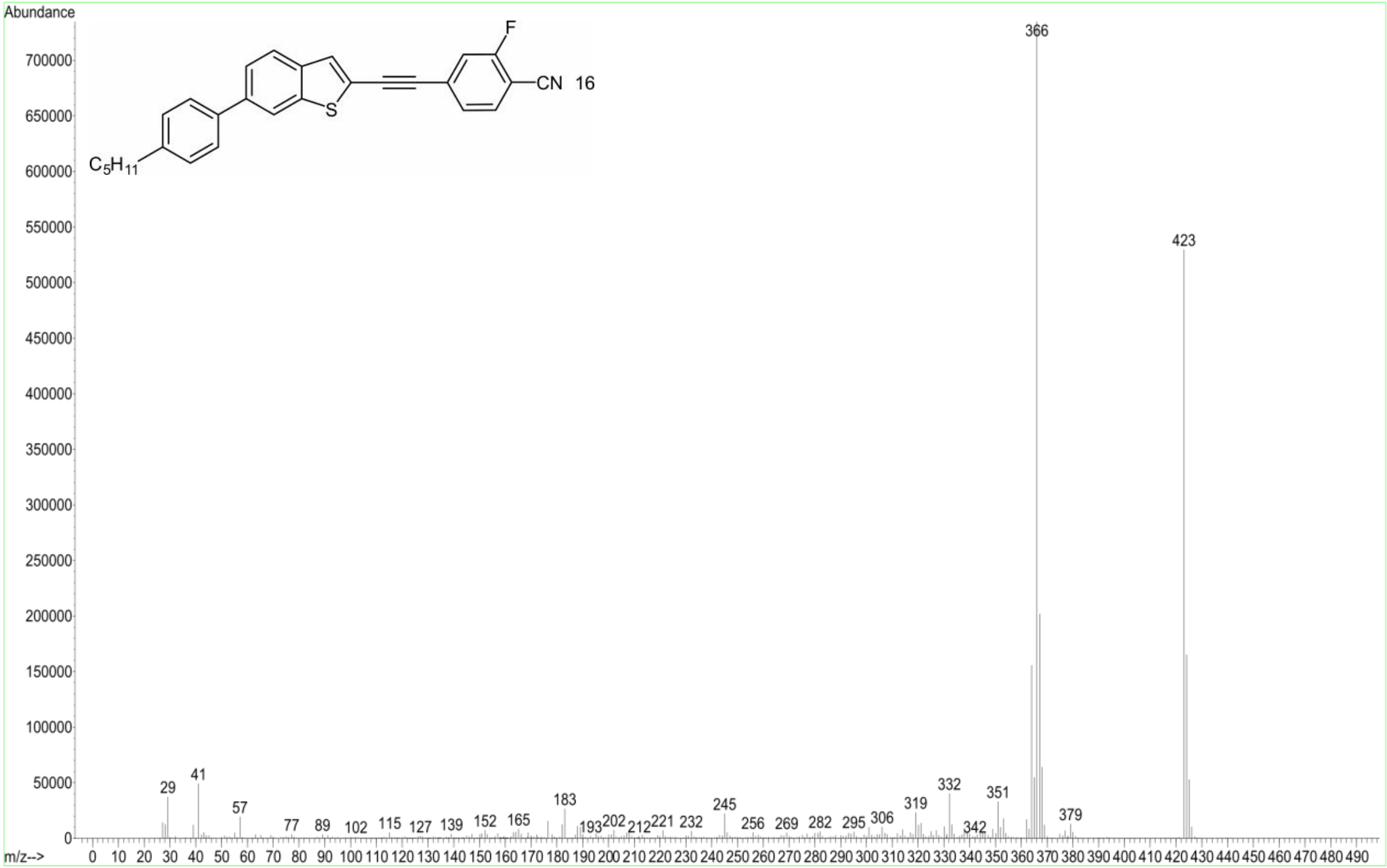

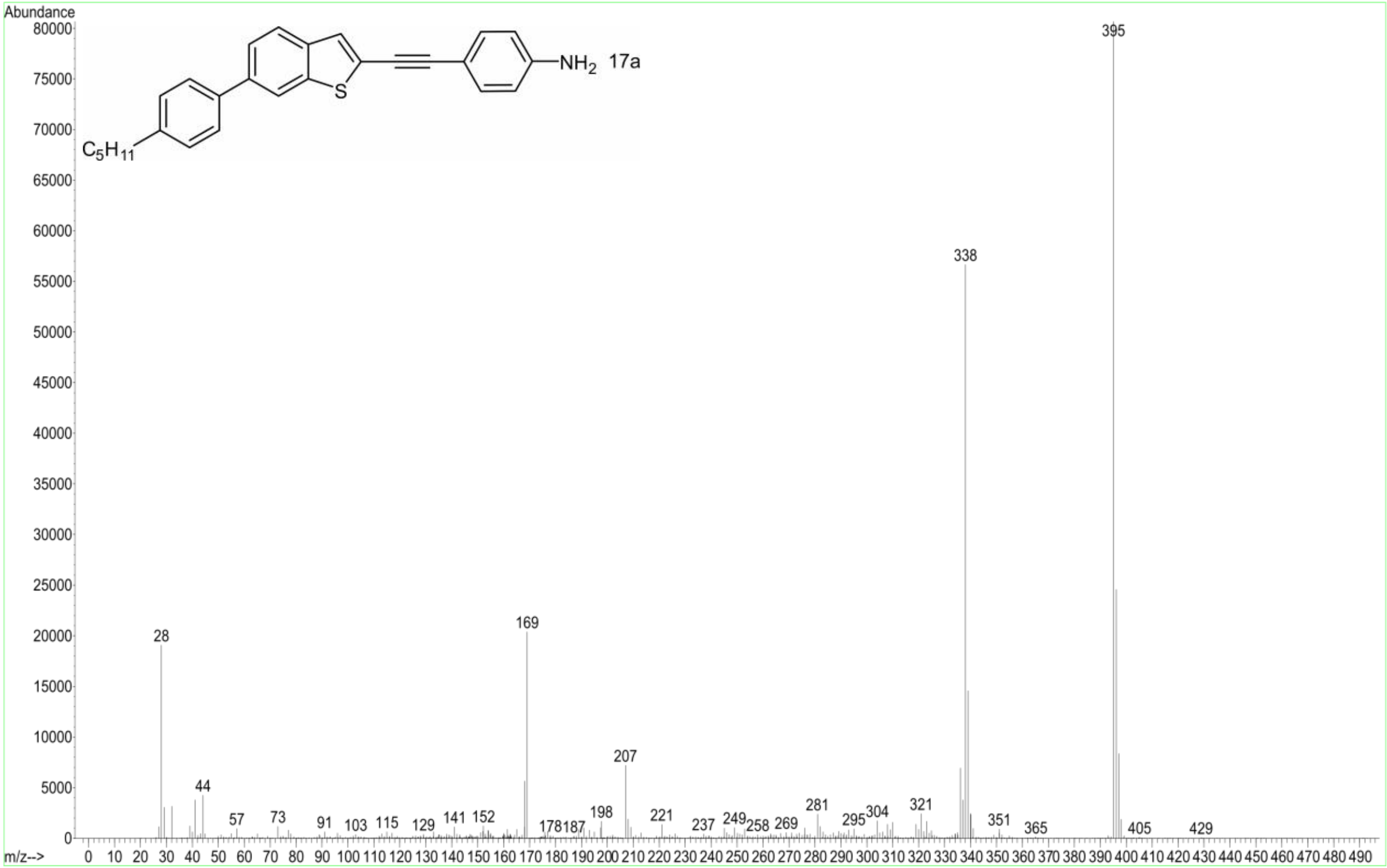

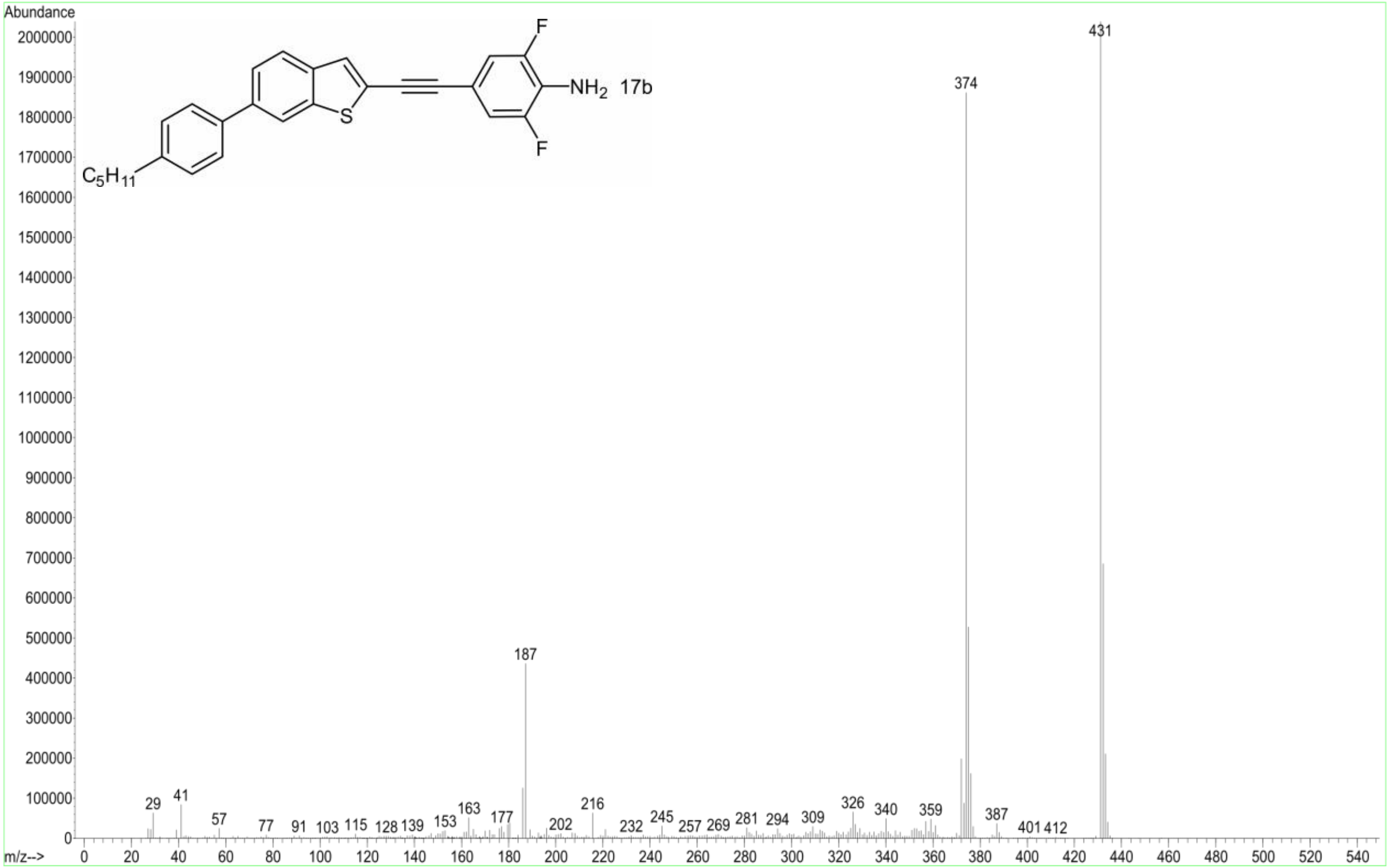

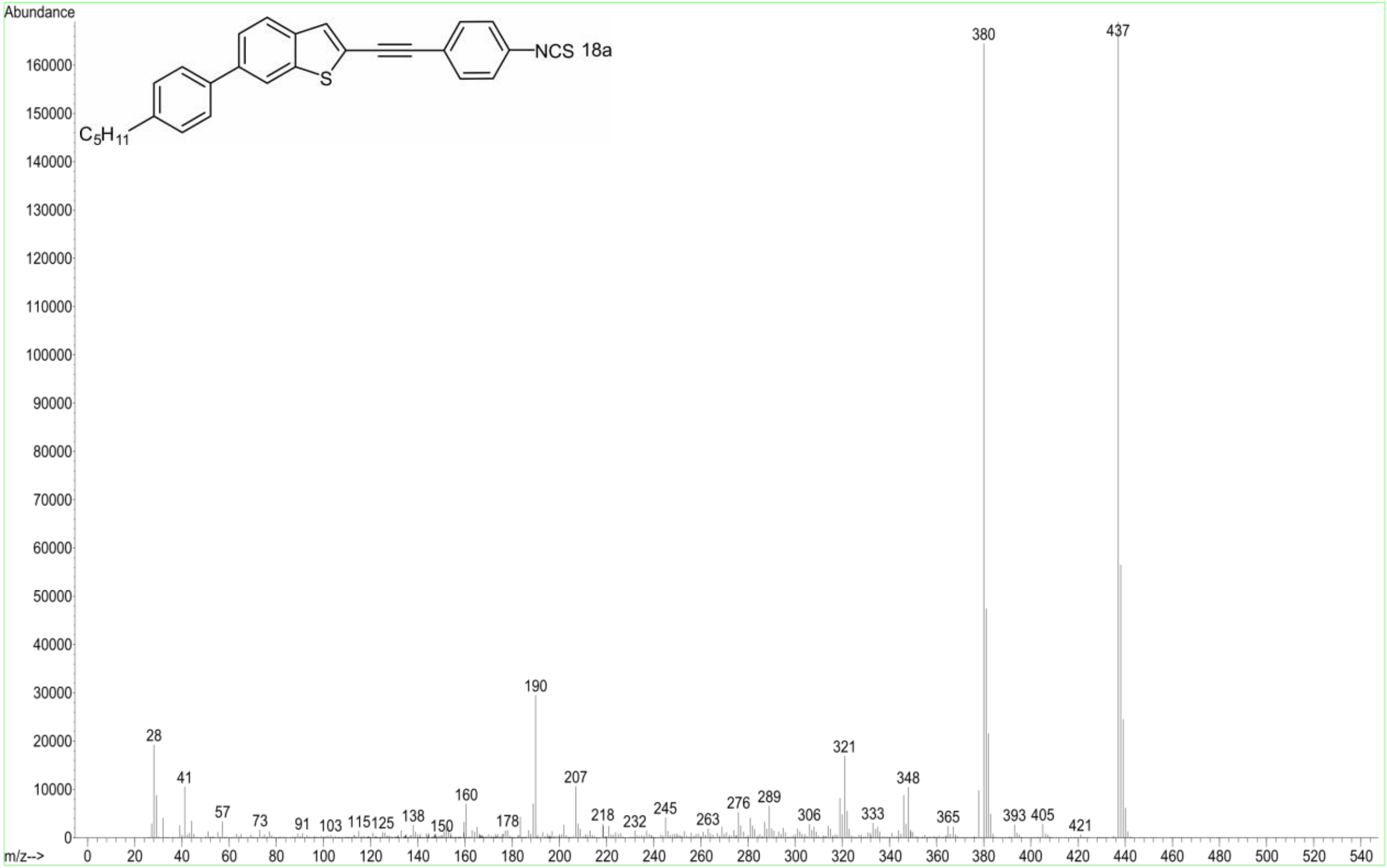

1

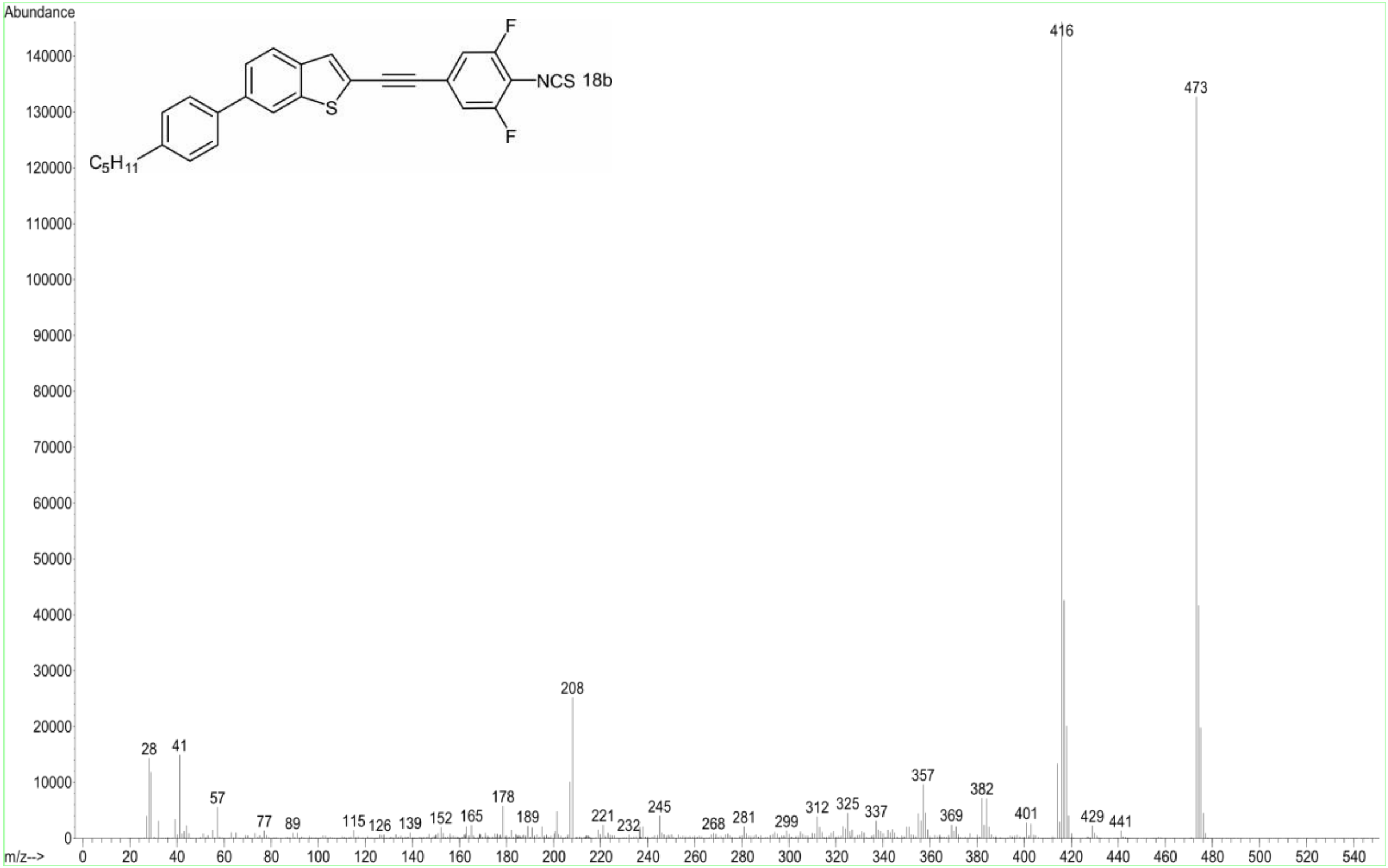

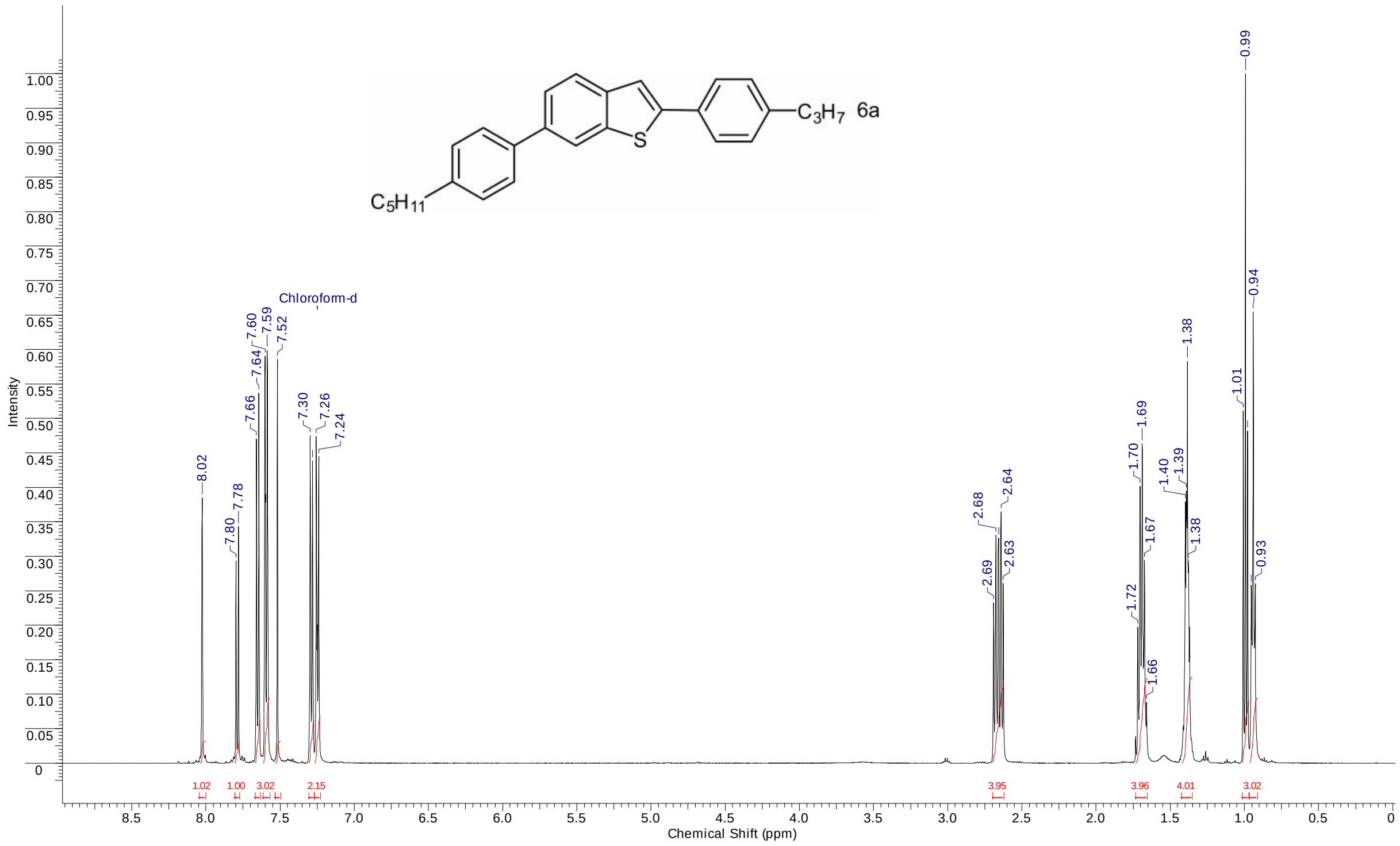

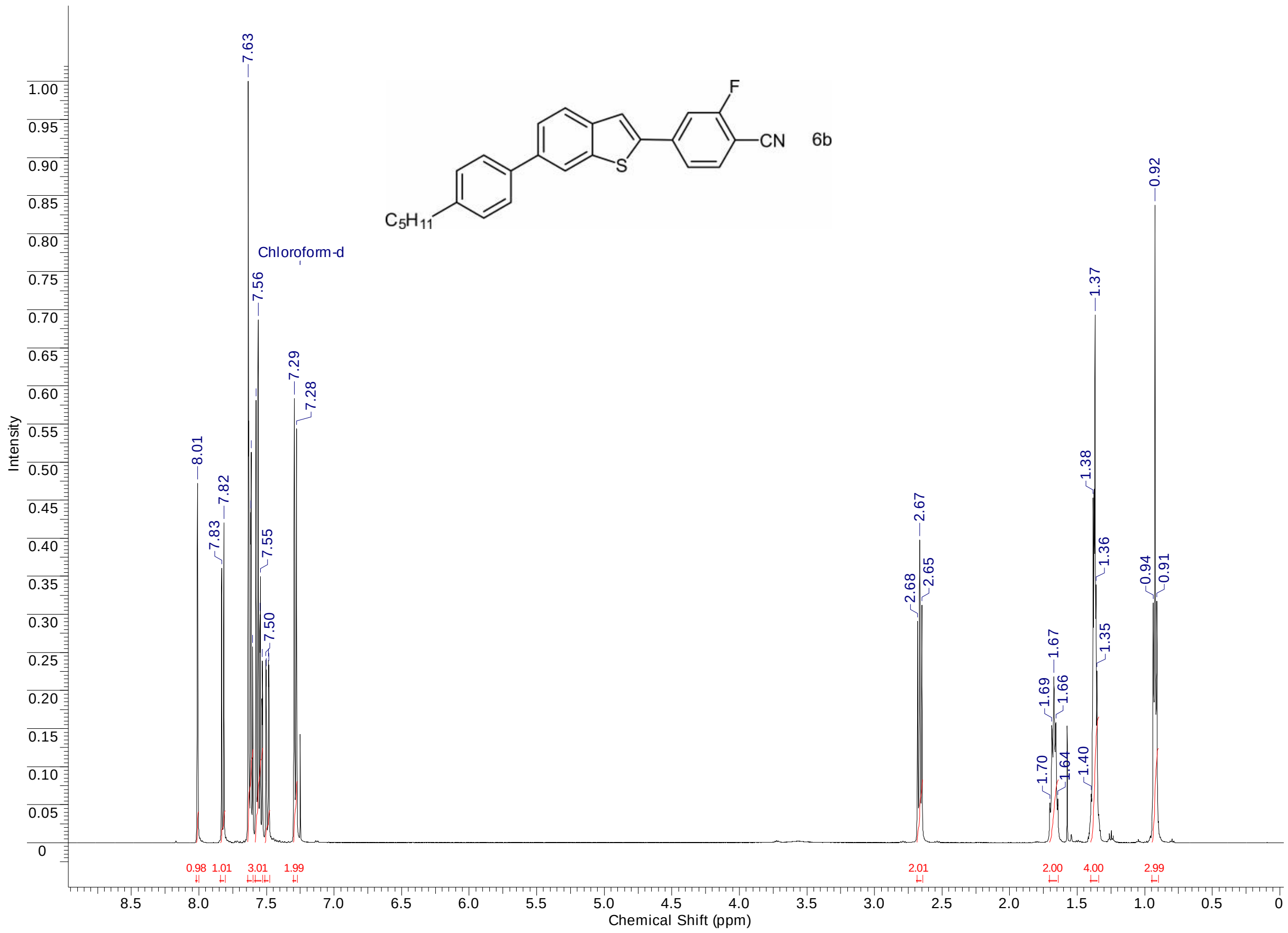

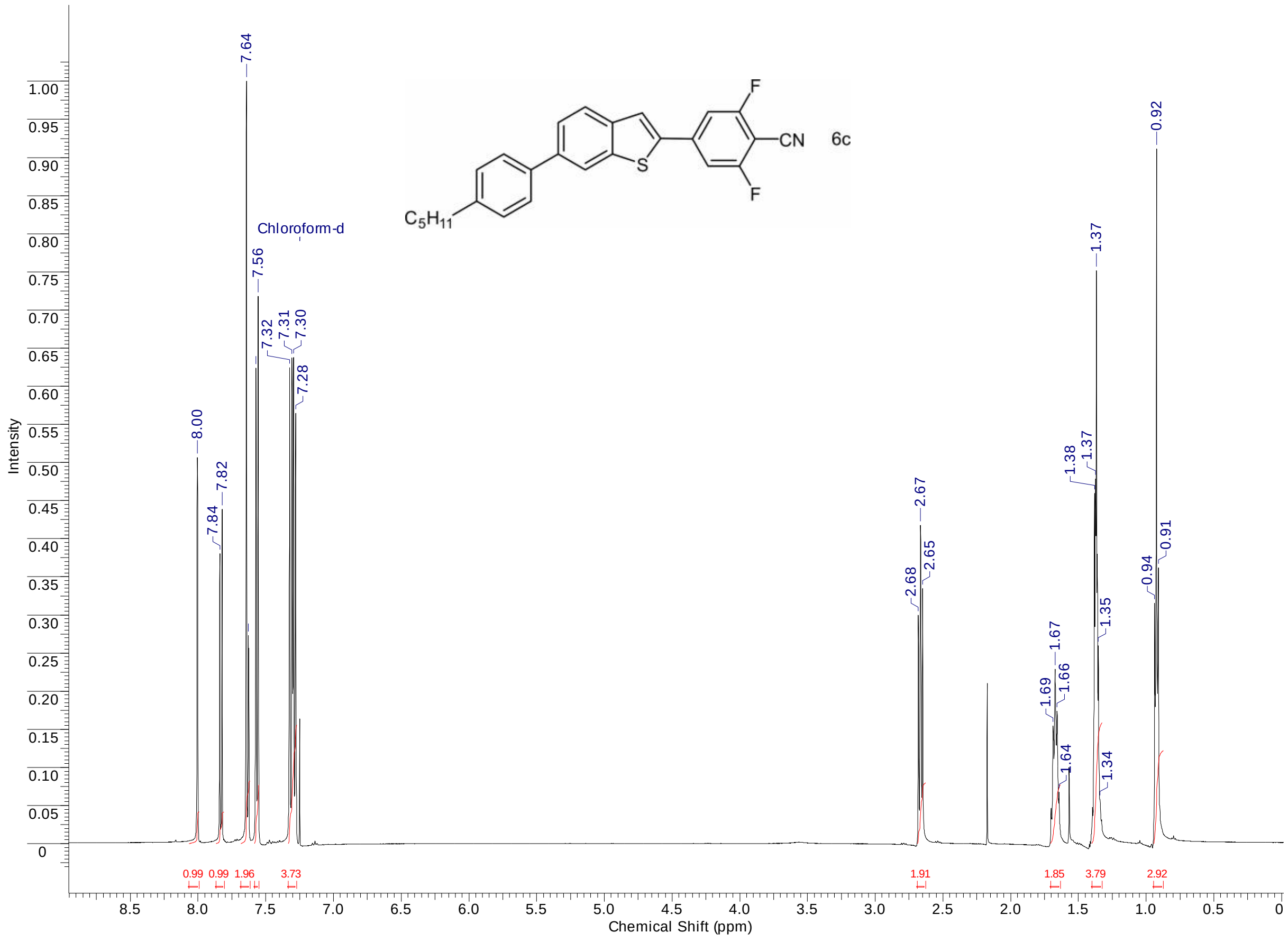

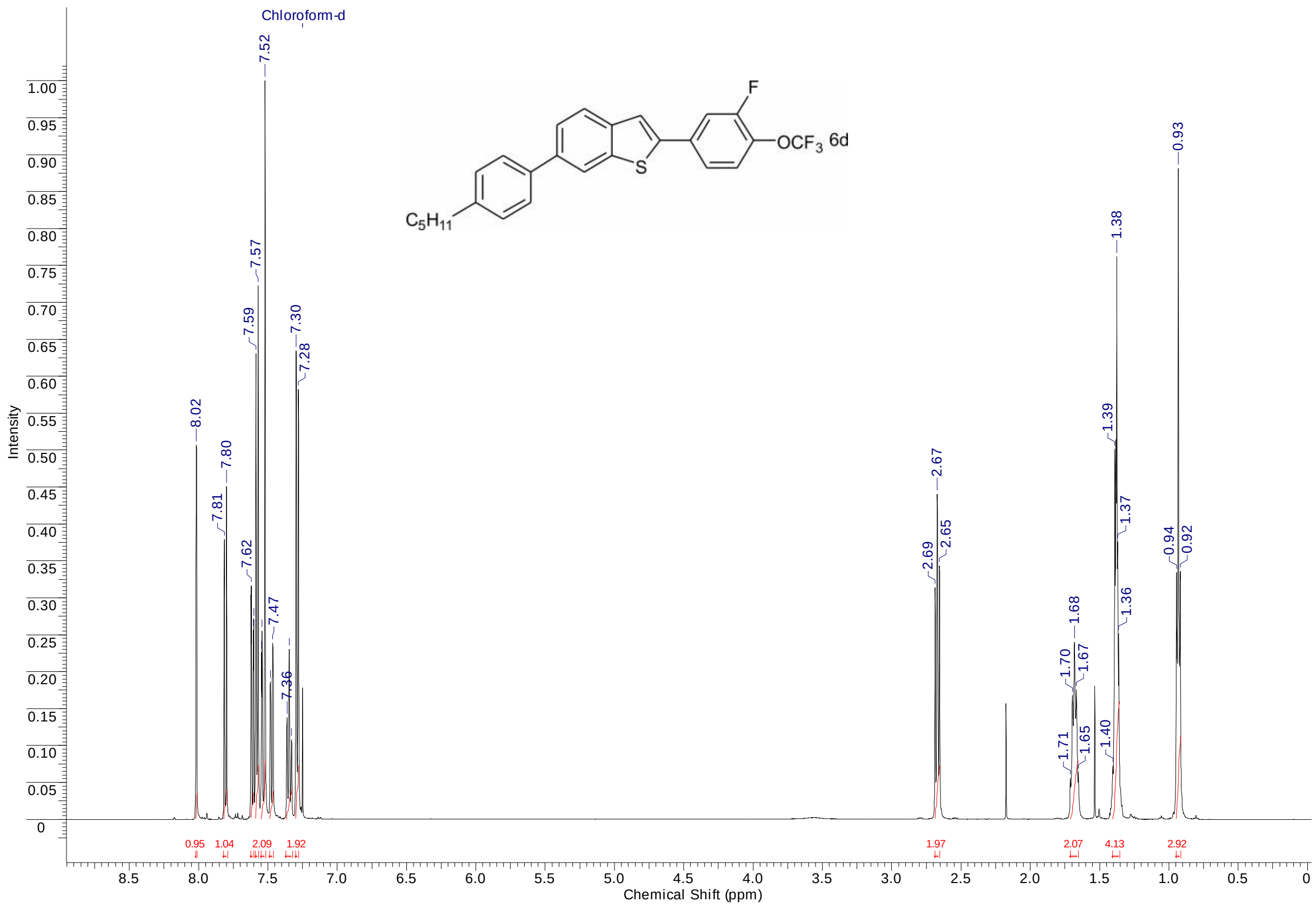

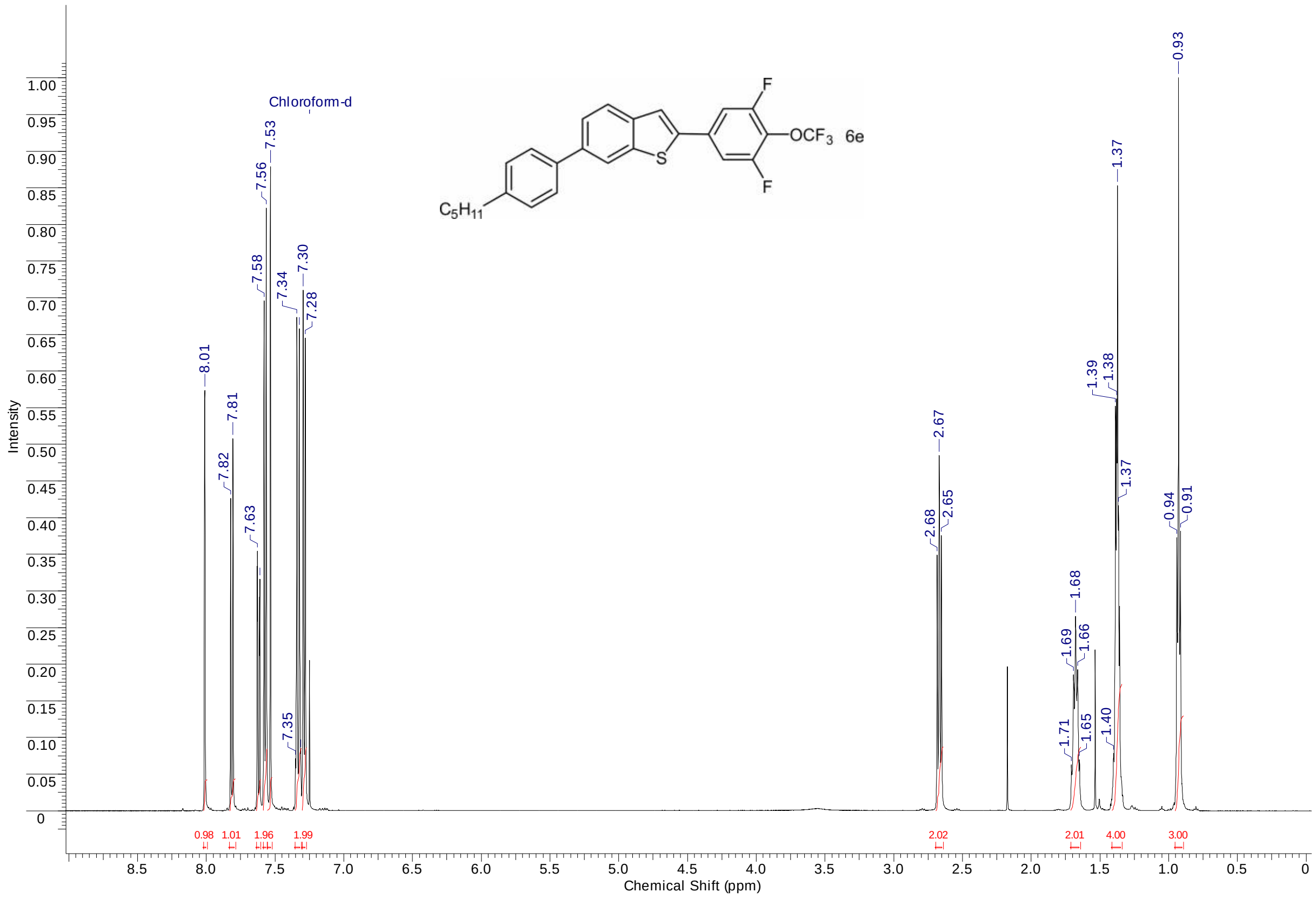
Chloroform-d
C5H11
S
F
OCF3 6e
F
Intensity
Chemical Shift (ppm)
8.01 7.82 7.81 7.63 7.58 7.56 7.53 7.35 7.34 7.30 7.28
2.68 2.67 2.65
1.71 1.69 1.68 1.66 1.65 1.40 1.39 1.38 1.37 1.37 1.37
0.94 0.93 0.91
0.98 1.01 1.96 1.99 2.02 2.01 4.00 3.00

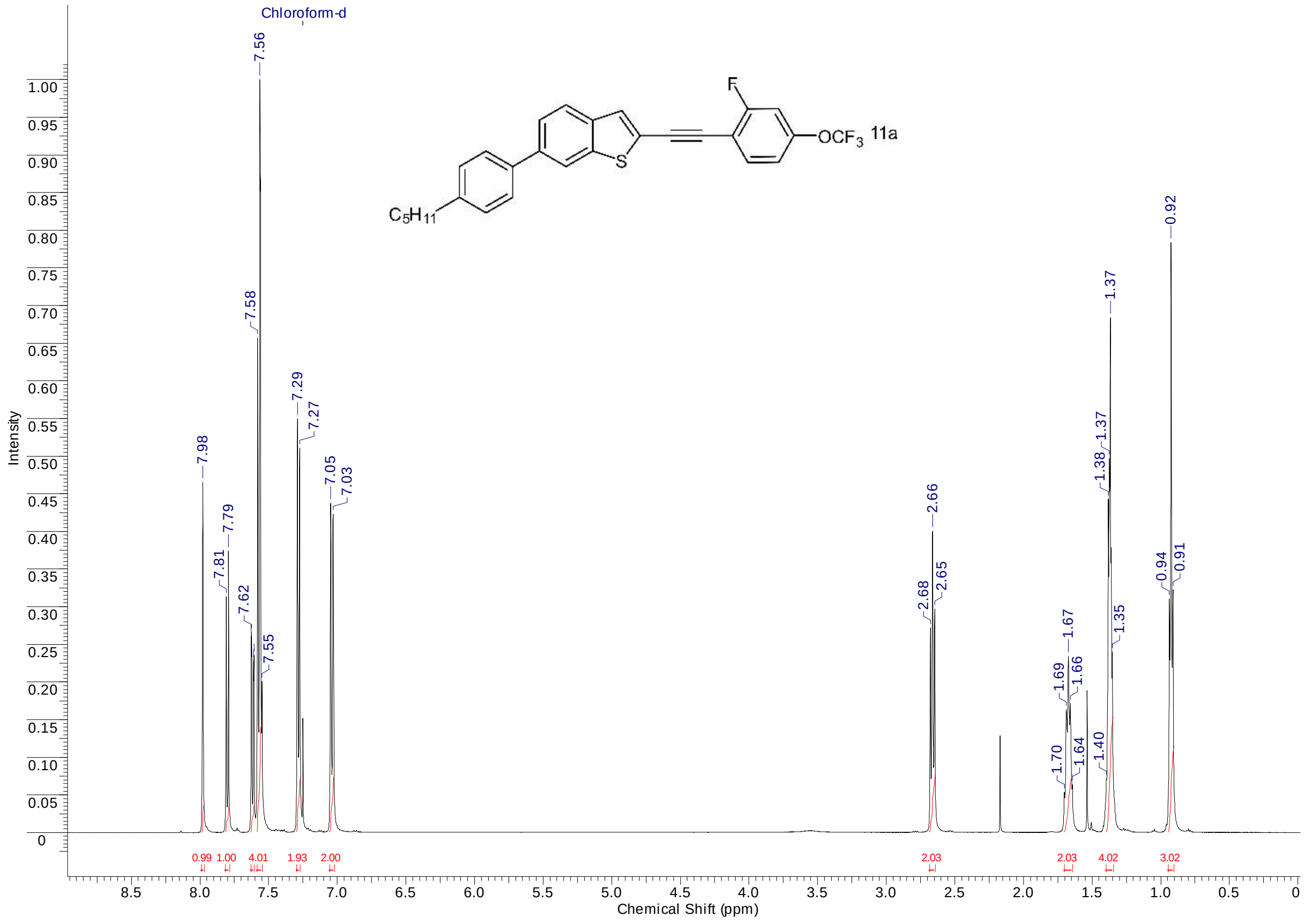
Chloroform-d
F
OCF3 11a
S
C5H11
Intensity
Chemical Shift (ppm)
7.98 7.81 7.79 7.62 7.58 7.56 7.55 7.29 7.27 7.05 7.03
2.68 2.66 2.65
1.70 1.69 1.67 1.66 1.64 1.40 1.38 1.37 1.37 1.35
0.94 0.92 0.91
0.99 1.00 4.01 1.93 2.00 2.03 2.03 4.02 3.02

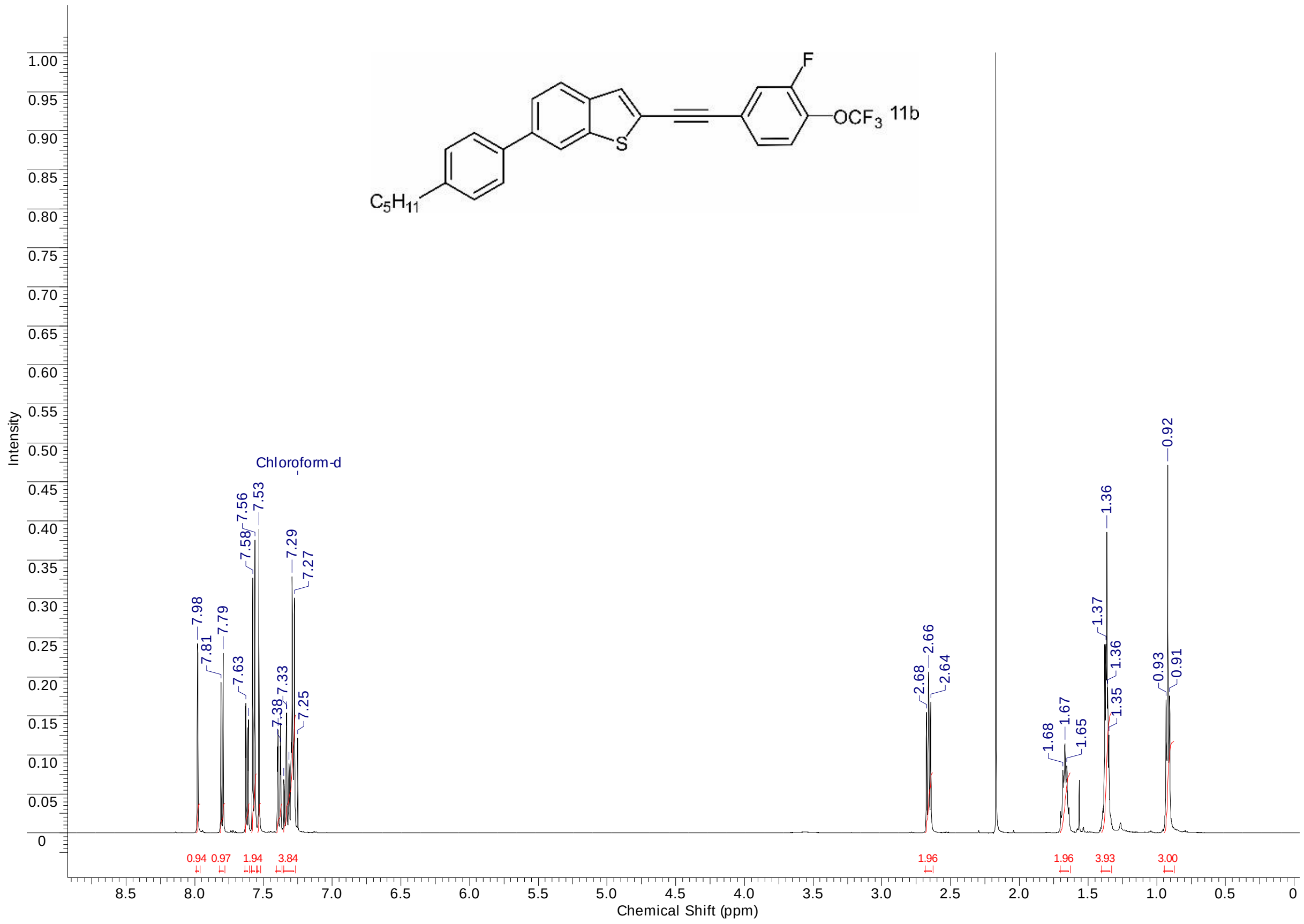
11b
Chloroform-d
Intensity
Chemical Shift (ppm)

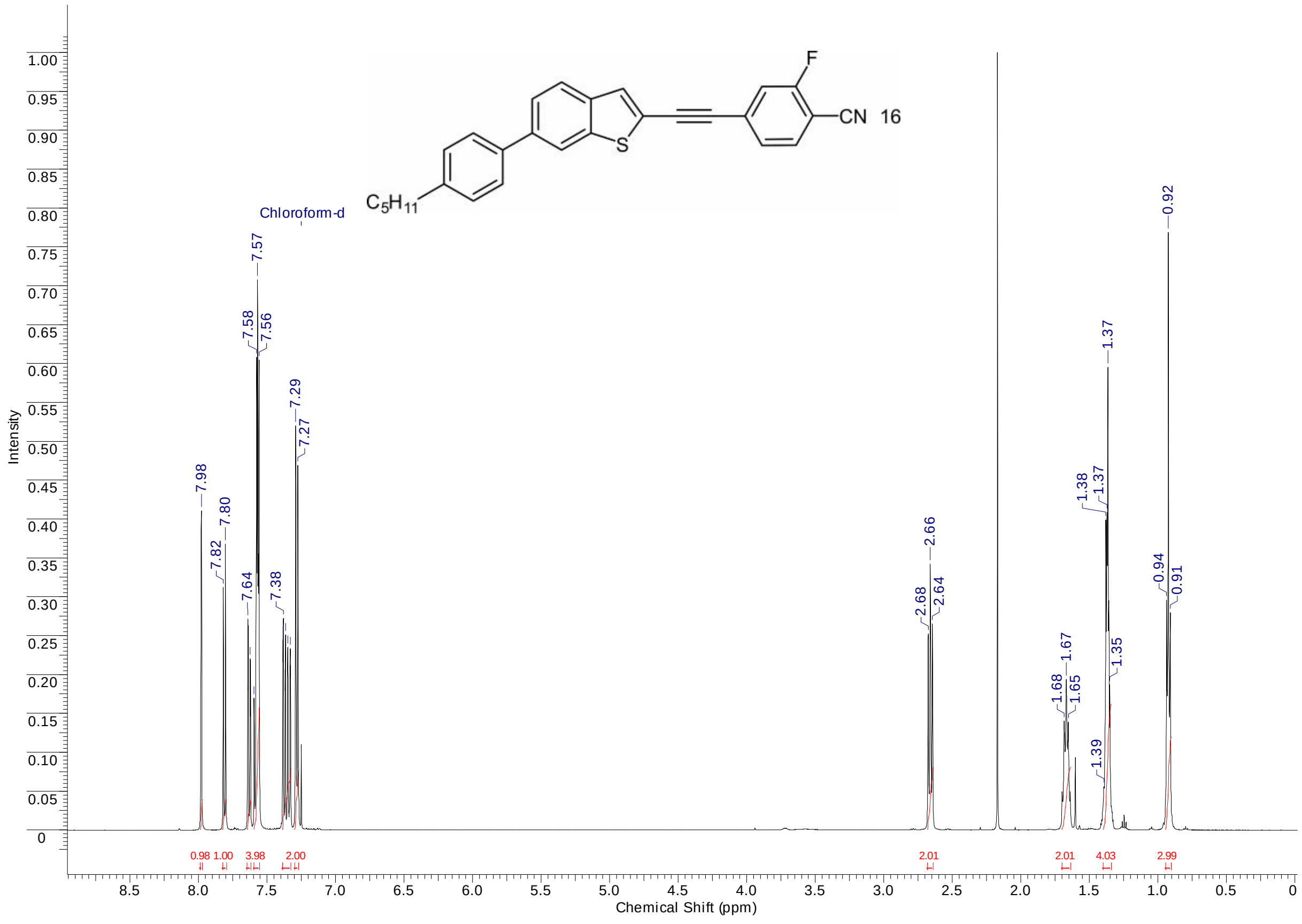
16
Chloroform-d
Intensity
Chemical Shift (ppm)

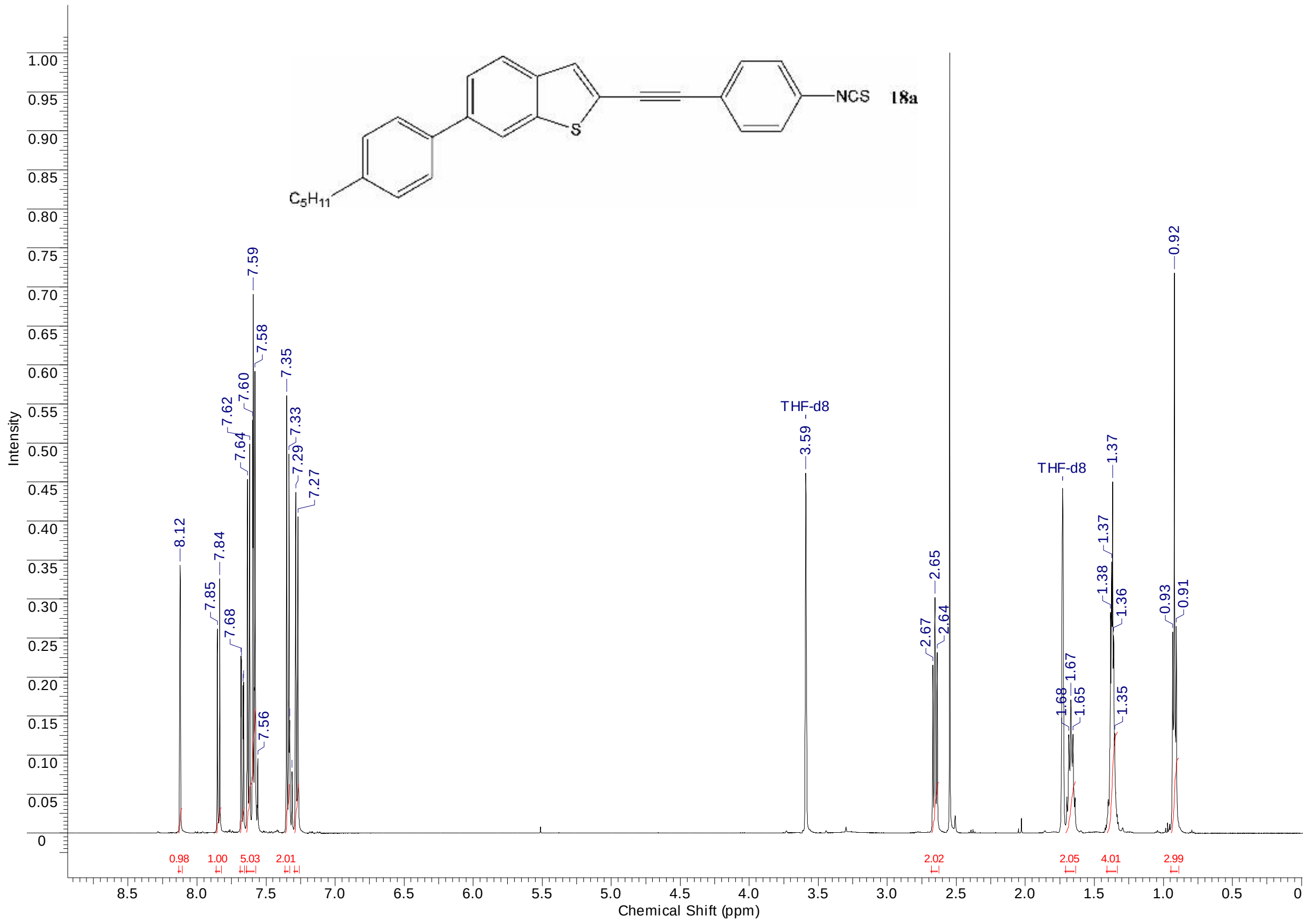

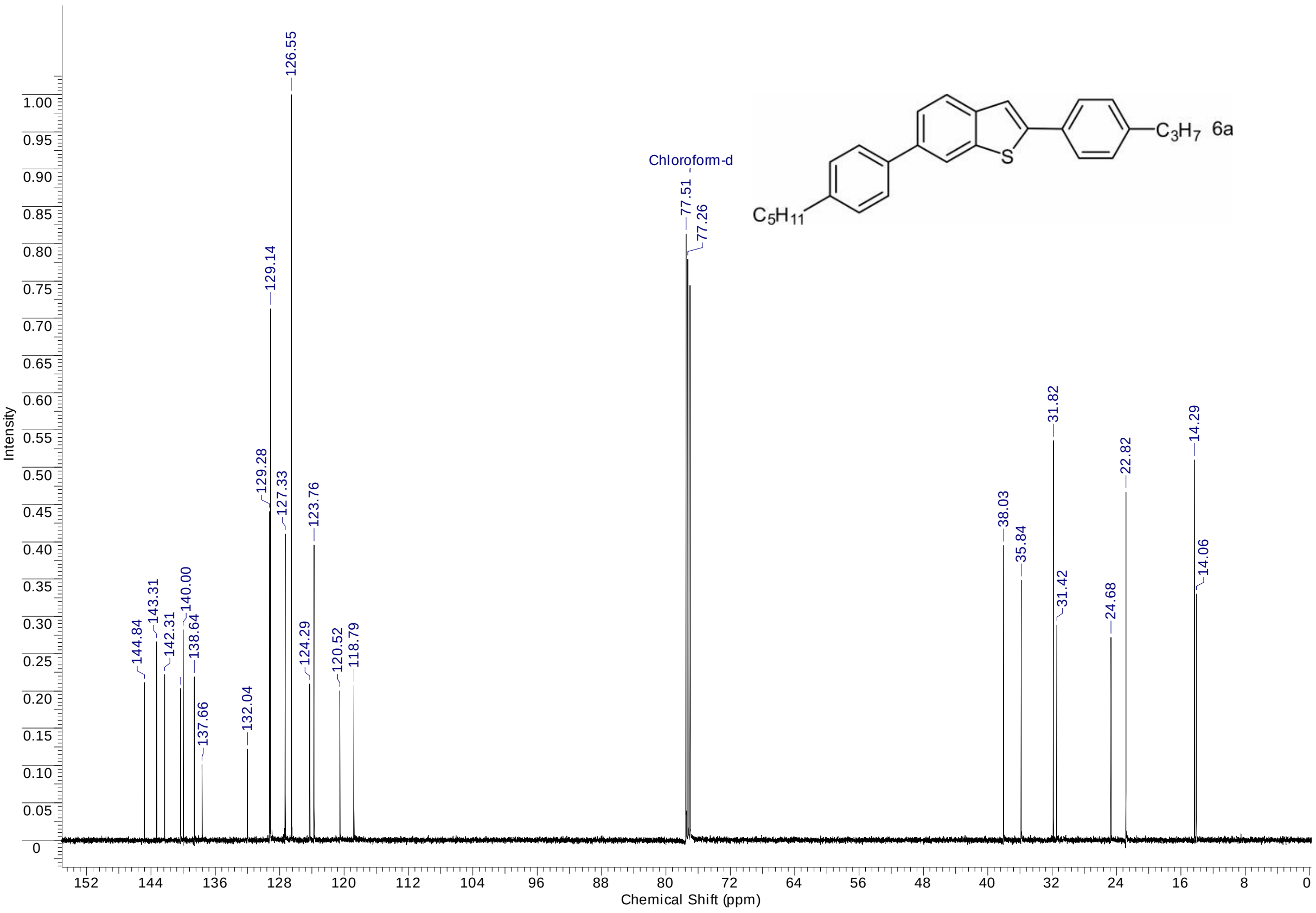

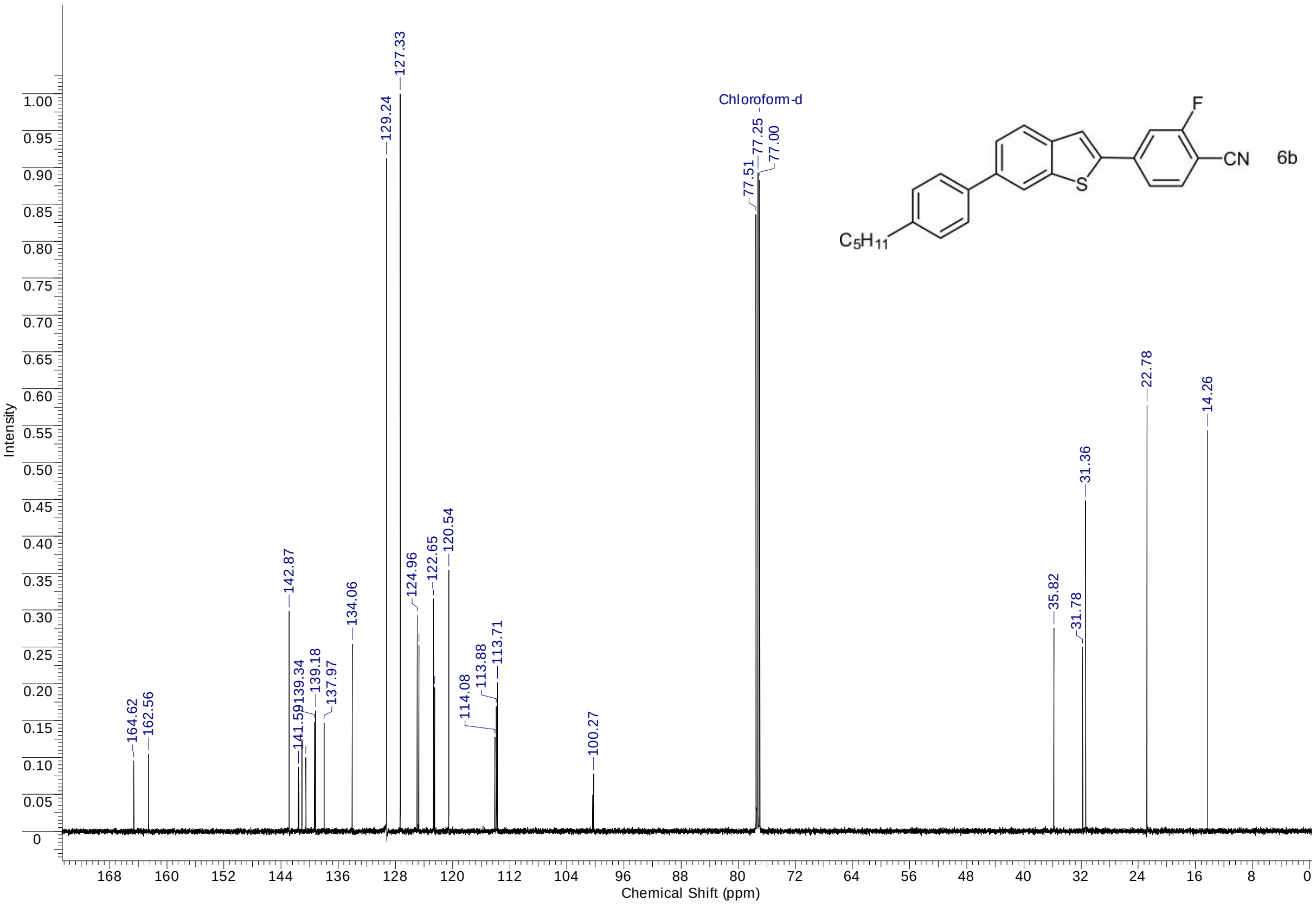

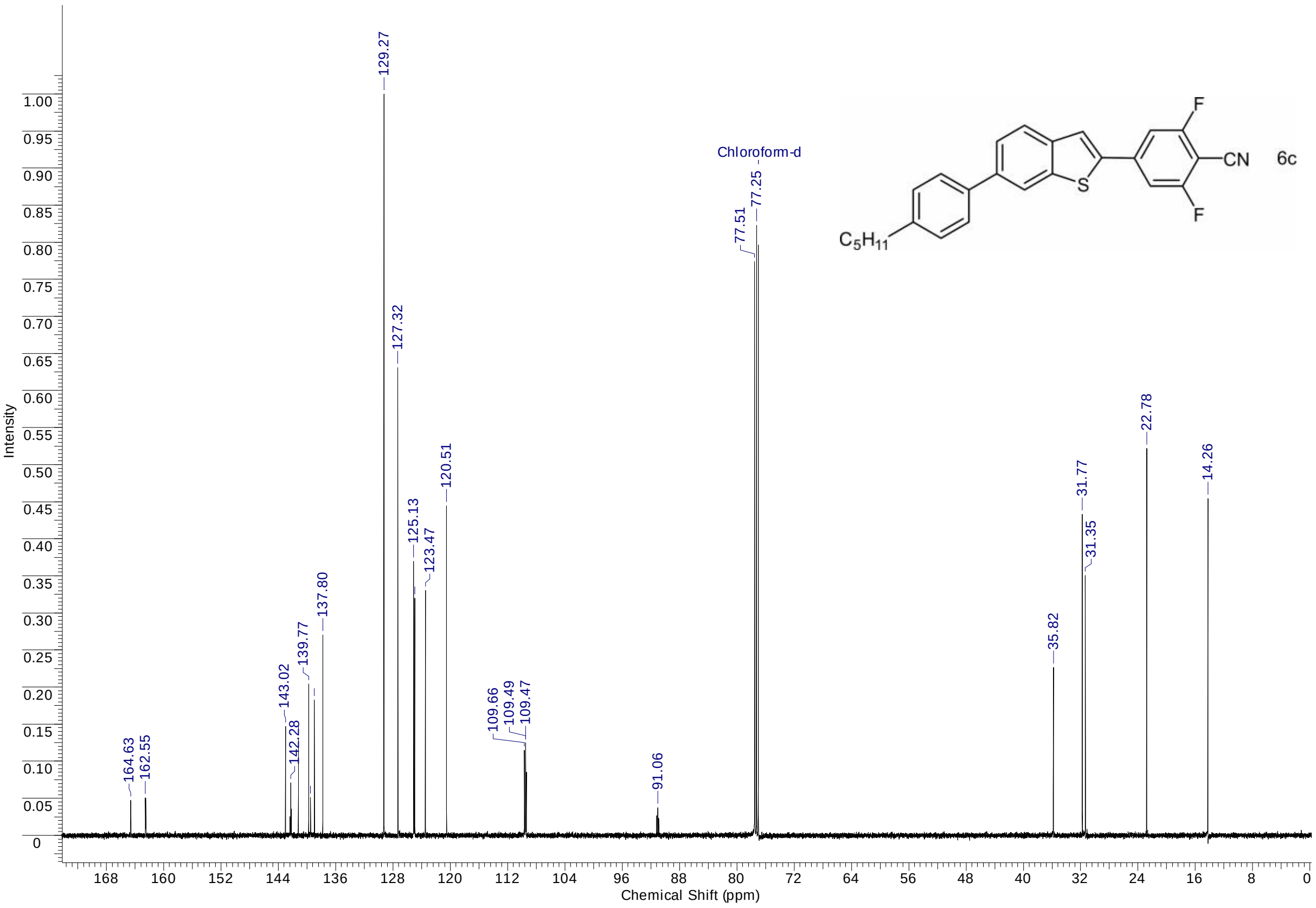

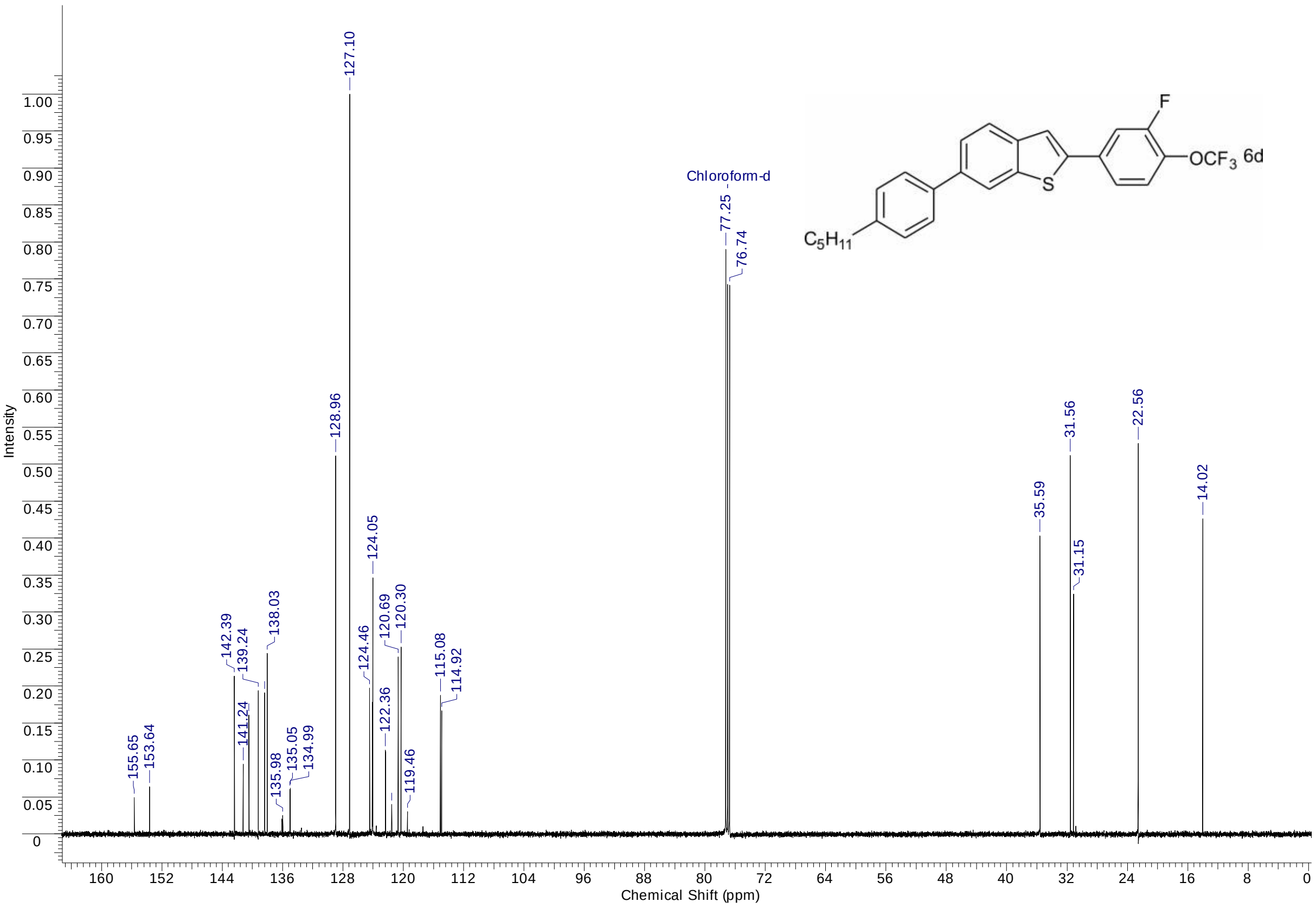

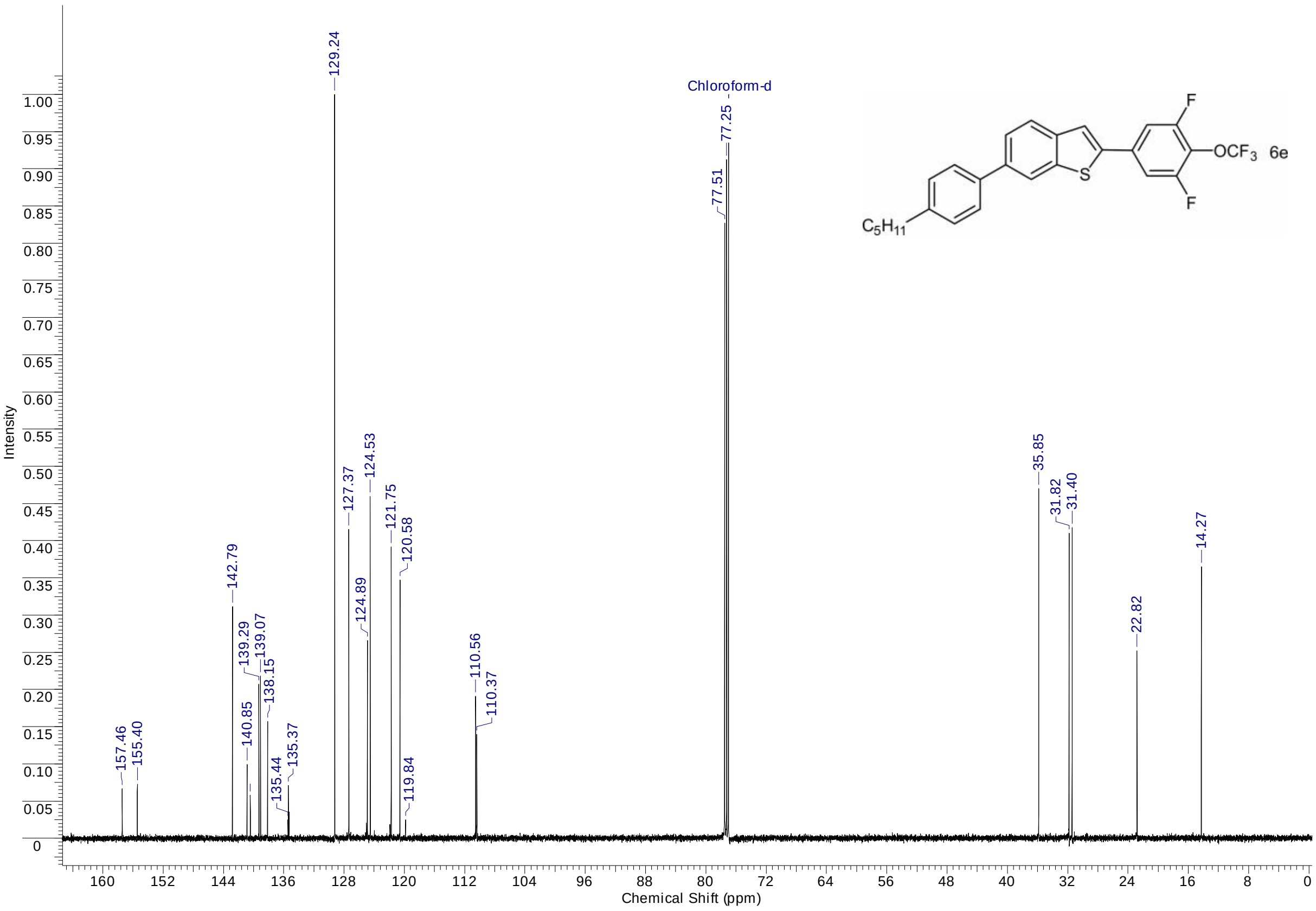

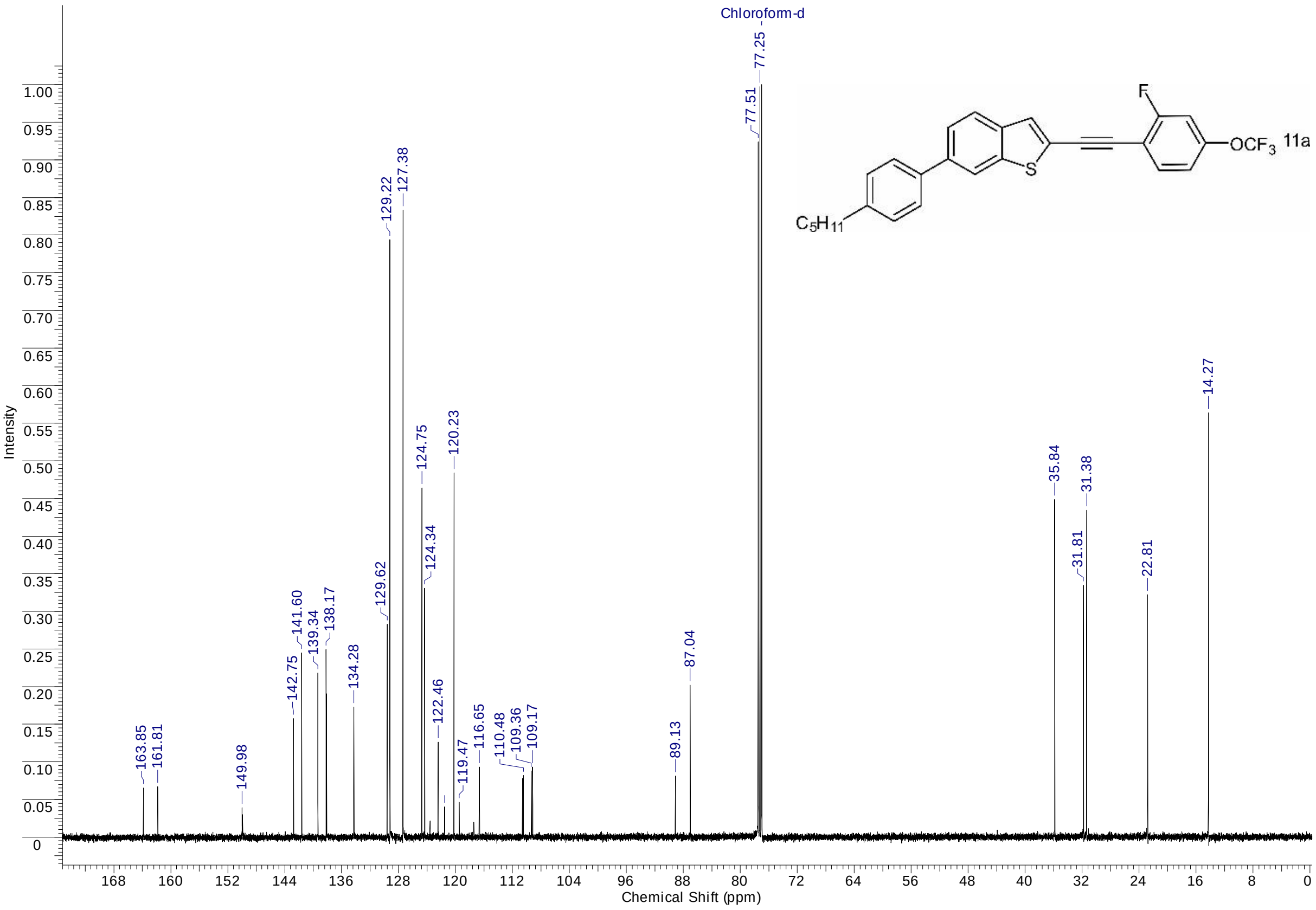

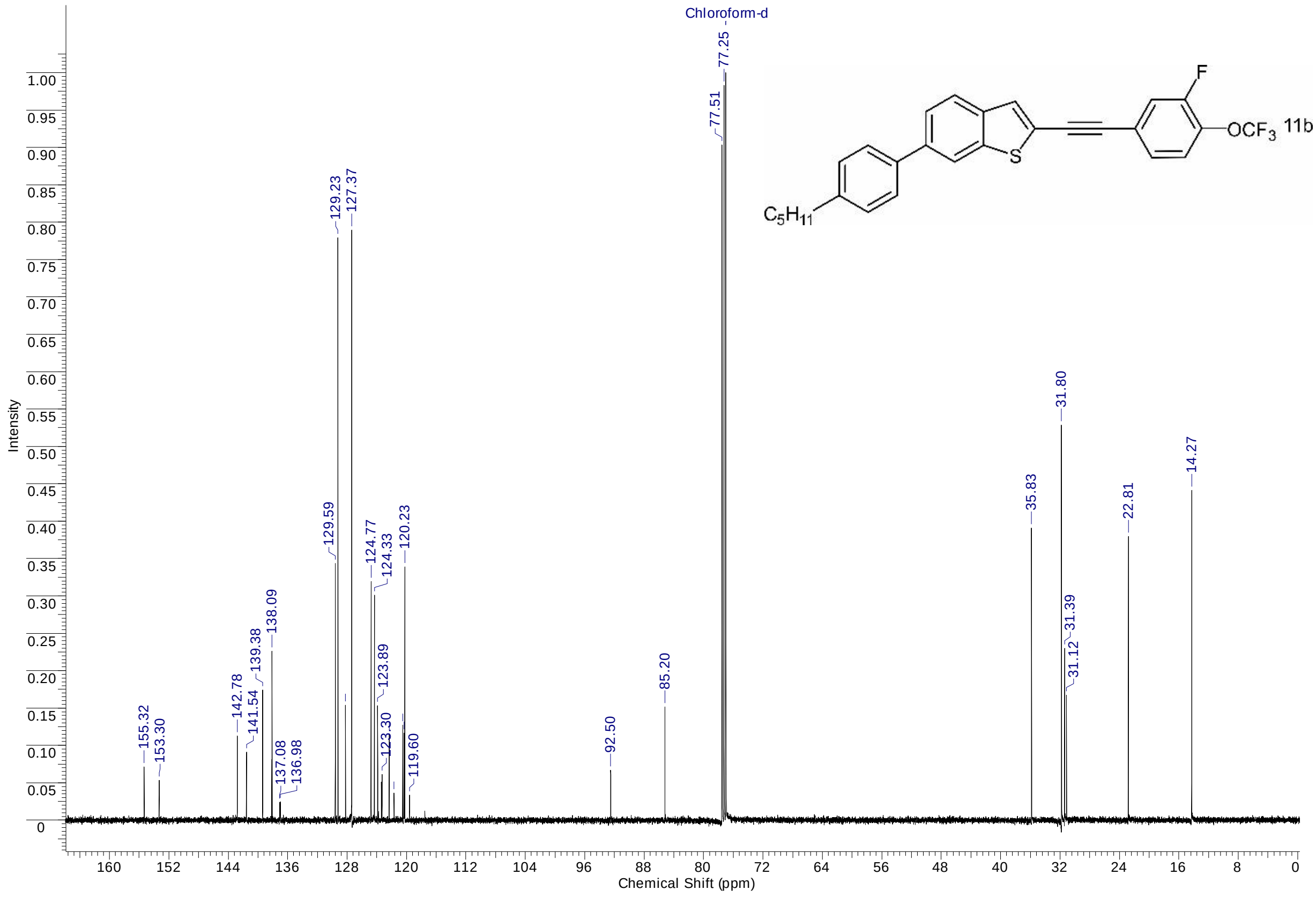

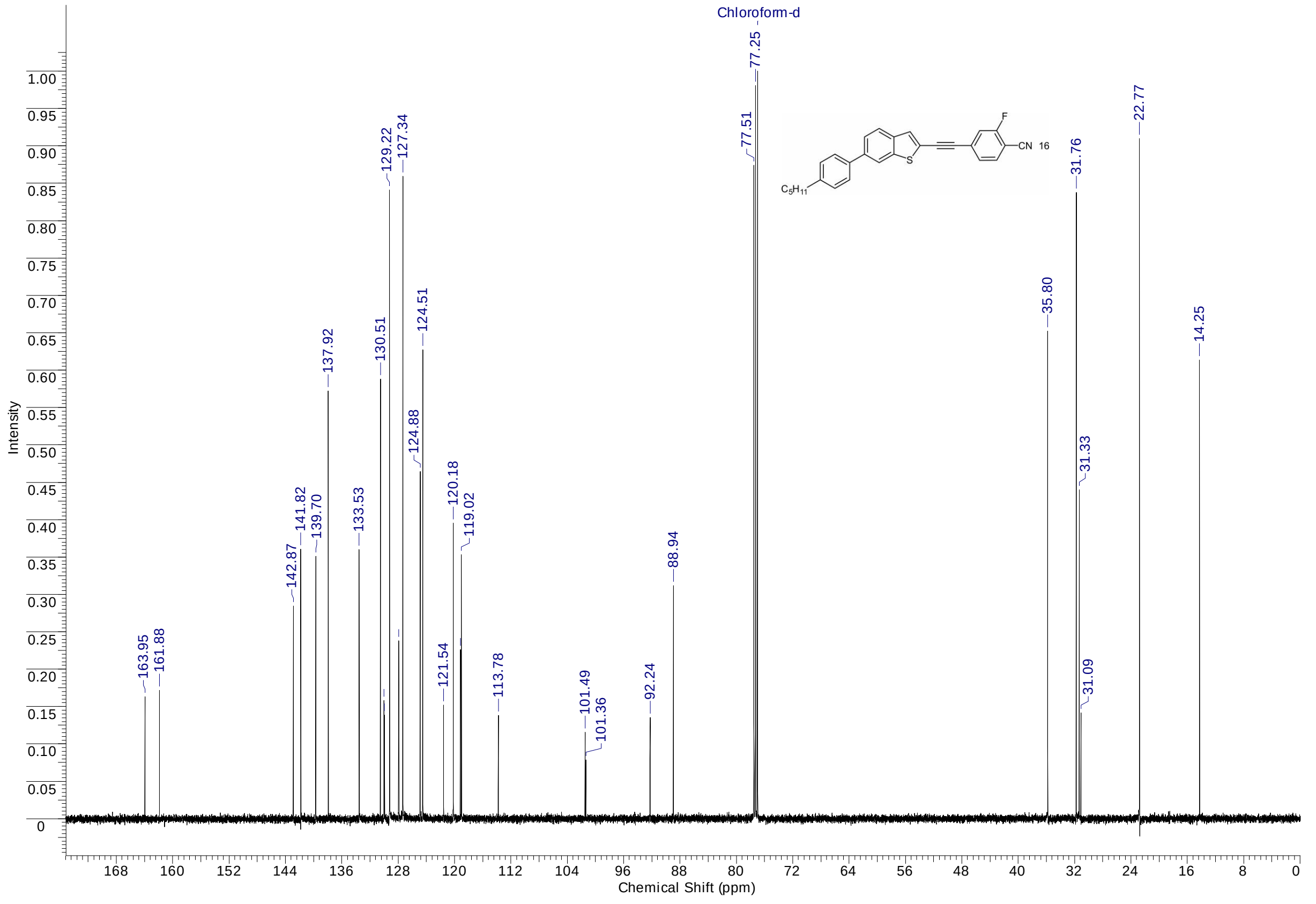

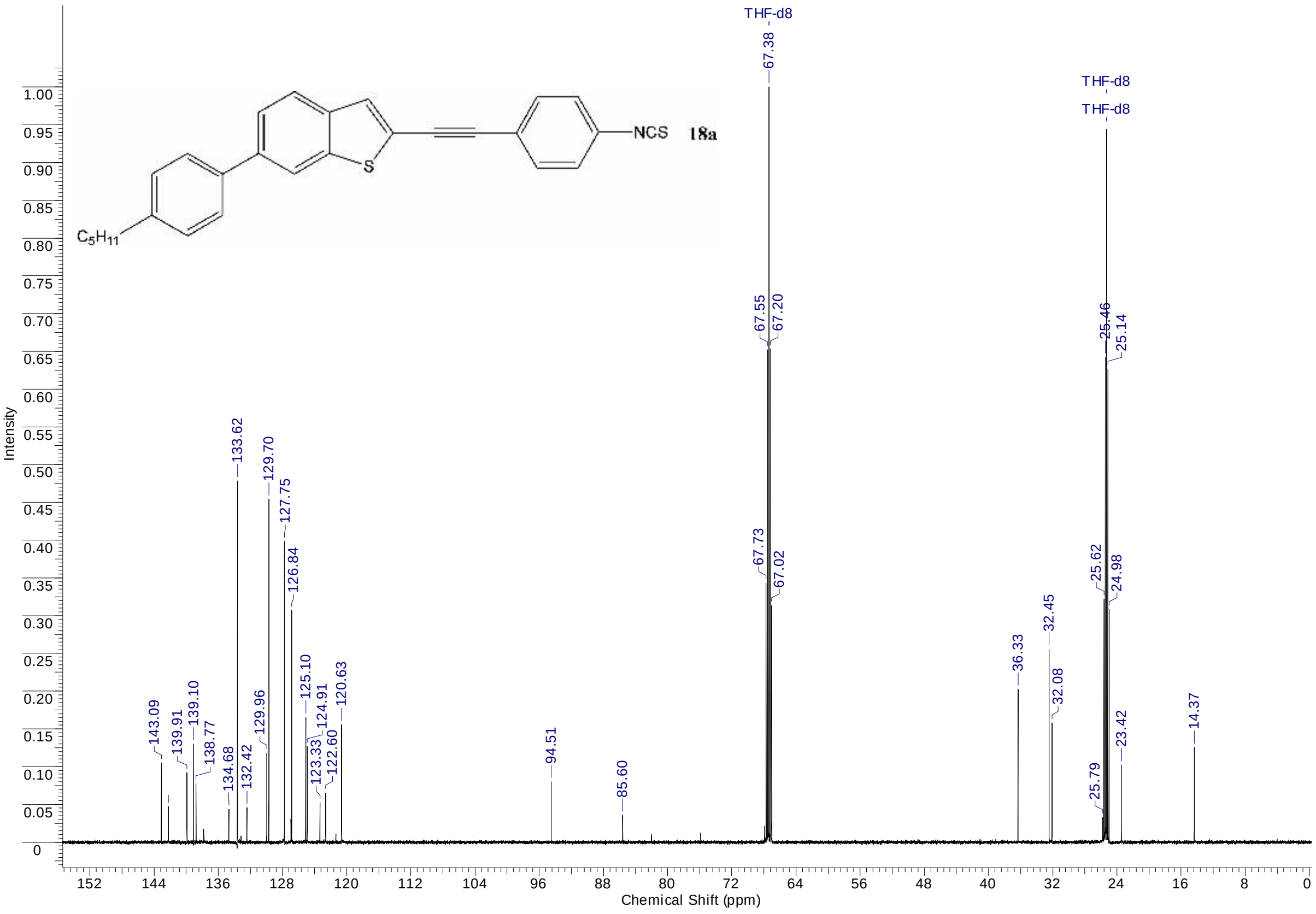

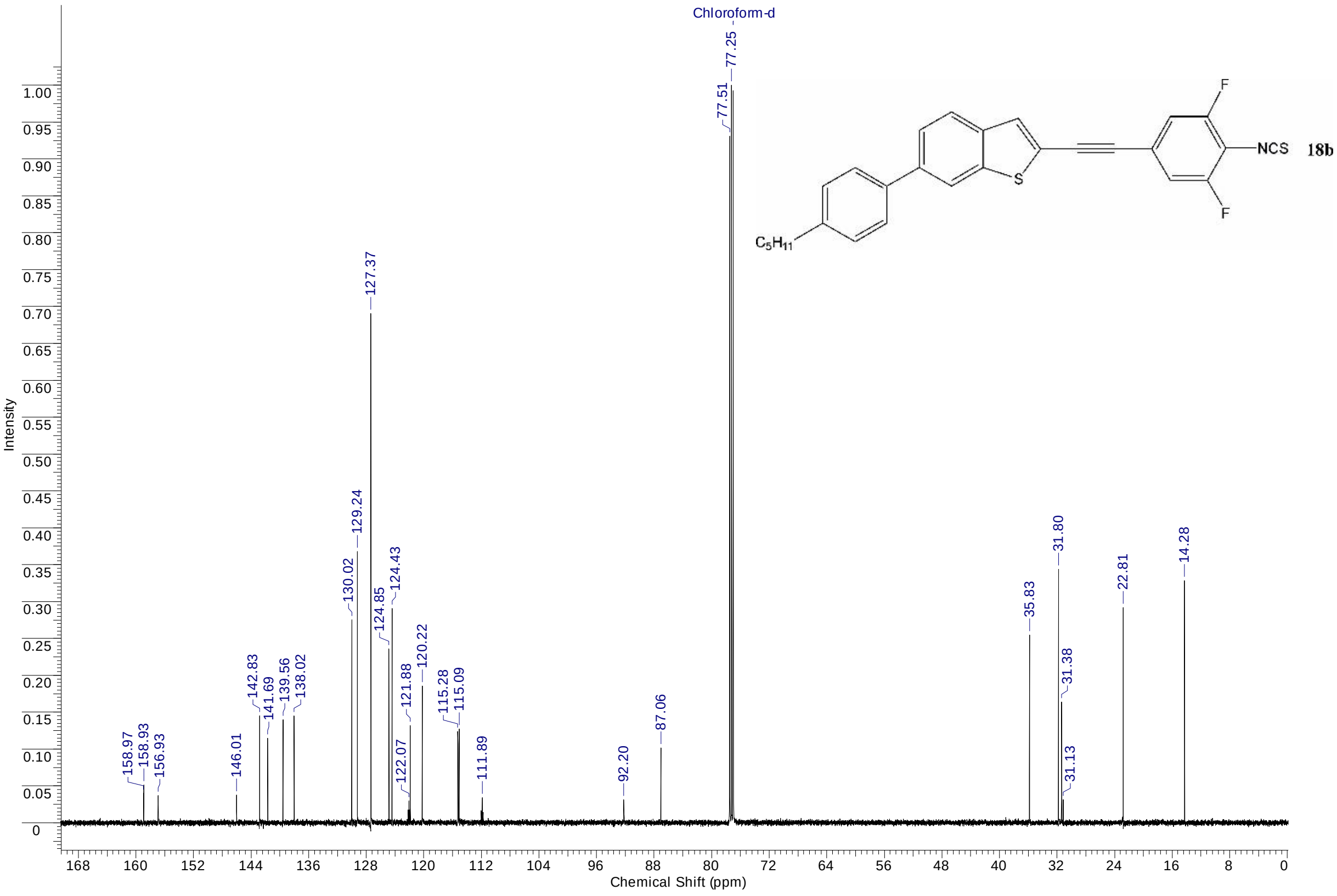

**DSC**

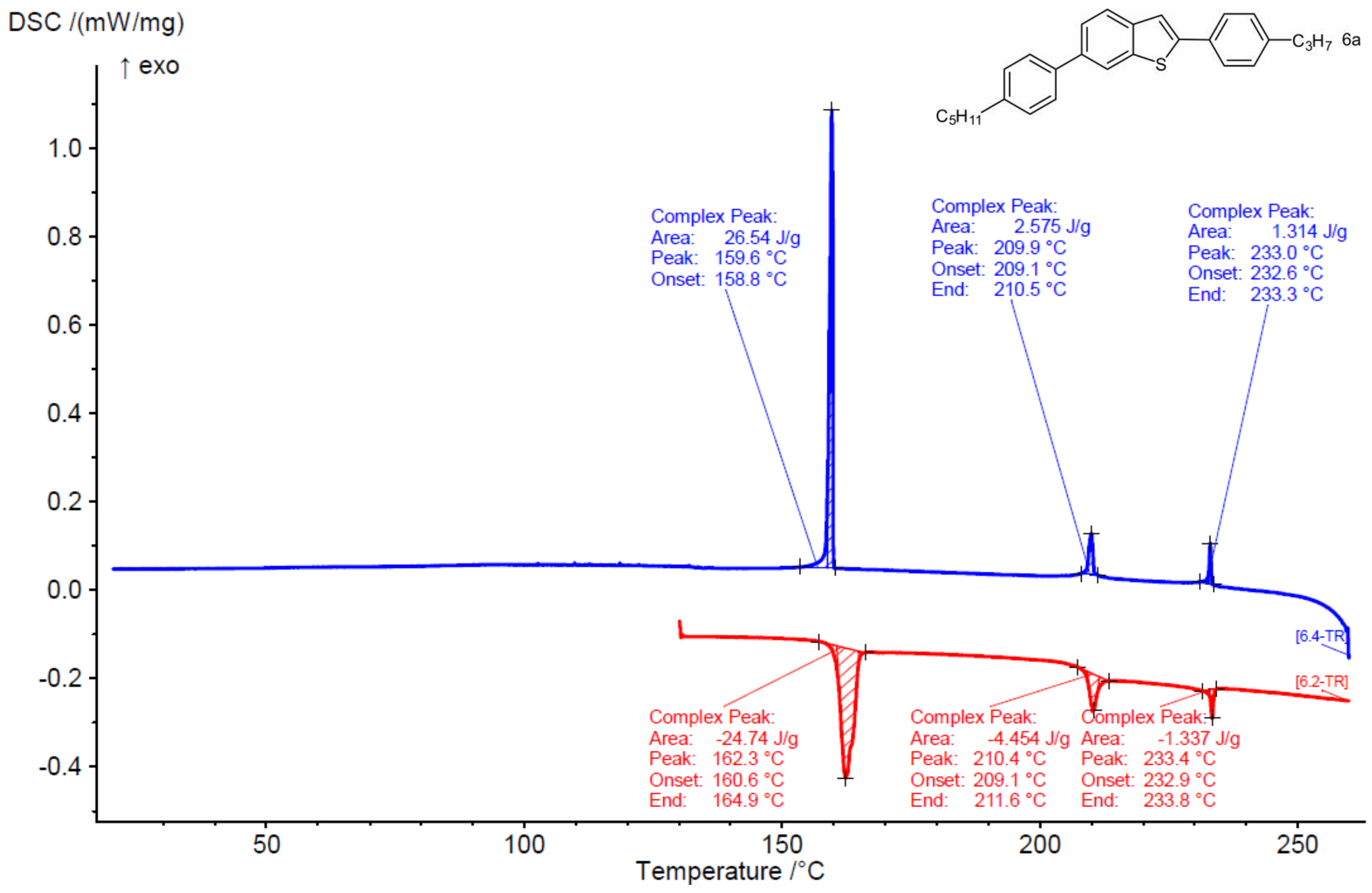

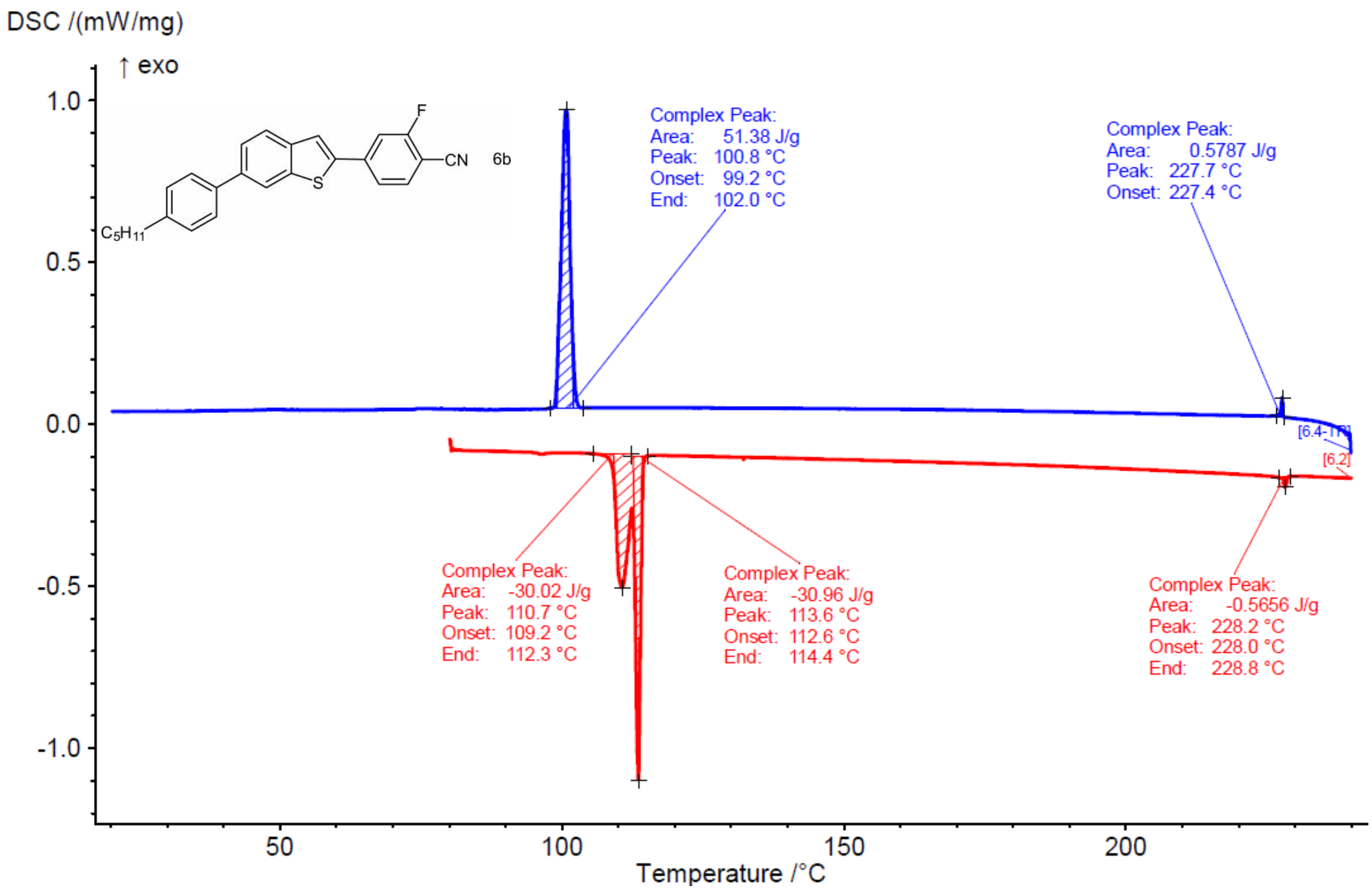

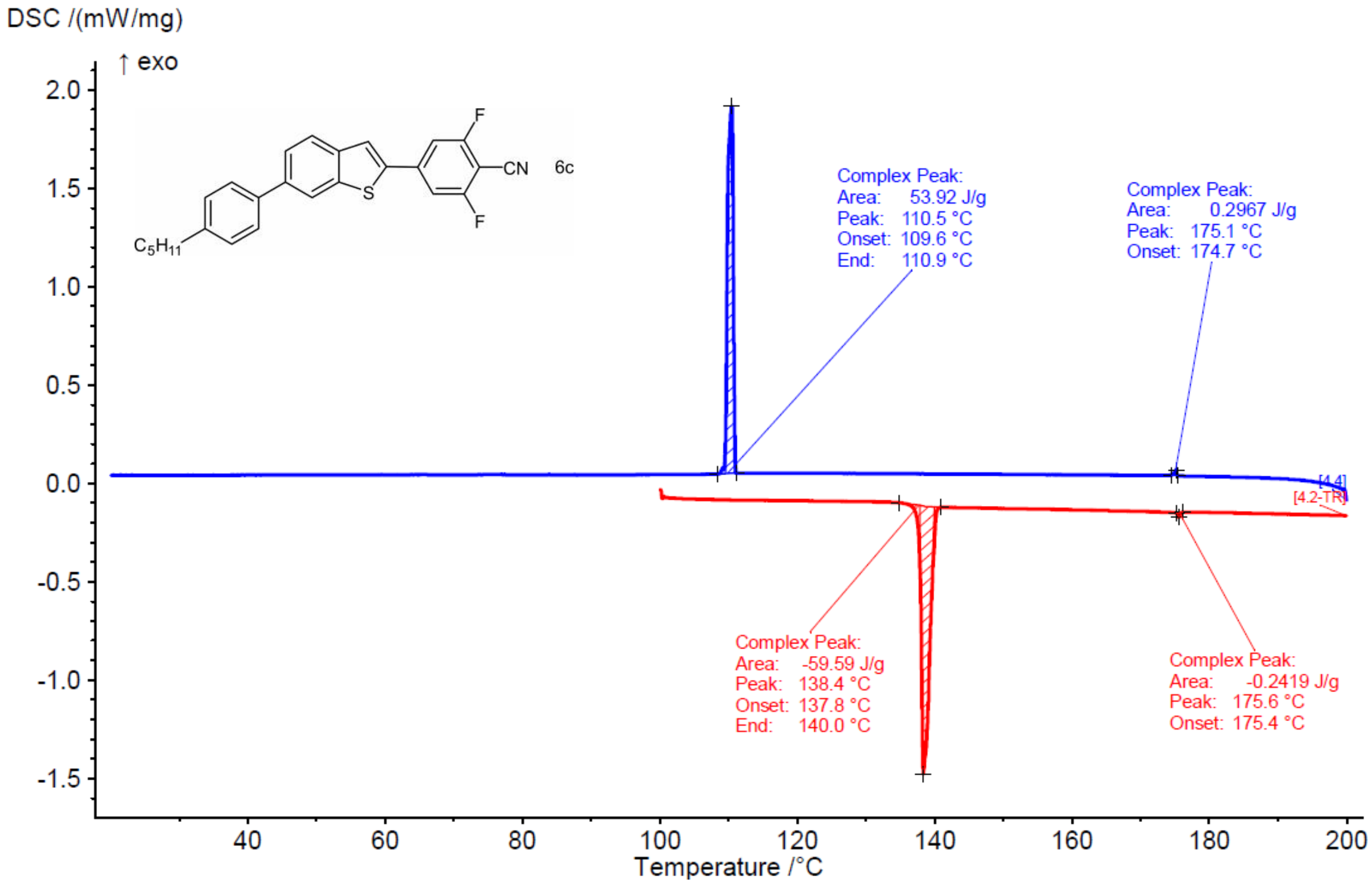

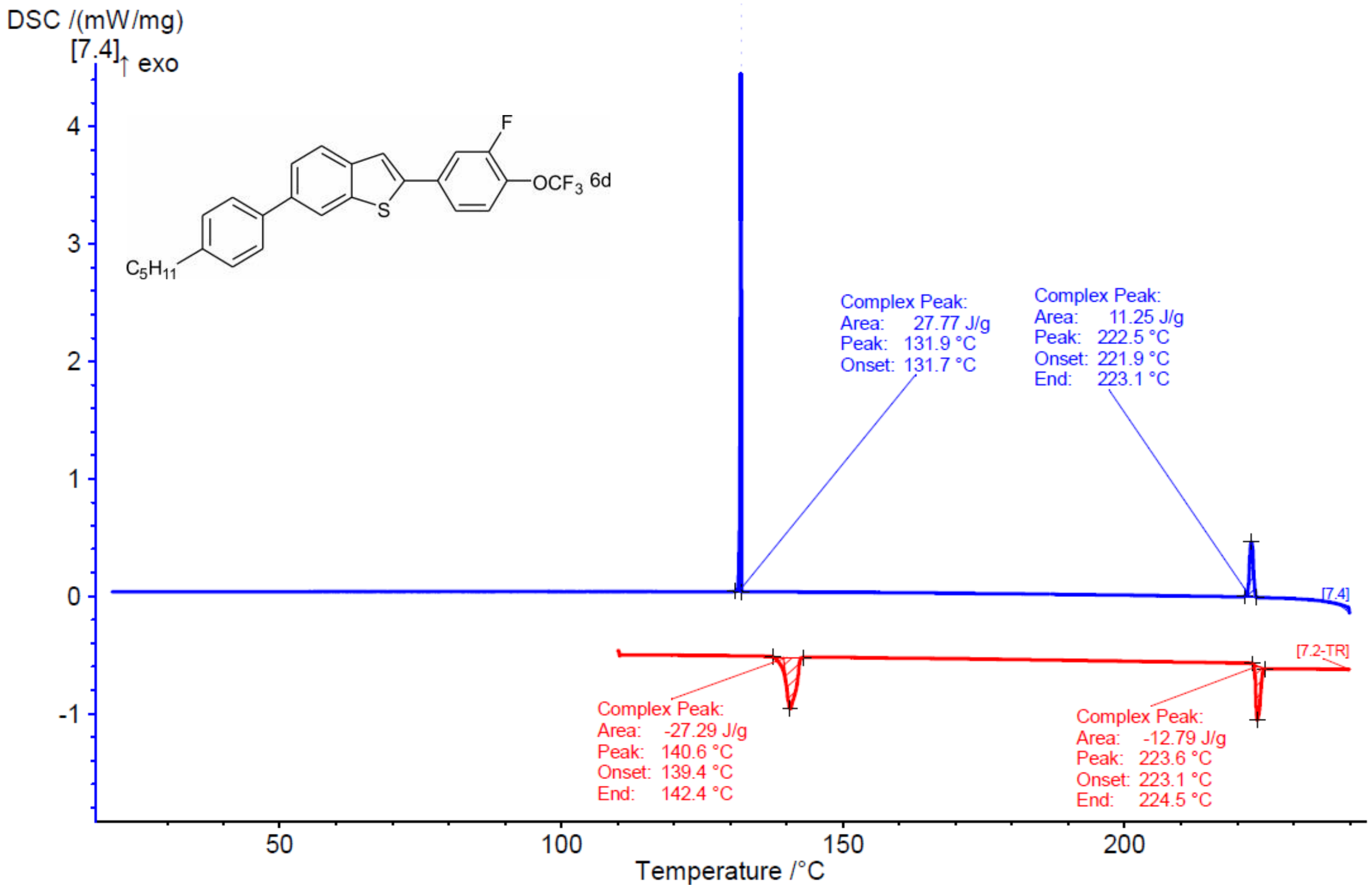

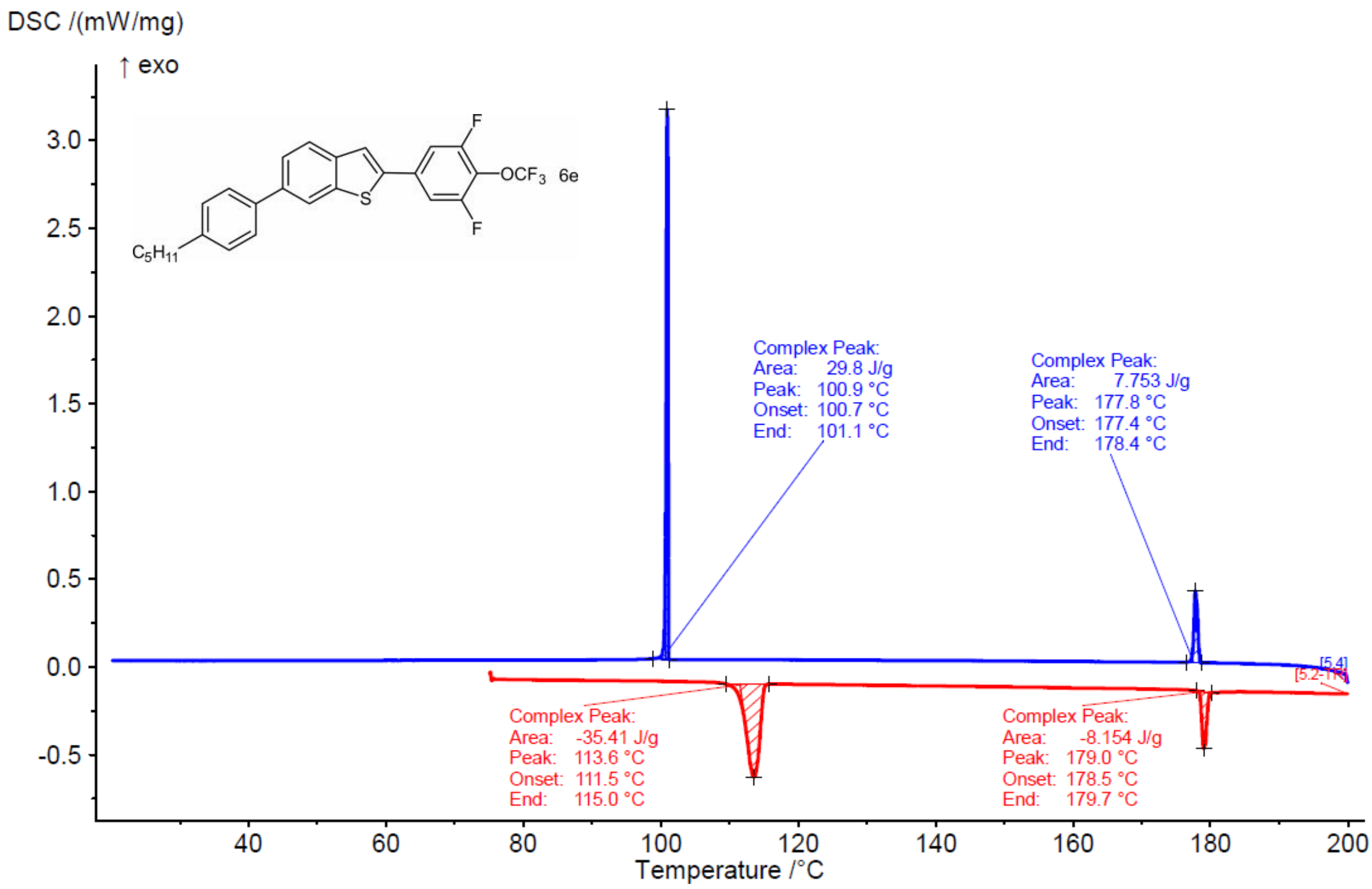

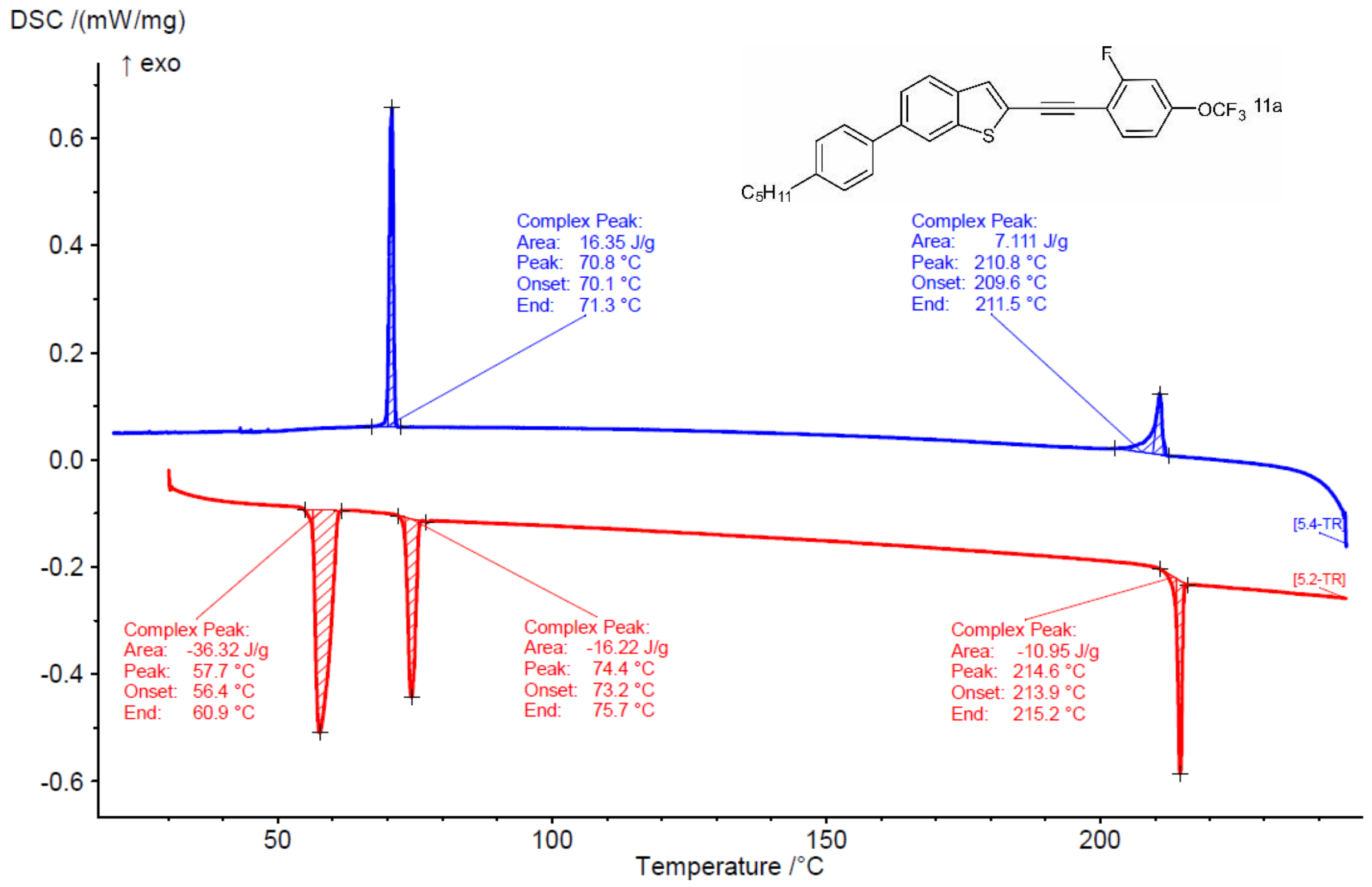

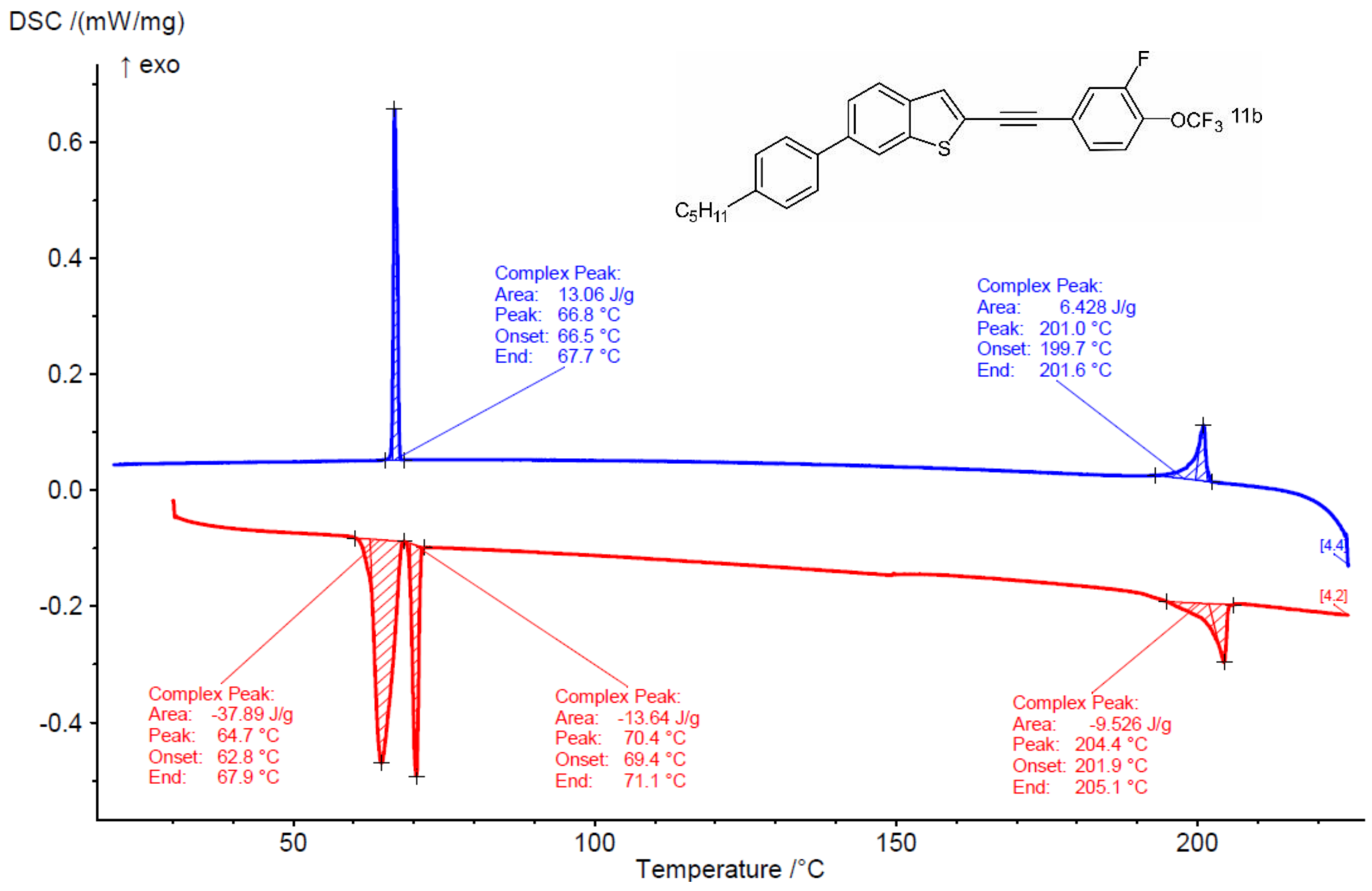

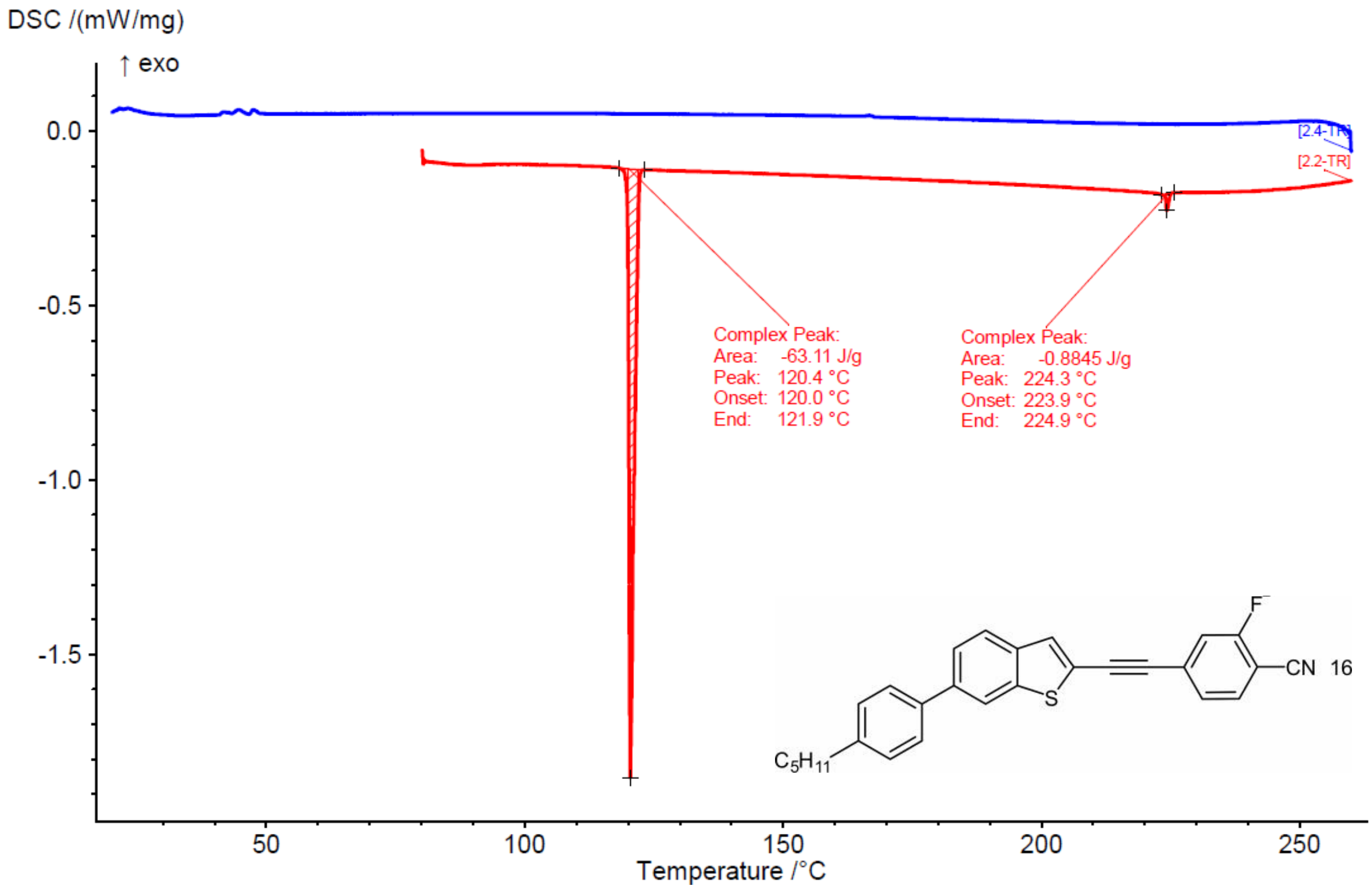

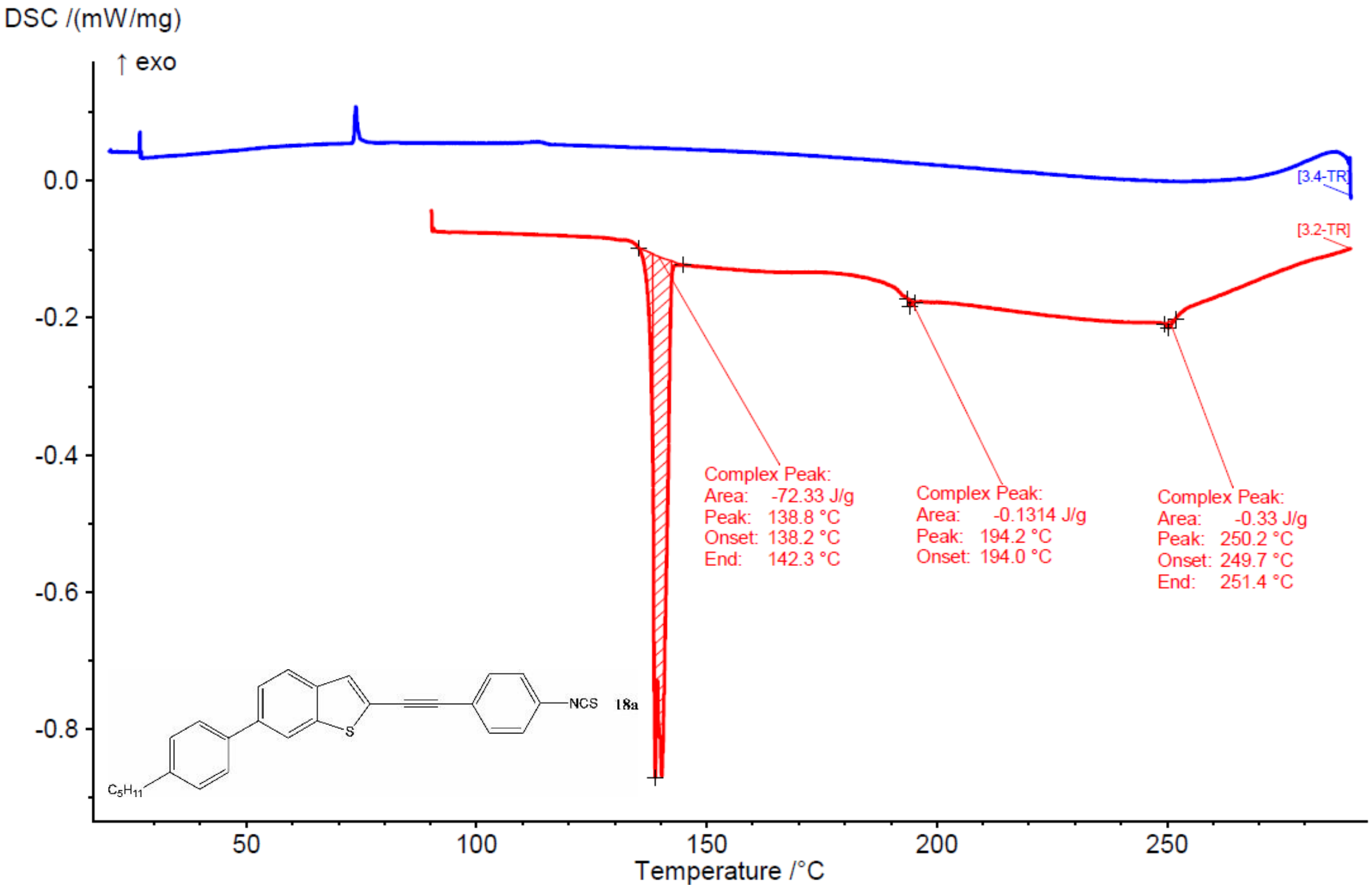

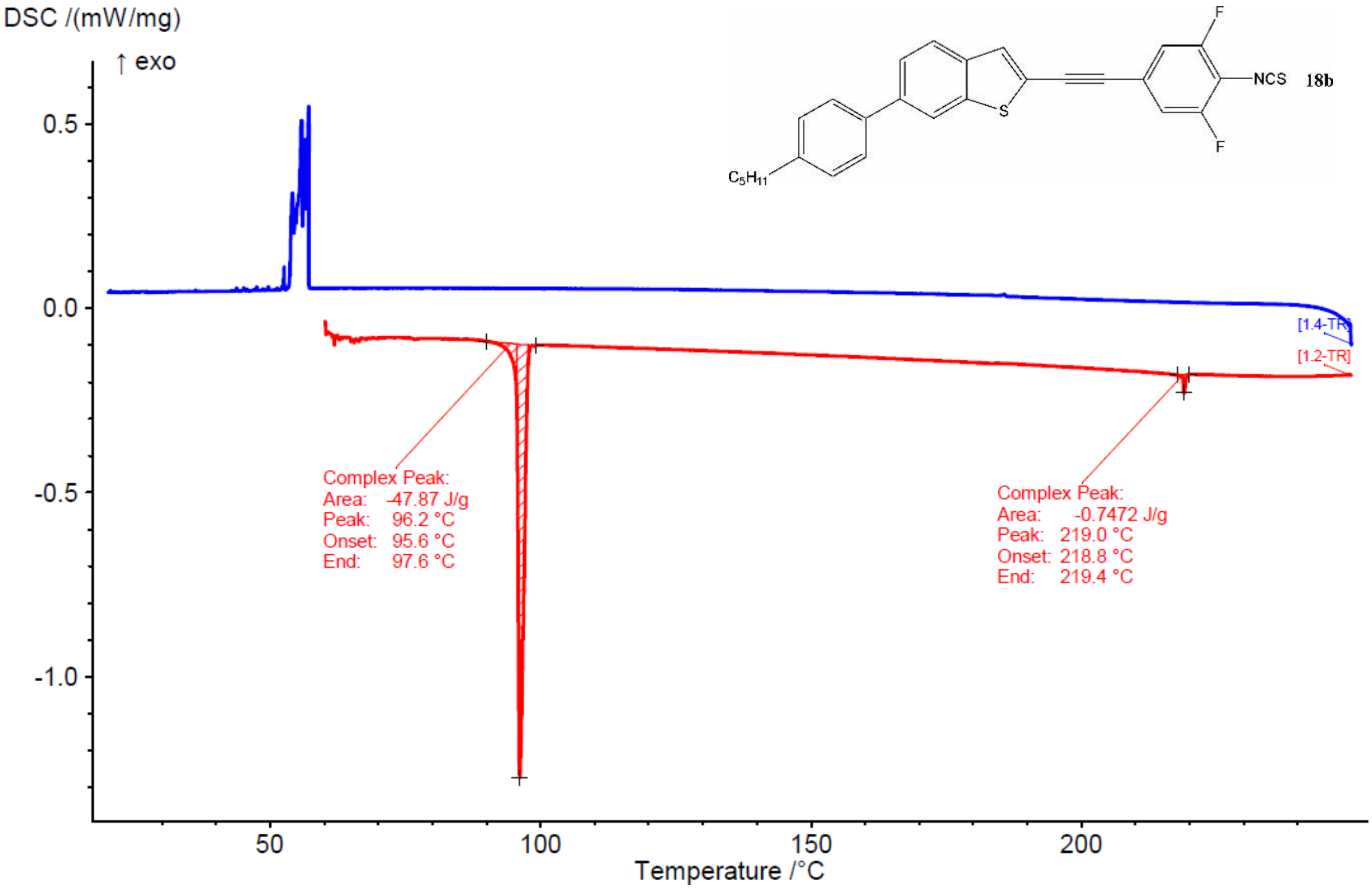

**Birefringence measurements**

| **Wavelength [nm]** | 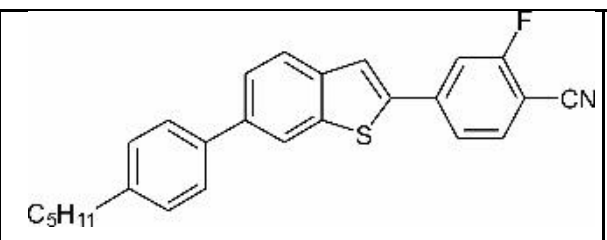 | 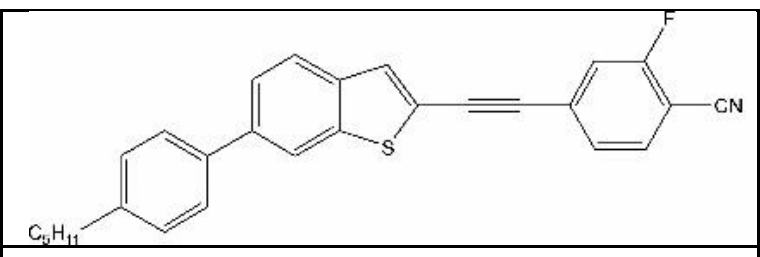 | 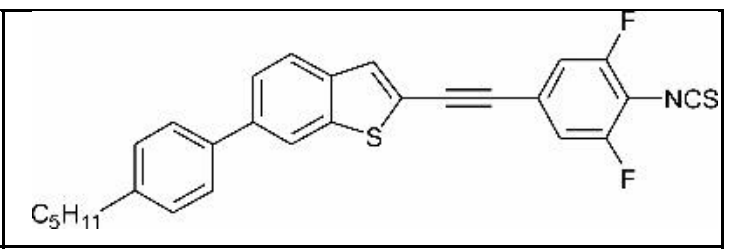 |
|---|---|---|---|
| 443 | $n_o = 1{,}6438$<br>$n_e = 2{,}4176$<br>$\Delta n = 0{,}6057$ | $n_o = 1{,}6554$<br>$n_e = 2{,}5247$<br>$\Delta n = 0{,}6788$ | $n_o = 1{,}6299$<br>$n_e = 2{,}4275$<br>$\Delta n = 0{,}7946$ |
| 636 | $n_o = 1{,}5737$<br>$n_e = 1{,}8606$<br>$\Delta n = 0{,}3431$ | $n_o = 1{,}5918$<br>$n_e = 2{,}0728$<br>$\Delta n = 0{,}4091$ | $n_o = 1{,}6042$<br>$n_e = 2{,}1658$<br>$\Delta n = 0{,}46$ |
| 1550 | $n_o = 1{,}5306$<br>$n_e = 2{,}8353$<br>$\Delta n = 0{,}2516$ | $n_o = 1{,}5566$<br>$n_e = 1{,}9451$<br>$\Delta n = 0{,}2834$ | $n_o = 1{,}4977$<br>$n_e = 1{,}9030$<br>$\Delta n = 0{,}3825$ |